\RequirePackage{rotating}
\documentclass[fleqn,usenatbib]{mnras}

\usepackage[T1]{fontenc}

\DeclareRobustCommand{\VAN}[3]{#2}
\let\VANthebibliography\thebibliography
\def\thebibliography{\DeclareRobustCommand{\VAN}[3]{##3}\VANthebibliography}

\usepackage{graphicx}	
\usepackage{amsmath}	
\usepackage{amssymb}	
\usepackage{newtxtext,newtxmath}
\usepackage{multirow}
\usepackage{lscape}
\usepackage{makecell}
\usepackage{subfigure}
\usepackage{float}
\usepackage{newtxtext,newtxmath}
\usepackage{lineno}

\usepackage{orcidlink}

\setcitestyle{notesep={ }}
\makeatletter
\newcommand{\labtext}[2]{%
  \@bsphack
  \csname phantomsection\endcsname 
  \def\@currentlabel{#1}{\label{#2}}%
  \@esphack
}

\setcitestyle{notesep={}}

\makeatother

\title[Impact of Gaps]{Distortions in Periodicity Analysis of Blazars II: The Impact of Gaps}
\author[P. Pe\~nil et al.]{
P. Pe\~nil,$^{1}$\orcidlink{0000-0003-3741-9764}\thanks{E-mail: ppenil@clemson.edu}
N. Torres$-$Alb\`a,$^{2, 3}$\orcidlink{0000-0003-3638-8943}\thanks{E-mail: nuria@virginia.edu}
A. Rico,$^{1}$\orcidlink{0000-0001-5233-7180}\thanks{E-mail: aricoro@clemson.edu}
S. Buson,$^{4, 5}$\orcidlink{0000-0002-3308-324X}
M. Ajello,$^{1}$\orcidlink{0000-0002-6584-1703}
A. Dom\'inguez,$^{6}$\orcidlink{0000-0002-3433-4610}
S. Adhikari,$^{1}$\orcidlink{0009-0006-1029-1026}
\\
$^{1}$Department of Physics and Astronomy, Clemson University, Kinard Lab of Physics, Clemson, SC 29634-0978, USA\\
$^{2}${GECO Fellow} \\
$^{3}$Department of Astronomy, University of Virginia, P.O. Box 400325, Charlottesville, VA 22904, USA \\
$^{4}$Deutsches Elektronen-Synchrotron DESY, Platanenallee 6, 15738 Zeuthen, Germany\\
$^{5}$Julius-Maximilians-Universit\"at W\"urzburg, Fakultät f\"ur Physik und Astronomie, Emil-Fischer-Str. 31, D-97074 W\"urzburg, Germany\\
$^{6}$IPARCOS and Department of EMFTEL, Universidad Complutense de Madrid, E-28040 Madrid, Spain\\
}

\date{Accepted 2025 October 20. Received 2025 October 4; in original form 2025 August 15}

\pubyear{2025}

\begin{document}
\label{firstpage}
\pagerange{\pageref{firstpage}--\pageref{lastpage}}
\maketitle
\begin{abstract}
Time series analysis is fundamental to characterizing the variability inherent in multi-wavelength emissions from blazars. However, a major observational challenge lies in the need for well-sampled, temporally uniform data, which is often hindered by irregular sampling and data gaps. These gaps can significantly affect the reliability and accuracy of methods used to probe source variability. This paper investigates the impact of such observational gaps on time series analysis of blazar emissions. To do so, we systematically evaluate how these gaps alter observed variability patterns, mask genuine periodic signals, and introduce false periodicity detections. This evaluation is conducted using both simulated and real observational data. We assess a range of widely used time series analysis methods, including the Lomb-Scargle periodogram, Phase Dispersion Minimization, and the recently proposed Singular Spectrum Analysis (SSA). Our results demonstrate a clear and significant degradation in period detection reliability when the percentage of gaps exceeds 50\%. In such cases, the period-significance relationship becomes increasingly distorted, often leading to misleading results. Among the tested methods, SSA stands out for its ability to yield consistent and robust detections despite high data incompleteness. Additionally, the analyzed methods tend to identify artificial periodicities of around one year, likely due to seasonal sampling effects, which can result in false positives if not carefully recognized. Finally, the periods detected with $\geq$3$\sigma$ confidence are unlikely to result from stochastic processes or from the presence of gaps in the analyzed time series.  
\end{abstract}


\section{Introduction}\label{sec:intro}

Blazars \citep[e.g.,][]{cavaliere_1989, wiita_lecture}, a subclass of active galactic nuclei \citep[AGNs, see e.g.,][]{urry1995}, are known for their intense emissions across the electromagnetic spectrum, from radio waves to high-energy gamma rays \citep[e.g.,][]{urry_variability, penil_mwl_pg1553}. These emissions are often characterized by extreme variability \citep[e.g.,][]{Giannios2009, otero_3C_371} and, in some cases, they have been proposed to display periodic behavior \citep[e.g.,][]{ackermann_pg1553, penil_2020}. Such periodicity can provide valuable insights into the internal processes of these objects, such as, for instance, the potential presence of a binary system of supermassive black holes (SMBH)  \citep[e.g.,][]{dey2018authenticating, Adhikari2023}, phenomena associated with the intrinsic properties of the jet \citep[e.g.,][]{villata_helical_jet, sarkar_curved_jet}, or with an origin from the accretion disk \citep[e.g.][]{gracias_modulation_disk, franchini_lense}. However, these patterns may also arise from stochastic phenomena that can mimic periodic emissions, which is an obstacle for accurate and certain detection \citep[e.g.,][]{covino_negation, rieger_2019}. 

Periodicity analysis in blazar data faces several challenges. One prominent issue is red noise, a correlated noise common in astrophysical signals that can distort the identification of true periodic signals \citep[][]{vaughan_2003}. Red noise often produces long-term trends or low-frequency variations that can mimic periodic behavior, increasing the risk of false detections \citep[][]{vaughan_criticism}. Another common phenomenon in blazars is stochastic flaring \citep[e.g.][]{Resconi2009, Nalewajko2013, Liodakis2018}, which can occur on top of an underlying periodic signal. In a recent study, we investigated the impact of flares on periodicity detection. We found that intense flaring events can distort periodic signals, either by mimicking periodic behavior or by diminishing the statistical significance of a genuine periodic signal, even to the point of completely disrupting it \citep[][]{penil_flares_2025}.

Another critical challenge in the periodicity analysis of blazar emissions lies in the presence of data gaps. These gaps may arise from observational constraints (such as telescope scheduling or poor weather conditions), the presence of upper limits (ULs) in the signal, or other interruptions in data collection. These situations can affect the reliability of periodicity detection, potentially masking true periodic signals or introducing spurious ones. Addressing this challenge requires robust statistical techniques and careful examination of observational data to ensure valid conclusions, without manipulating the data in ways that could artificially enhance or produce false periodicities. This study aims to assess the impact of data gaps on periodicity searches in blazar emission data. By simulating data gaps of varying durations and frequencies, we analyze their influence on common periodicity detection methods such as the Lomb-Scargle periodogram (LSP), based on Lomb's algorithm \citep{lomb_1976} and Scargle's extension \citep{scargle_1982}, and Phase Dispersion Minimization (PDM), which analyzes phase variations in time series data \citep{pdm_stellingwerf}. Additionally, we explore the capabilities of the Singular Spectrum Analysis \citep[SSA, ][]{ssa_greco, SSA_algorithm}, a method that has been recently proposed to be more robust than others when searching for periodicity in blazars \citep[][]{alba_ssa, penil_flares_2025}. Furthermore, we explore potential mitigation strategies to improve the robustness of periodicity analysis in the presence of incomplete datasets. Specifically, we employ the approach of \citet{Adhikari2023, Adhikari2024}, consisting of simulating pure noise light curves (LCs) with the same gap structure as the original LC, to study the potential stochastic origin of the period-significance relationship obtained. 

By applying various gap scenarios, we aim to quantify how specific patterns of data interruption distort periodic signals and to identify conditions under which periodicity detection remains reliable. The findings of this study are expected to enhance data analysis methodologies in astrophysical research, particularly in the context of blazar variability studies.

The paper is structured as follows. Section \textsection{\ref{sec:gaps_in_lcs}} presents a study of the ULs in the $\gamma$-ray emissions of the  \textit{Fermi}-LAT AGNs (Large Area Telescope). Section \textsection{\ref{sec:methodology}} explains the methodology used to assess the impact of gaps on the search for periodicity. In Section \textsection{\ref{sec:results}}, we present the different tests performed on the impact of the gaps according to such tests. In Section \textsection{\ref{sec:usecase}}, we analyze real $\gamma$-ray data from a sample of blazars and conduct a detailed study based on the test results. Finally, we conclude with a summary of the primary findings and conclusions in Section \textsection{\ref{sec:summary}}.

\section{Gaps in Lightcurves} \label{sec:gaps_in_lcs}

In this study, we focus on $\gamma$-ray data, as our prior analyses have primarily been conducted within this wavelength domain \citep[e.g.,][]{penil_2020, alba_ssa, penil_flares_2025, penil_2022}. This choice ensures methodological consistency in our periodicity analysis, since it is essential to evaluate the statistical properties, such as gap distribution, within the same observational domain. Using $\gamma$-ray LCs ensures that our simulation setup reflects the real observational conditions under which periodic signals are being searched, thereby enhancing the relevance and reliability of our tests.

To design our tests, we first analyze the distribution of ULs in real-time-series data, treating them as missing data and, consequently, as gaps in the LCs. Specifically, we examine the sample analyzed in \citet{penil_trends} to infer the gap distribution in real gamma-ray LCs of blazars. Our dataset consists of 3,300 \textit{Fermi}-LAT LCs, including AGN listed in the Fourth \textit{Fermi}-LAT AGN Catalog \citep[4FGL,][]{4fgl_catalog}, enabling us to characterize the frequency and extent of missing data. The LCs of such samples are 28-day binned LCs above an energy of 0.1~GeV, the same time bin LCs as used in \citet{alba_ssa}, \citet{penil_flares_2025}, and \citet{penil_2022}. The 28-day binning allows us to capture and analyze long-term variations in the $\gamma$-ray data while reducing the impact of short-term variability. The exposure time is approximately 12 years, from August 2008 to December 2020. 

\begin{figure*}
	\centering
	\includegraphics[scale=0.22]{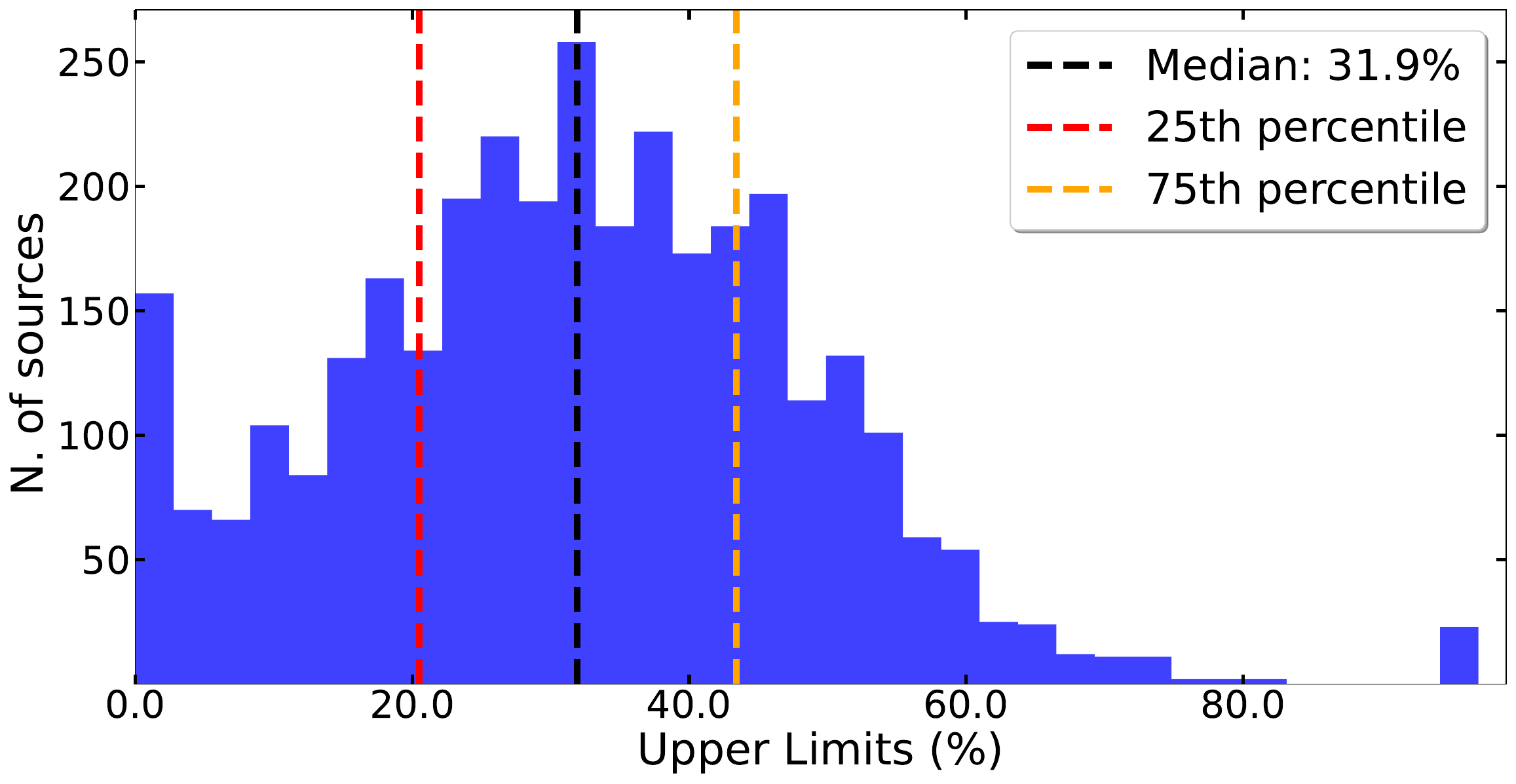}
        \includegraphics[scale=0.22]{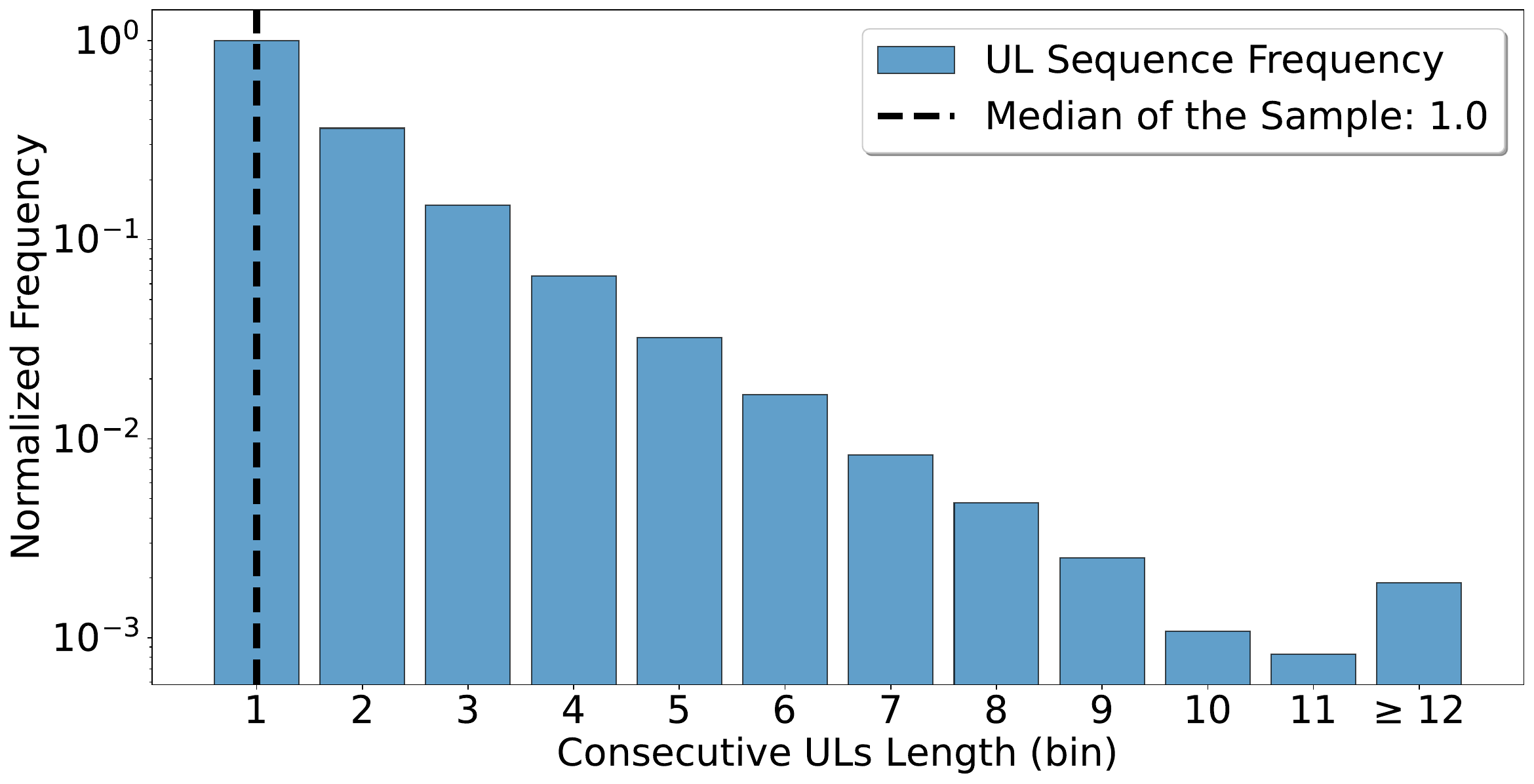}
	\caption{\textit{Right}: Distribution of the number of ULs (interpreted as observational gaps in this paper) of the sample of 3,300 AGN analyzed in \citet{penil_trends}. The median of this distribution is $\approx$32\%, denoted by the dotted vertical line. The majority of the objects are in the range 20\%-50\% (denoted by the 25th and 75th percentiles). \textit{Left}: Distribution of consecutive gaps of the LCs in our sample, using a bin size of 28 days. The histogram illustrates that the majority of ULs appear as isolated points, with a maximum observed consecutive UL sequence of 161. Additionally, most UL sequences remain short, while consecutive UL lengths of $\geq$12 are grouped in the final bin to highlight extended sequences.} \label{fig:gaps distribution}
\end{figure*}

The resulting gap distribution is presented in Figure~\ref{fig:gaps distribution}. The analysis shows that the median percentage of gaps (i.e., missing data points, typically UL) is $\approx$32\%. Most sources exhibit a percentage of gaps between 0\% and 50\%, indicating that moderate levels of gaps are common in $\gamma$-ray LCs. There is a distinct peak at 0\%, which corresponds to bright sources that are consistently detected in each 28-day bin. On the other end of the distribution, a smaller subset of sources shows gap percentages approaching 90\%, reflecting cases of faint variable blazars, for which detections likely only happen when they flare. This heterogeneous distribution of gap percentages highlights the importance of accounting for gap presence when evaluating the robustness of periodicity detection methods.

Based on these findings, our simulations incorporate a fraction of artificially induced gaps that spans this observed distribution. Specifically, we simulate missing data at percentages ranging from 10\% to 90\% to evaluate the impact of gaps on periodicity search methods. 

However, gaps are not necessarily randomly distributed, as a source that goes into a more quiescent state may do so for longer than approximately one month (which is the temporal binning of our LCs). Therefore, Figure \ref{fig:gaps distribution} presents the distribution of consecutive ULs (gaps) in the same sample. The results indicate that the majority of ULs appear as isolated points, with most consecutive UL sequences falling within the range of 1 to 7 ULs. The maximum observed sequence of 161 consecutive ULs (out of a total 166 points) suggests again that some weaker sources are only detected when flaring. For these sources, no periodicity analysis can be done, with previous studies often removing sources with more than 50\% of ULs, ensuring sufficient signal for a meaningful analysis.  

\section{Methodology} \label{sec:methodology}

Assessing the influence of a given factor on a property requires controlled scenarios. Here, we present a similar approach to that applied in \citet{penil_flares_2025} to evaluate the impact of flares in observed periods.

We conduct a series of tests using LCs that incorporate a range of intrinsic properties (i.e., periodic and non-periodic; different periods) and different types of gaps. By simulating various scenarios, ranging from randomly distributed gaps to more structured patterns, we aim to assess how the presence and nature of the gaps affect the reliability of periodicity detection. This approach helps us to systematically evaluate the extent to which observational gaps can distort or mask periodic signals, resulting in fake detections. 

\subsection{Type of LCs}

The tests are performed using two different types of artificial LCs. In the first case, we analyze purely stochastic LCs exhibiting red-noise properties. To simulate this, we generate the red-noise signals using the method proposed by \citet{timmer_koenig_1995} (hereafter, TK). This approach generates red noise with a given power spectral density (PSD), allowing us to produce stochastic variability.

The second type of LCs are sinusoidal signals contaminated by random noise according to the model: 

\begin{equation} \label{eq:model_fitting}
\phi(t) = O + A\sin \bigg(\frac{2 \pi t}{T} + \theta \bigg).
\end{equation}

The parameters considered in our methodology include offset ($O$), amplitude ($A$), period ($T$), and phase ($\theta$). Since our goal is to evaluate different types of time series, such as those associated with $\gamma$-ray or optical data, we characterize the sinusoidal signal using generic values. Specifically, we set the offset to $6\times 10^{-8}$ ph cm$^{-2}$ s$^{-1}$ and the amplitude to $5.5\times 10^{-8}$ ph cm$^{-2}$ s$^{-1}$, which corresponds approximately to the values of the LC of PG 1553+113 \citep[][]{penil_2020, penil_flares_2025}. The phase is a random value in the range [-$\pi$, $\pi$]. Regarding the period, we consider two values: 2 and 3 years, as most of the blazars in the \citet{penil_2020} and \citet{alba_ssa} samples exhibit periodicities falling within the 2$-$3~yr range. To adapt to the real data, we examine 5-6 cycles for the 2-year period and 3-4 cycles for the 3-year period. Therefore, the LCs have a duration of $\approx$12 years. 

We contaminate the signal with red noise generated with the approach of TK. To generate this red noise, we randomly sample power-law indices in the range [0.7–1.2], consistent with the values reported in \citet{bhatta_s5_0716} and \citet{penil_2022}. Moreover, Gaussian noise, distributed as $N(0, Std)$, is added to introduce stochastic variability to the sinusoidal signal points. The standard deviation ``Std'' for this noise is chosen to ensure a detection significance of $\approx$5-5.5$\sigma$ before any gap is introduced \citep[][]{penil_flares_2025}. 

We generate 150,000 artificial signals using the TK to determine the significance of the results obtained from these methods. These artificial signals share the same observational properties as the original sinusoidal signal, including mean, standard deviation, sampling intervals, and observing time. This significance refers to the local significance, not the global significance \citep[i.e., correcting for the number of trials, see e.g.,][]{gross_vitells_trial}. Taking into account global significance is not necessary for our tests because we evaluate the impact of gaps under controlled conditions rather than conducting a broad search for periodic signals in a real sample.

The test generates a distribution of periods and their associated significance using the selected methods (see $\S$\ref{sec:methods}). To assess how the period-significance relationship evolves across different gap configurations (see $\S$\ref{sec:results}), we compute the median and standard deviation of this distribution. These statistical measures provide a robust characterization of the typical behavior of the detected signals and the variability introduced by gaps. In addition to these metrics, the test also identifies the most frequently detected period, along with the corresponding median significance. This allows us to determine not only the overall stability of the detections but also whether certain periods are systematically favored under specific gap patterns, offering deeper insight into potential artifacts introduced by incomplete sampling.

This approach allows us to determine whether purely noisy LCs and sinusoidal signals can produce significant detections and whether the signal’s period can be reliably recovered. Additionally, we analyze how significance varies with different gap structures. Examining these distributions provides valuable insight into the robustness of each method when gaps are present in an LC.

\subsection{Type of Gap Tests}\label{sec:types_gaps}

To evaluate the impact of data gaps on LCs, we propose a series of tests incorporating different gap introduction strategies. These tests simulate real observational scenarios, such as periodic observations from ground-based telescopes and random gaps representing ULs. By systematically introducing gaps in various ways, we can assess how data loss affects the ability to perform robust periodicity searches. 

\subsubsection{Randomly Introduced Gaps}

\begin{figure*}
	\centering
	\includegraphics[scale=0.22]{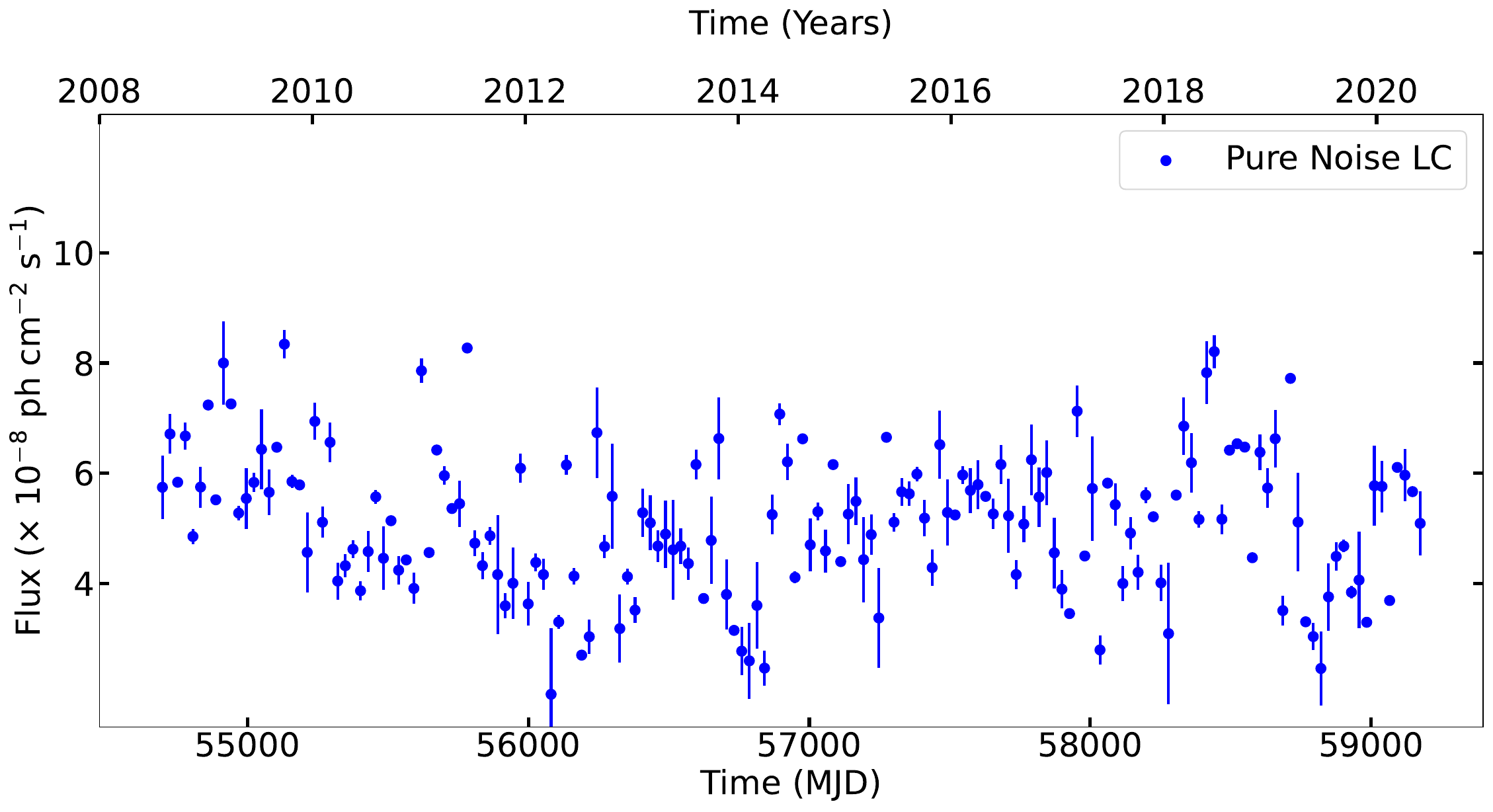}
	\includegraphics[scale=0.22]{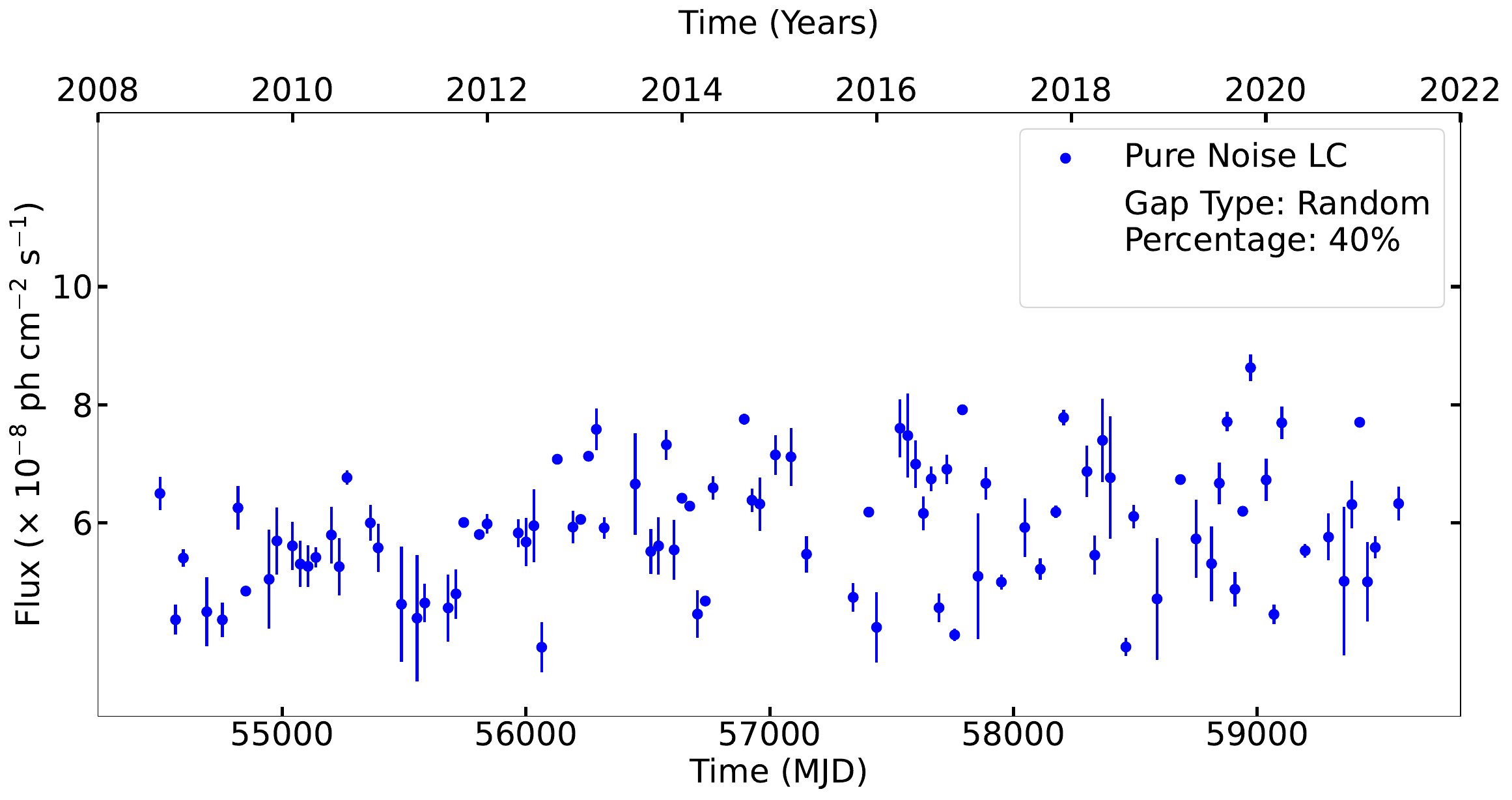}
        \includegraphics[scale=0.22]{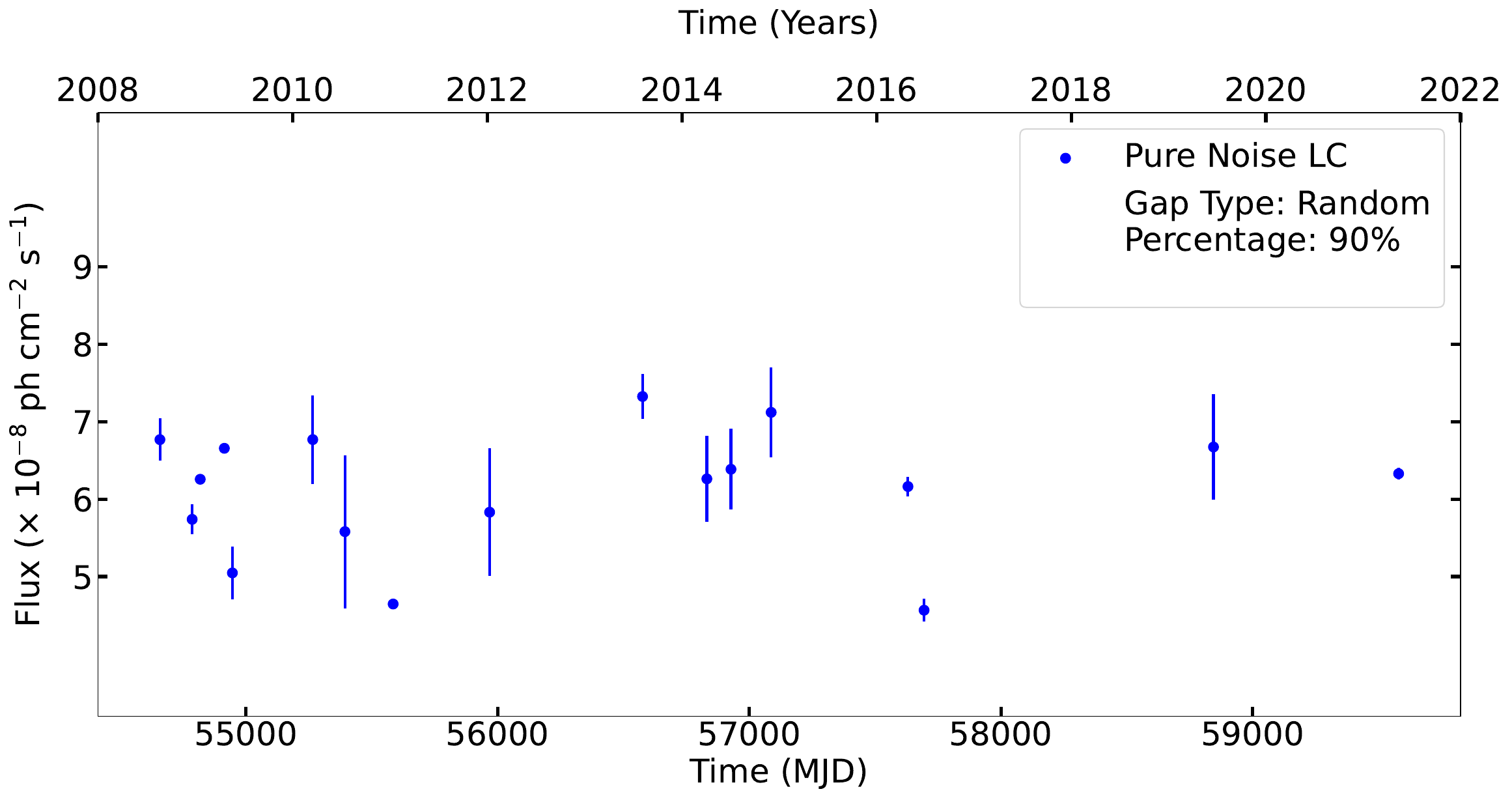}
	\caption{Examples of noise LCs and random gaps distribution. \textit{Top Left}: Pure noise. \textit{Top Right}: 40\% of gap distribution. \textit{Bottom}: 90\% of gap distribution.} \label{fig:examples_noise_lcs}
\end{figure*}

Randomly introduced gaps simulate data loss from instrumental failures, environmental conditions, or sensitivity limits (see Figure~\ref{fig:examples_noise_lcs}). This also includes cases where intrinsically variable sources temporarily fall below the detection threshold, resulting in ULs. The percentage of missing data will vary, allowing us to examine how different levels of data loss impact the detectability and accuracy of LC analysis. Additionally, the positions in the LC of such gaps are not correlated; they are introduced randomly in the LC. This approach ensures that data gaps occur without a predefined pattern, representing real-world scenarios where data collection is sporadic. 

This test scenario is denoted in Figures \ref{fig:examples_noise_lcs} and tables (see Appendix \ref{sec:appendix}) as ``Random''. 

\subsubsection{Annual Variability with Periodic Gap Distribution}
To simulate realistic observational constraints, we introduce periodic gaps into the LCs yearly, replicating conditions typically encountered in ground-based observations. Such constraints typically arise from seasonal visibility limits or telescope scheduling, which restrict observations to specific periods of the year. As a result, data acquisition often follows a regular annual pattern, with alternating intervals of observation and inactivity.

In our simulations, we vary the duration of these annual gaps to span between 10\% and 90\% of the data points within a single year. This reflects scenarios where a source may only be visible for a few months per year, depending on its position in the sky and the geographical location of the observatory.

During each year of simulated observability, the pattern of gaps remains consistent across the LC time. This setup enables us to explore how structured, periodic sampling gaps, common in observing campaigns, impact the detection and characterization of periodic signals.

This test scenario is denoted in Figures \ref{fig:example_annual_distribution} as ``Periodic Annual Distribution'' and in tables  (see Appendix \ref{sec:appendix}), as ``Periodic''. 

\subsubsection{Annual Variability with Random Gap Distribution}
\begin{figure*}
	\centering
	\includegraphics[scale=0.22]{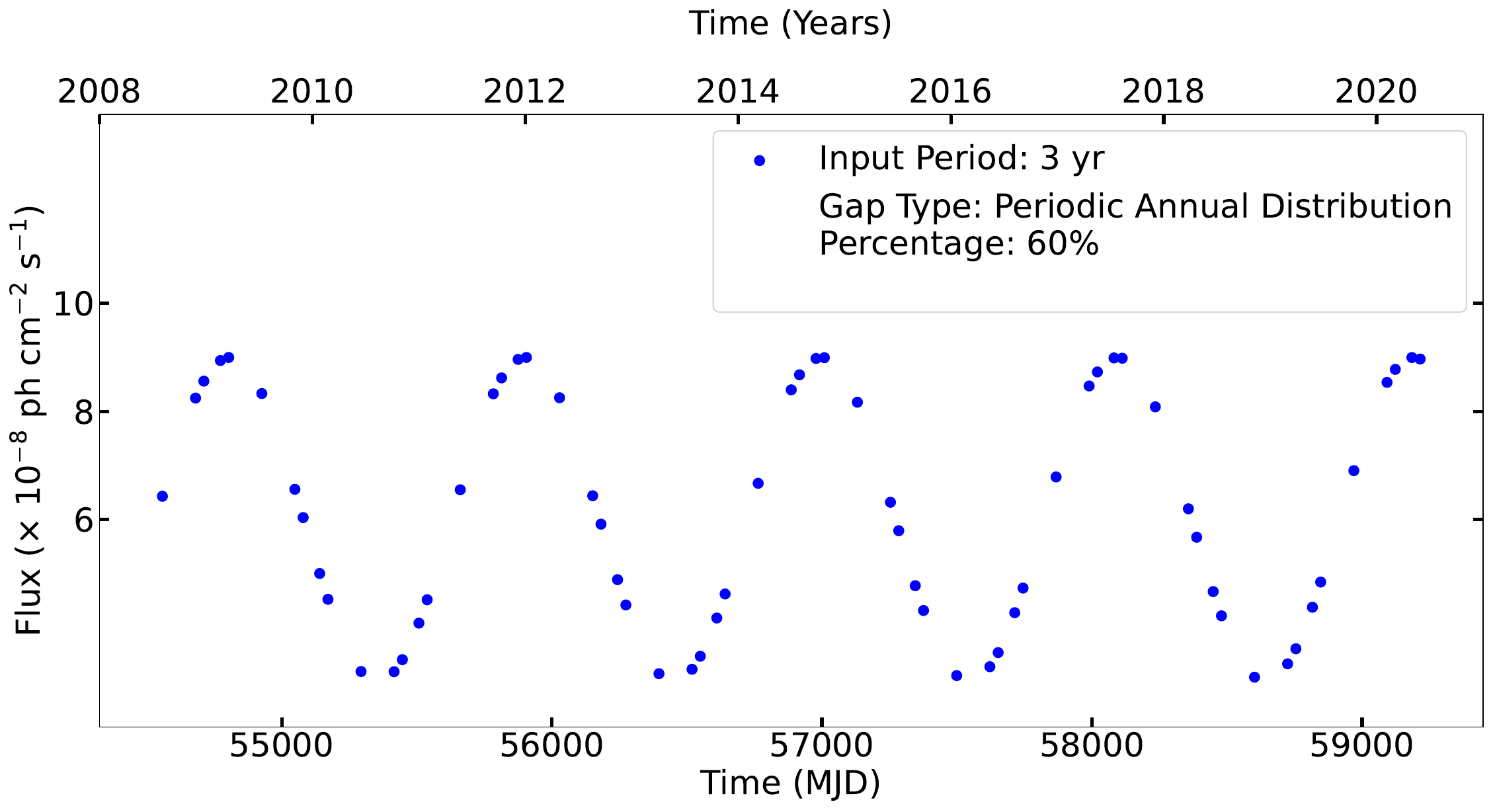}
        \includegraphics[scale=0.22]{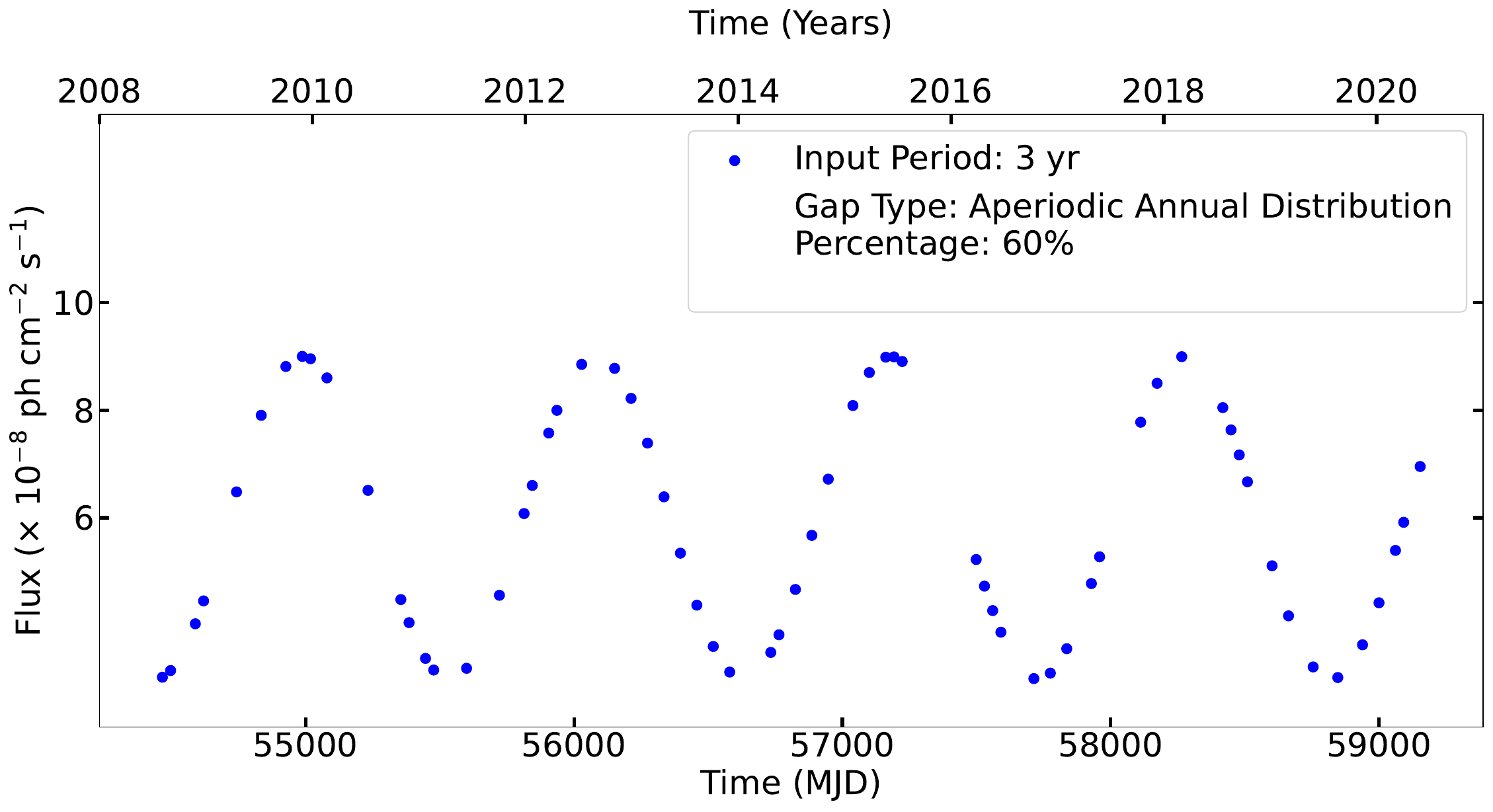}
	\caption{Examples of LCs to illustrate the different annual variability distribution using as a test. The LCs are pure sinusoidal without adding any noise to help see the differences in the gap distribution. We consider a percentage of gaps of 60\%. \textit{Left}: Annual variability with periodic gap distribution. \textit{Right}: Annual variability with random gap distribution, denoting in the legend as ``Aperiodic Annual Distribution''.} \label{fig:example_annual_distribution}
\end{figure*}

This test also introduces gaps yearly to simulate annual observation patterns. However, unlike the previous scenario where the gap structure is fixed and repeated each year, here the gaps are injected randomly each year, resulting in a different distribution of missing data across time (see Figure \ref{fig:example_annual_distribution}). This setup reflects the variability observed in many real observational campaigns, where scheduling constraints, proposal approvals, and changing seasonal conditions result in inconsistent yearly coverage.

Many ground and space-based monitoring programs operate on annual cycles, with observations tied to proposal deadlines and target visibility windows. By allowing the gap pattern to vary from year to year, we aim to more realistically replicate the stochastic nature of data collection in such programs. This approach enables us to evaluate how inconsistent observational coverage impacts the reconstruction of variability patterns and the robustness of periodicity detection.

This test scenario is denoted in Figures \ref{fig:example_annual_distribution} as ``Aperiodic Annual Distribution'' and in tables  (see Appendix \ref{sec:appendix}), as ``Aperiodic''. 

\section{Testing the Impact of Gaps} \label{sec:results}

In this section, we evaluate the impact of gaps using the previously described methodology, applied to a subset of the most commonly used methods for periodicity searches.

\subsection{Methods} \label{sec:methods}
\begin{figure*}
	\centering
        \includegraphics[scale=0.21]{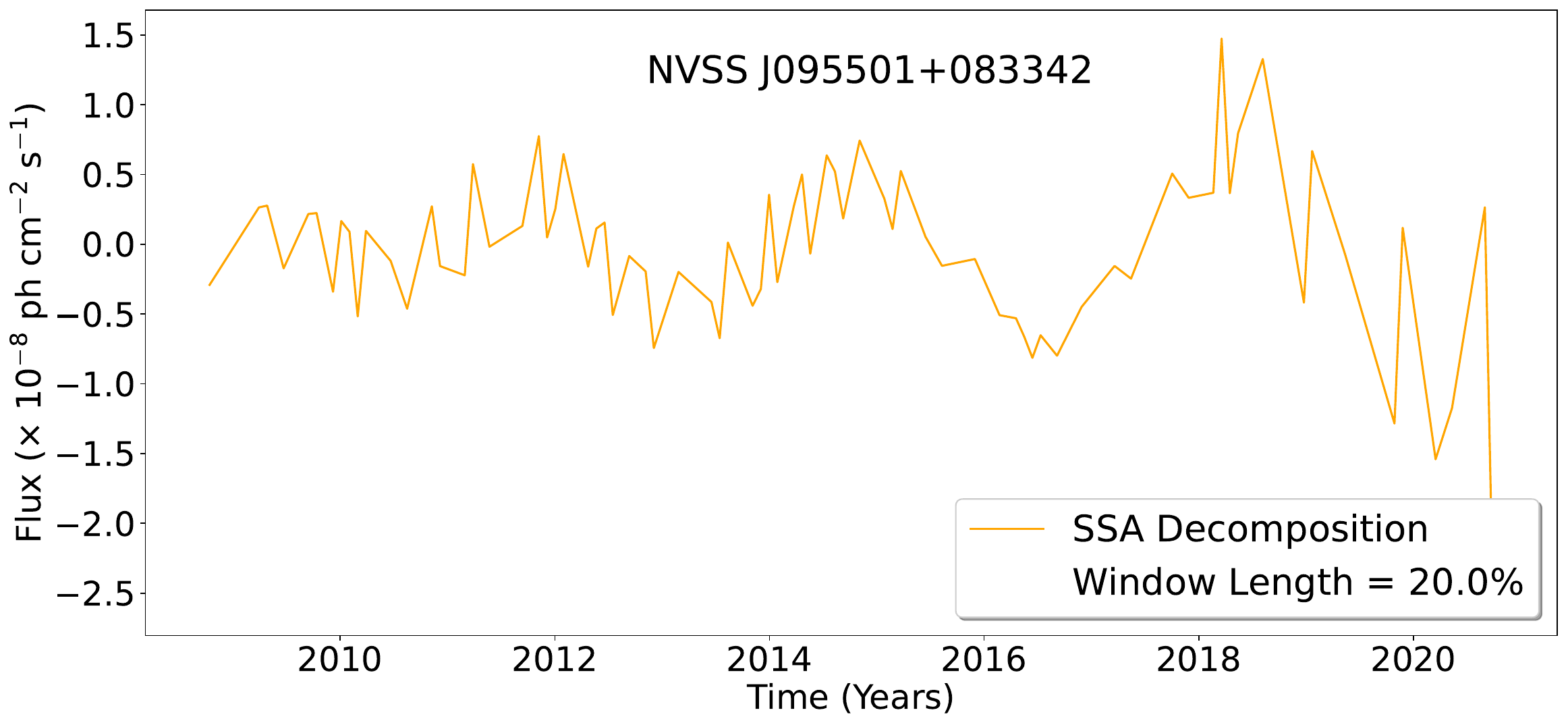}
        \includegraphics[scale=0.21]{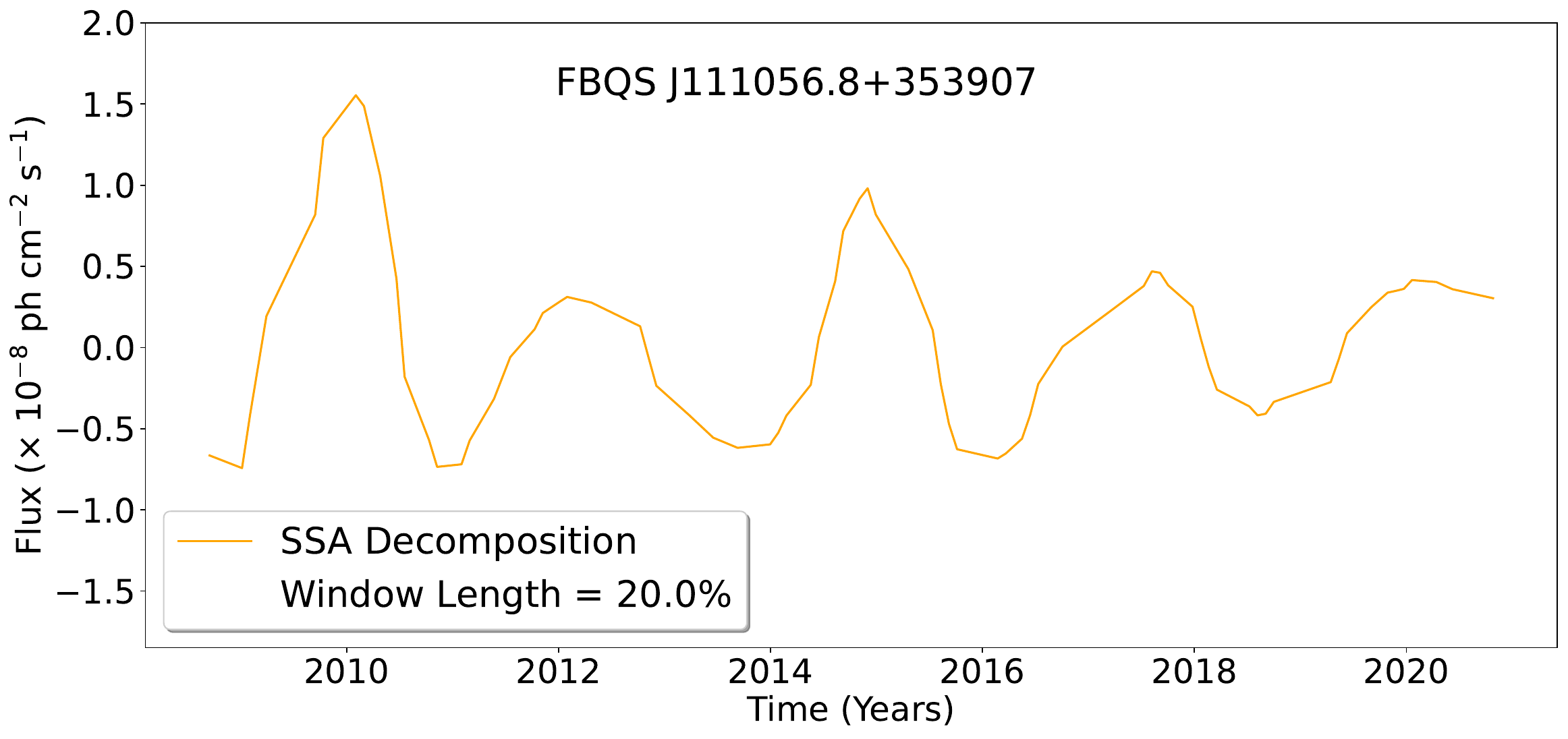}
        \includegraphics[scale=0.21]{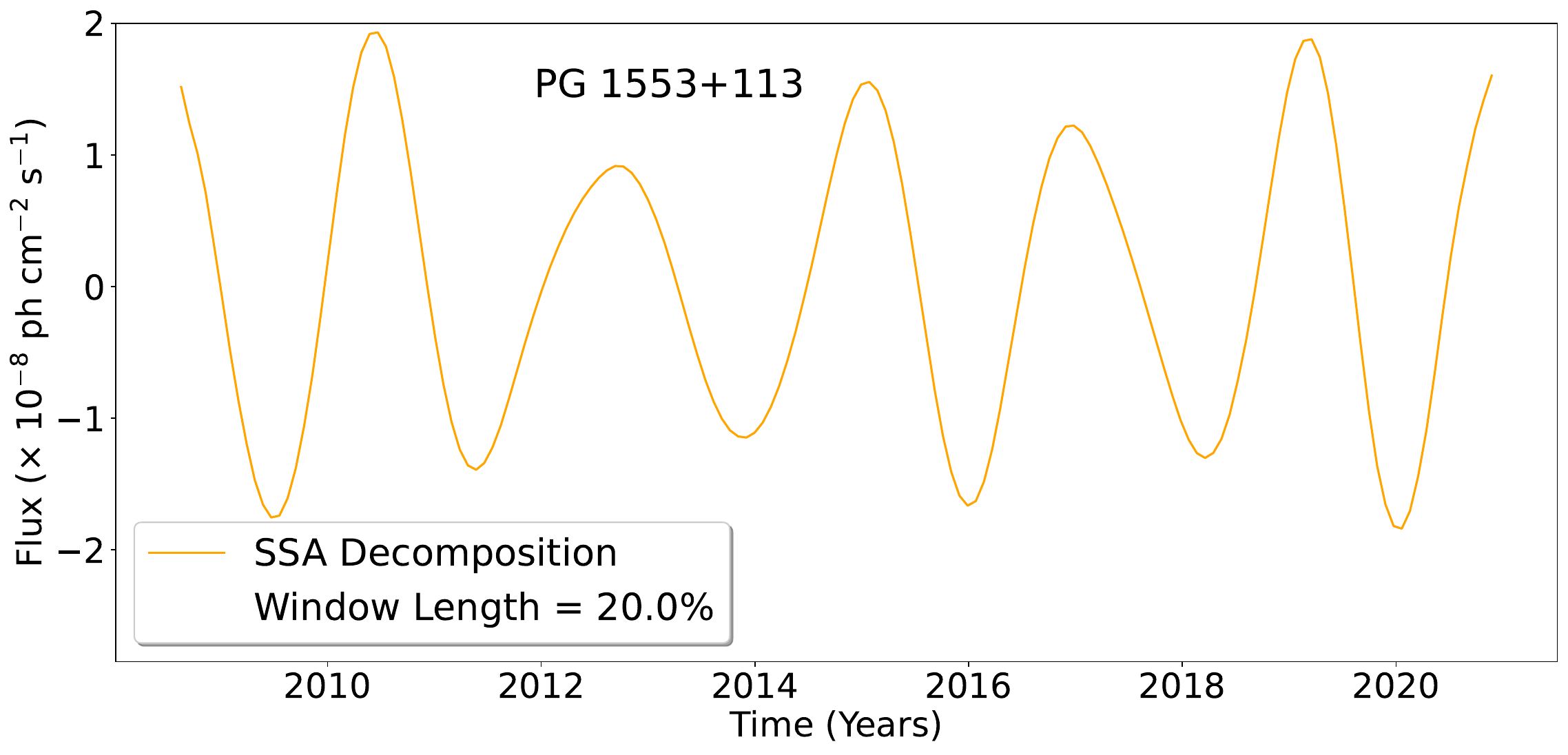} 
	\caption{SSA decomposition showing the underlying oscillatory structure using a window length of 20\%. \textit{Top}: NVSS J095501+083342 and FBQS J111056.8+353907 (see Figure \ref{fig:example_use_cases}). \textit{Bottom}: PG 1553+113 (see Figure \ref{fig:example_use_cases}). The flux axis shows negative values because it represents only the oscillatory component, excluding the overall emission behavior of the source. As a result, the oscillatory component is centered around zero, reflecting deviations from the mean rather than the total flux.} 
    \vspace{-0.3cm}
 \label{fig:ssa_blazars}
\end{figure*}

As we said in $\S$\ref{sec:intro}, the evaluation of the periodicity is conducted using three methods: LSP, PDM, and SSA. SSA is primarily a preprocessing technique designed to facilitate periodicity analysis; it is not a standalone method for detecting periodic signals. SSA functions by decomposing a time series into a sum of fundamental components, allowing the reconstruction of the original series while effectively separating the random noise. With SSA, we aim to extract the oscillatory patterns in LCs, reducing the influence of stochastic effects such as noise and trends (see Figure \ref{fig:ssa_blazars}). Then, SSA has been assessed for its robustness in identifying periodic patterns when a flare is present in the LC. The results of \citet{penil_flares_2025} demonstrate its ability to recover periodicity despite this stochastic phenomenon. In this paper, we further investigate SSA's performance in the presence of gaps. 

Classic SSA assumes that the input time series is regularly sampled. For comprehensive descriptions of the methodology, we refer the reader to \citet{golyandina_ssa} and \citet{alba_ssa}, which provide detailed theoretical and practical insights. To address incompleteness, several extensions of SSA have been developed in which missing values are reconstructed iteratively through singular value decomposition (SVD). In these approaches, the gaps are initially filled with a provisional guess, after which a trajectory matrix is constructed and decomposed via SVD to obtain a low-rank approximation that captures the dominant temporal structures of the series. The reconstructed signal is then used to update the missing entries, and this cycle of decomposition and reconstruction is repeated until convergence \citep[][]{golyandina2007}, thereby completing the LC in a manner consistent with its inherent dynamics. An alternative scheme addresses missing data without provisional filling by introducing a weighted version of the trajectory matrix \citep[][]{golyandina2020singular}. In this framework, each element of the matrix is accompanied by a weight that indicates whether the corresponding time-series value is observed or missing. The decomposition is then carried out using weighted SVD, so that only the valid entries contribute to the factorization, while gaps are effectively excluded from the optimization criterion. This procedure makes it possible to retain the structure of complete lagged vectors while still incorporating the partial information from incomplete ones.  This procedure preserves the essential temporal properties of the original LC, enabling a recovery of dominant and recurrent temporal components of the signal, even in the presence of gaps.

In the present work, we apply SSA directly to irregular LCs without prior interpolation, treating missing entries as absent, according to the second framework previously presented. This choice follows the broader practice of methods devised for regular sampling --e.g., ARIMA/ARFIMA models \citep{feigelson_arima} and autocorrelation-based techniques \citep{acf_irregular}--to moderately irregular LCs under comparable assumptions.

Our aim is to evaluate how SSA behaves under these conditions and to assess its ability to identify robust oscillatory components in irregular LCs. Specifically, we apply SSA to artificial LCs with controlled gap distribution, focusing on the extraction of the oscillatory component (Figure \ref{fig:ssa_blazars}). The resulting oscillatory component is then analyzed with the LSP to determine its period and significance, following the strategy of \citet{alba_ssa}. This combined SSA-LSP framework allows us to systematically test the capacity of SSA to serve as a preprocessing tool for periodicity searches in irregularly sampled blazar LCs.

For our tests, we discarded the weighted wavelet Z-transform \citep[WWZ, ][]{foster_wwz}. Although the WWZ is well-suited for analyzing irregularly sampled time series \citep[][]{foster_wwz}, its high computational cost makes it impractical for our study. WWZ requires extensive calculations to generate time-frequency representations, significantly increasing processing time, mainly when applied to large datasets or numerous simulations. Given our focus on evaluating period detection across multiple scenarios, more computationally efficient methods are necessary to ensure scalability and reproducibility. 

Beyond the WWZ, several other strategies have been developed to address irregularly sampled light curves. For instance, Bayesian period searches implemented with MCMC, such as the Bayesian Kepler periodogram, build on the Gregory-Loredo framework and have been applied to long, uneven time series \citep[e.g.,][]{bayesian_mcmc}. Gaussian process regression has also been adopted \citep[e.g.,][]{gaussian_process}. In addition, continuous-time autoregressive moving average (CARMA) models have been introduced as a statistical framework for irregular time series, and have been applied to AGN LCs to search for periodic patterns \citep[e.g.,][]{carma_kelly}. More recently, deep learning–based frameworks have been proposed to reconstruct missing segments and recover periodic signals on irregular astronomical time series \citep[e.g.,][]{machine_learning}. The widely used methods, such as the LSP and PDM, remain standard tools for detecting periodicities in unevenly sampled data. Building on this foundation, SSA provides a novel and complementary framework to perform the analysis of periodicity. Our work aims to extend the comparison and the characterization of these techniques by explicitly incorporating the gap effects, since their response under different scenarios where flares can distort periodic signals or produce fake periodic features was systematically analyzed in \citet{penil_flares_2025}.

\subsection{Pure Noise LCs: Results} \label{sec:pure_noise_lcs}
As a control scenario, we initially assess the distribution of period and significance for pure noise LCs without introducing any gaps. Then, gaps are injected into these pure noise LCs according to the tests presented in $\S$\ref{sec:types_gaps}. 

\begin{figure*}
	\centering
	\includegraphics[scale=0.24]{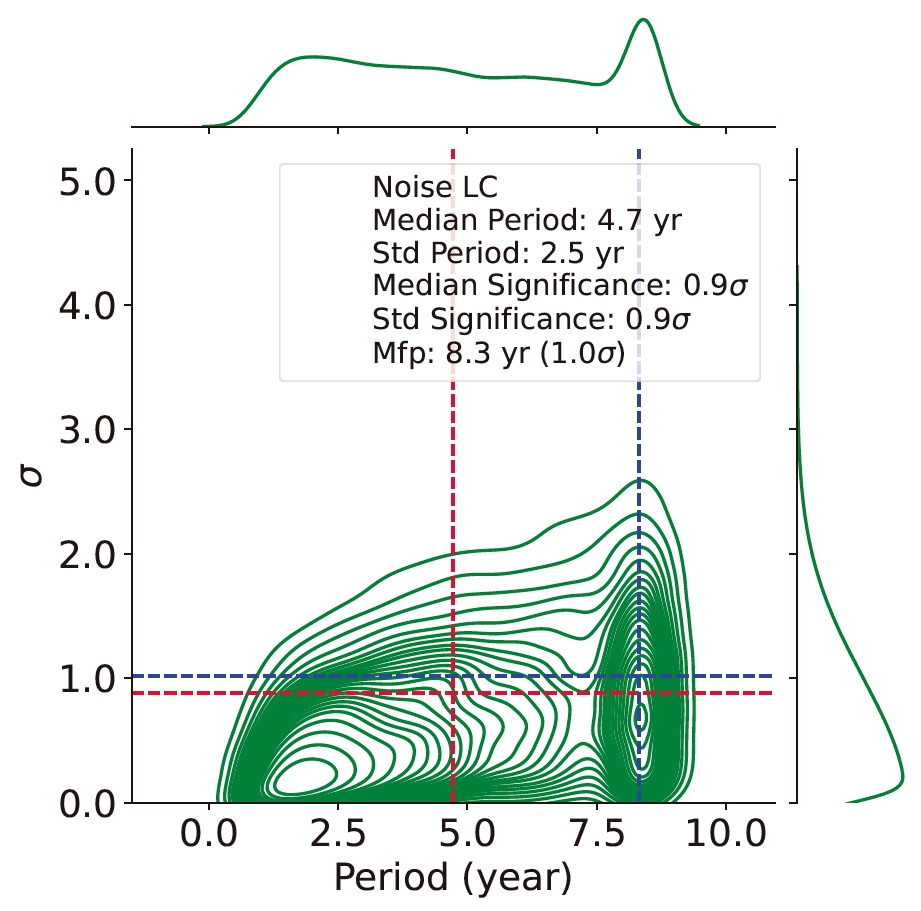}
        \includegraphics[scale=0.24]{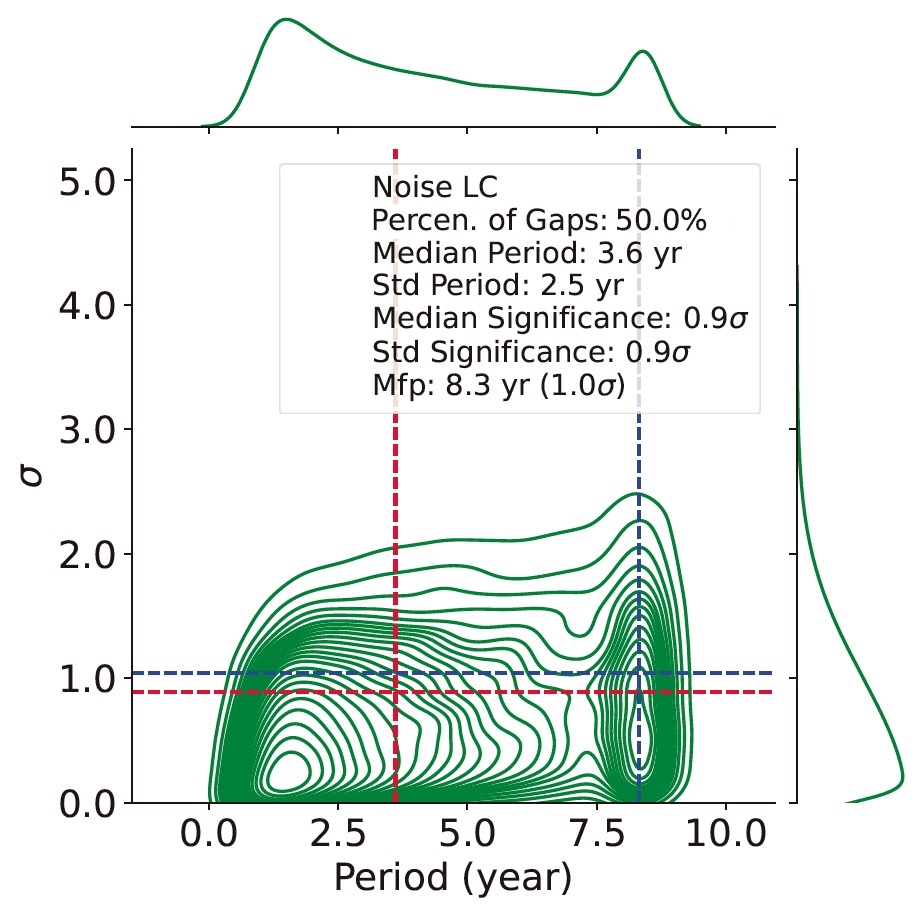}
        \includegraphics[scale=0.24]{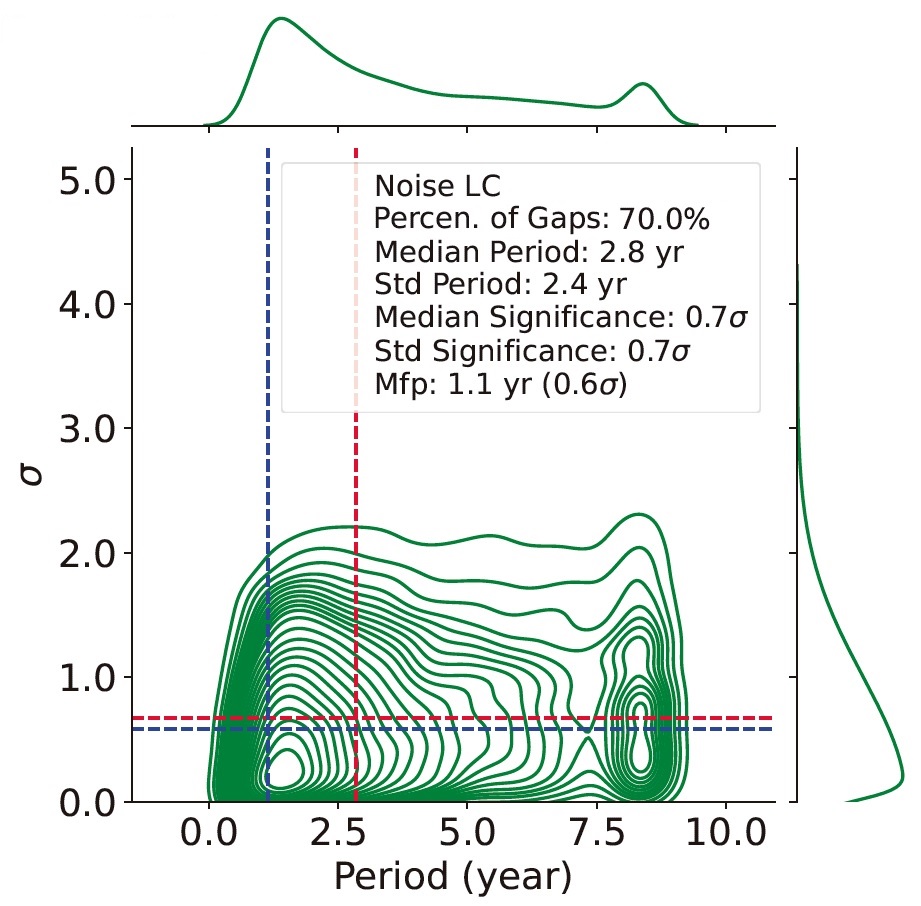}
        \caption{Distributions for the period and significance for pure noise LCs, for LSP. \textit{Left}: no gaps. \textit{Center}: 50\% of randomly distributed gaps. \textit{Right}: 70\% of randomly distributed gaps. The results denote the evolution of the period according to the increase in the percentage of gaps. When there are no gaps, the period distribution is approximately homogeneous in the period range, with an excess in the higher period expected for the red noise LCs. With the introduction of gaps, this disruption is modified to a lower period. The dotted red vertical and horizontal lines indicate the median values for both the period and the significance of the test. The blue dotted vertical line highlights the most frequently occurring period in the tests (and the associated significance), emphasizing its prominence in the distribution. The ``Percen. of Gaps'' refers to the percentage of gaps injected in the LC. The ``Median Period'' represents the median of all periods resulting from the test, and the ``Std Period'' is the standard deviation of such periods' distribution. The ``Median Significance'' represents the median of the significance distribution associated with the test, and the ``Std Significance'' is the standard deviation of this significance distribution. ``Mfp'' represents the most frequent period resulting from the test.} \label{fig:testing_pure_noise_lsp}
\end{figure*}

\subsubsection{LSP} \label{sec:lsp_pure_noise}
The results, presented in Figure \ref{fig:testing_pure_noise_lsp}, reveal that for the LSP, the median detected period is 4.7 years (denoted by the dotted red vertical line), with a significance of 0.9$\sigma$, which aligns with typical expectations for a stochastic noise scenario \cite{vaughan_2003}.

When the gaps are injected, the results of the different tests are consistent across all cases, independently of the type of gap strategy considered in the test ($\S$\ref{sec:types_gaps}) and the percentage of gaps considered. The detailed results are presented in Table \ref{tab:noise_random}. The presence of gaps does not significantly affect period estimation, which is expected, given the lack of a genuine underlying periodic signal. The distribution of detected periods remains constant, with no significantly preferred period in the range between 1 and 8.5 yr, except for an increase at the end of the distribution corresponding to the 8.5-year period. This behavior is linked to the edge effects of the defined period range used for the period search test [0.9-8.5 years] and the pure noise origin of the LCs. These periods are typically obtained with significances $\leq$1.0$\sigma$, with most around 0.8$\sigma$, and therefore would be unlikely to be confused with real detections. 

Additionally, as shown in the significance distribution in Figure~\ref{fig:testing_pure_noise_lsp}, we observe a clear trend: the probability of obtaining a significant detection (defined here as exceeding the 3$\sigma$ threshold) remains constant with the percentage of gaps in the LC, around 0.21\% in all the tests. This behavior indicates that a higher percentage of gaps is unlikely to artificially inflate the significance of detected signals, even in purely stochastic LCs. 

To further characterize the distribution, we identify the most frequently detected period in our simulations (in Figure~\ref{fig:testing_pure_noise_lsp}, this value is denoted by a dotted blue vertical line). This value not only indicates the most commonly occurring periods but also provides their associated significance levels, offering a more comprehensive perspective on the statistical behavior of the methods.  

The most frequent period detected is 8.5 years (1.0$\sigma$) for gap distributions $\leq$60\%. This can be attributed to the edge effects, as mentioned above \citep[see also e.g., ][]{vaughan_2003, vaughan_criticism}. This most-frequent-period value remains until $\geq70\%$ gaps, where the most commonly obtained period transitions to 1.1 years (0.8$\sigma$). The presence of this alternative period suggests that gap-induced effects may influence the results. The period of $\approx$1.1 years is also the second most frequently detected period in cases $<70\%$ gaps. The frequency difference between the 8.5-year and 1.1-year periods diminishes as the percentage of gaps increases. In the absence of gaps, the relative occurrence of the 8.5-year period is 76\% higher than that of the 1.1-year period. For a 60\% gap scenario, the last case where the 8.5-year period remains the most frequent, this rate decreases to 21\%. Therefore, as the percentage of gaps increases, the likelihood of falsely detecting a $\approx$1.1-year period grows, potentially leading to incorrect conclusions about periodic behavior and becoming increasingly susceptible to spurious detections of shorter periods.

These artificial $\sim$1.1-year signals may be attributed to the interaction between the temporal structure of the gaps and the observational window function. In particular, large or regularly spaced gaps can introduce spurious periodicity detections, especially near the observational cadence or gap pattern. Consequently, when the data are significantly incomplete, these effects can dominate the periodogram and bias the detection toward false, short periodicities. 

Regarding the detection of significant periods (i.e., $\geq$3$\sigma$), the LSP shows a low rate of false positives, 0.21\%, with no instances reaching the 5$\sigma$ threshold. These results remain unchanged during the different gap tests.

These results highlight the robustness of LSP in handling purely stochastic data, regardless of data completeness. The consistent absence of 5$\sigma$ detections across all gap scenarios demonstrates that LSP is highly conservative in noise-dominated conditions and does not easily infer false periodic signals.

\subsubsection{PDM}\label{sec:pdm_pure_noise}
\begin{figure*}
	\centering
	\includegraphics[scale=0.24]{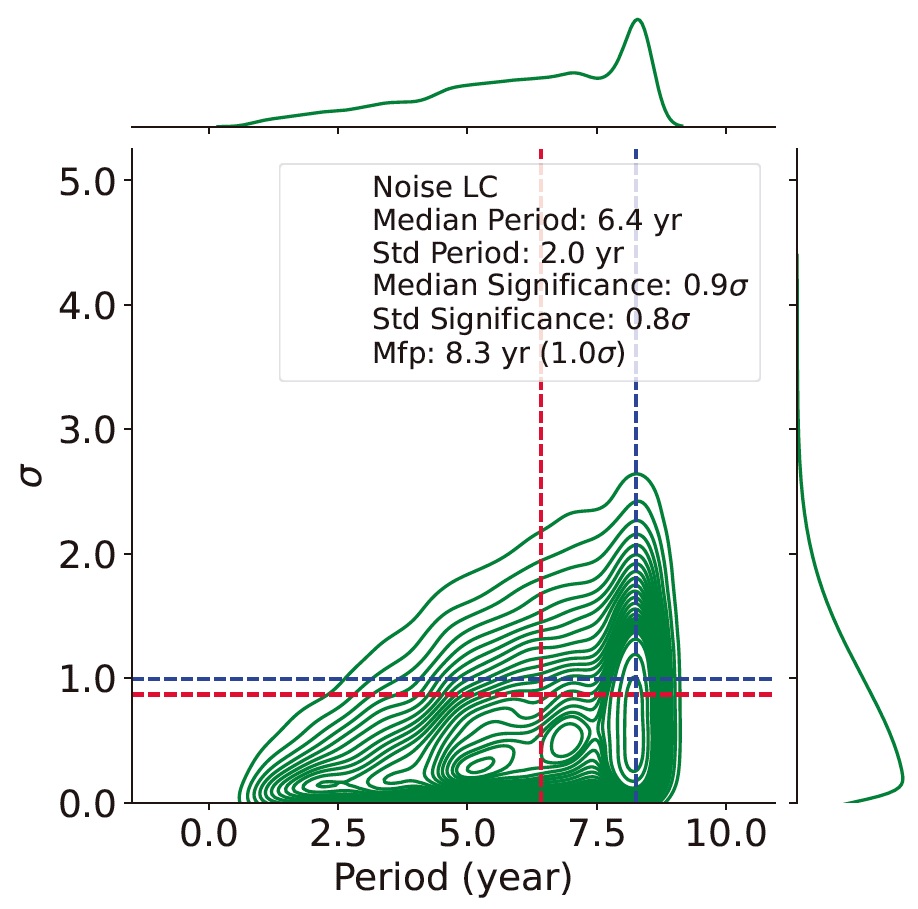}
        \includegraphics[scale=0.24]{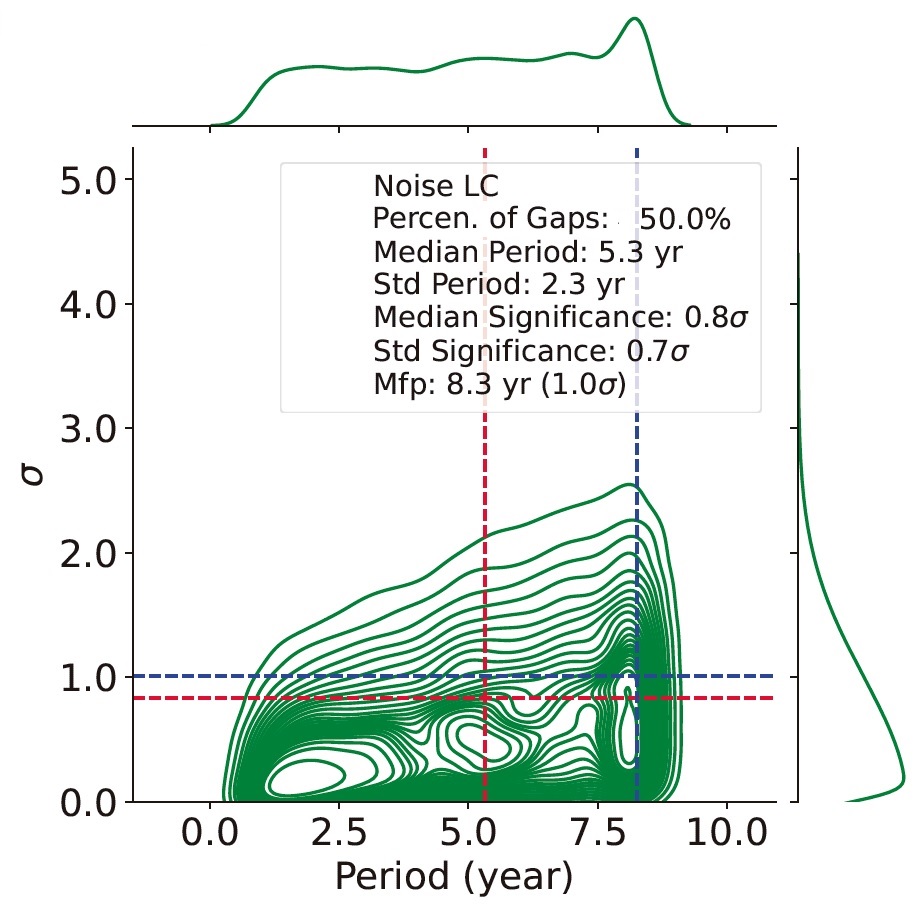}
        \includegraphics[scale=0.24]{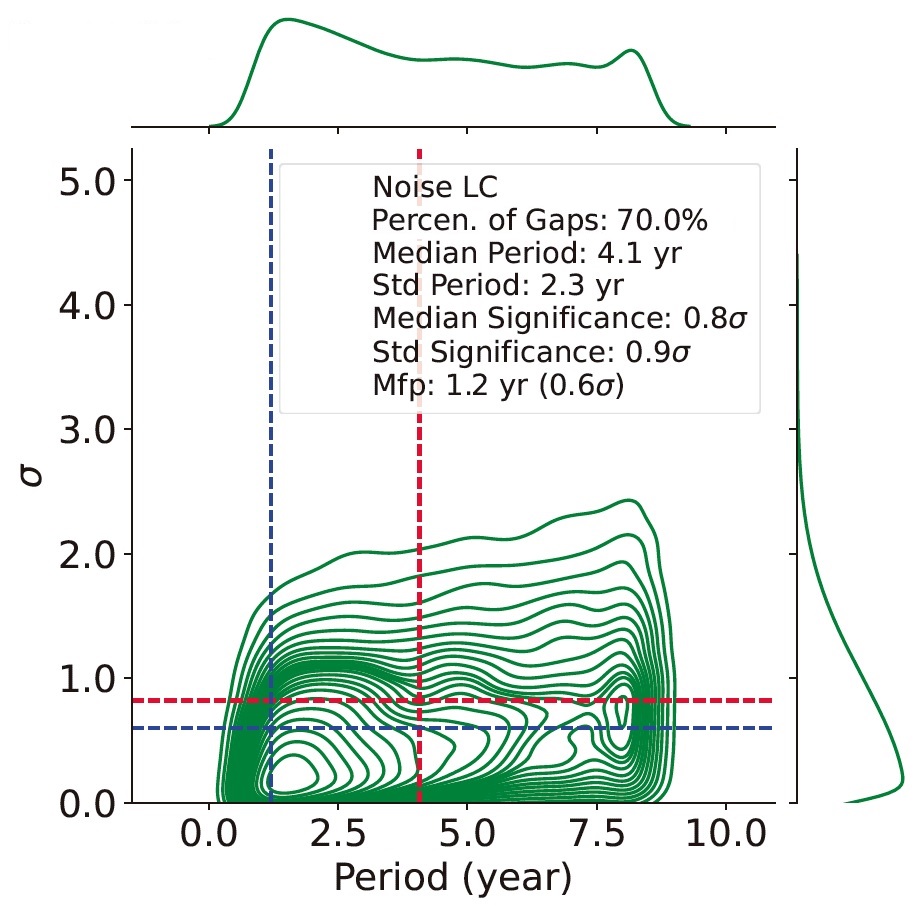}
        \caption{Distributions for the period and significance for pure noise LCs, for PDM. \textit{Left}: no gaps. \textit{Center}: 50\% of randomly distributed gaps. \textit{Right}: 70\% of randomly distributed gaps. The results show how the period evolves as the percentage of gaps increases. When there are no gaps in the LC, the distribution follows an increasing trend, peaking at 8.5 years, the upper limit of the period search range. As the percentage of gaps rises, the distribution becomes more uniform across the entire period range. In terms of significance, the distribution remains relatively stable with increasing gaps, with low significance ($\approx$1$\sigma$). The dotted red vertical and horizontal lines indicate the median values for both the period and the test's significance. The blue dotted vertical line highlights the most frequently occurring period in the tests (and the associated significance), emphasizing its prominence in the distribution. The ``Percen. of Gaps'' refers to the percentage of gaps injected in the LC. The ``Median Period'' represents the median of all periods resulting from the test, and the ``Std Period'' is the standard deviation of such periods' distribution. The ``Median Significance'' represents the median of the significance distribution associated with the test, and the ``Std Significance'' is the standard deviation of this significance distribution. ``Mfp'' represents the most frequent period resulting from the test.} \label{fig:testing_pure_noise_pdm}
\end{figure*}

Regarding PDM, Table \ref{tab:noise_random} shows results similar to those obtained for LSP. When there are no gaps in the LC, the median period is 6.5 years (0.8$\sigma$), which aligns with the red noise scenario (Figure \ref{fig:testing_pure_noise_pdm}). In this case, the period distribution shows a linear increase from the shortest to the longest period, peaking at 8.5 years, which is linked to the stochastic nature of the LC. As the presence of gaps increases, the period distribution shifts to 1.1 years. When no gaps are present, the relative occurrence of the 8.5-year period is 80\% higher than that of the 1.2-year period. For a scenario with 60\% gaps, the last case where the 8.5-year period remains the most frequent, this rate decreases to 31\%.  Additionally, when no gap is present in the LC, regarding the detection of significant periods, the results are similar to LSP; for PDM, this percentage is  0.24\% (with no 5$\sigma$ detections). 

\subsubsection{SSA} \label{sec:alternative_methods}
\begin{figure*}
	\centering
	\includegraphics[scale=0.24]{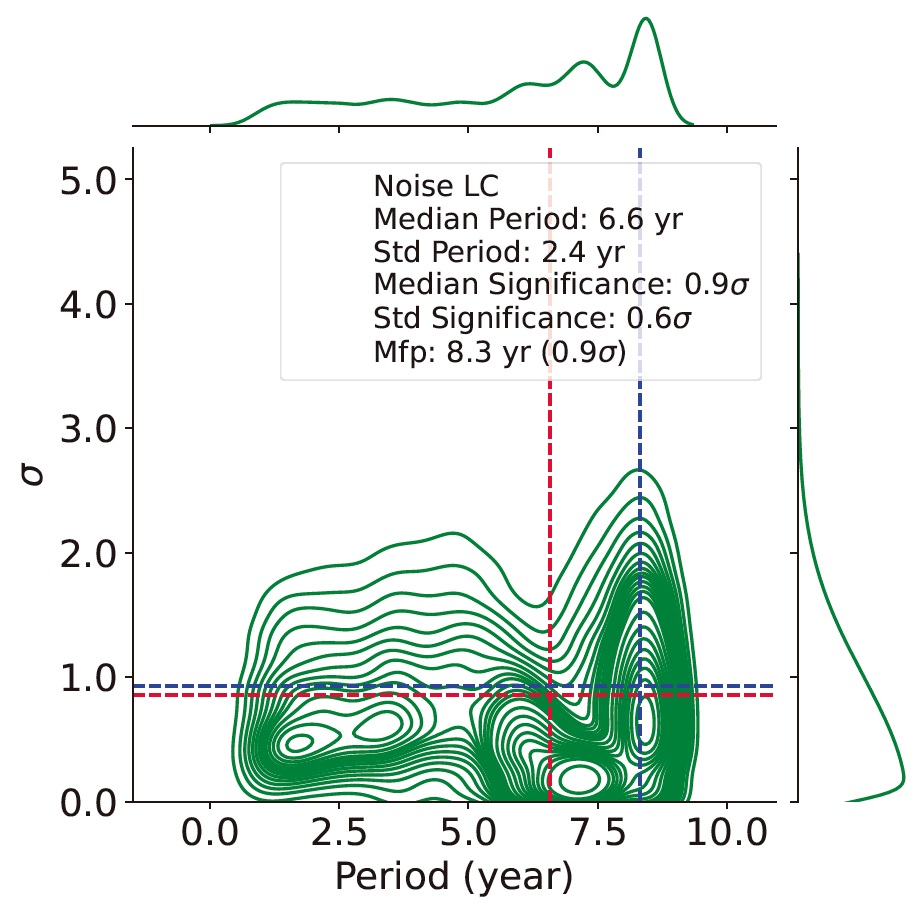}
        \includegraphics[scale=0.24]{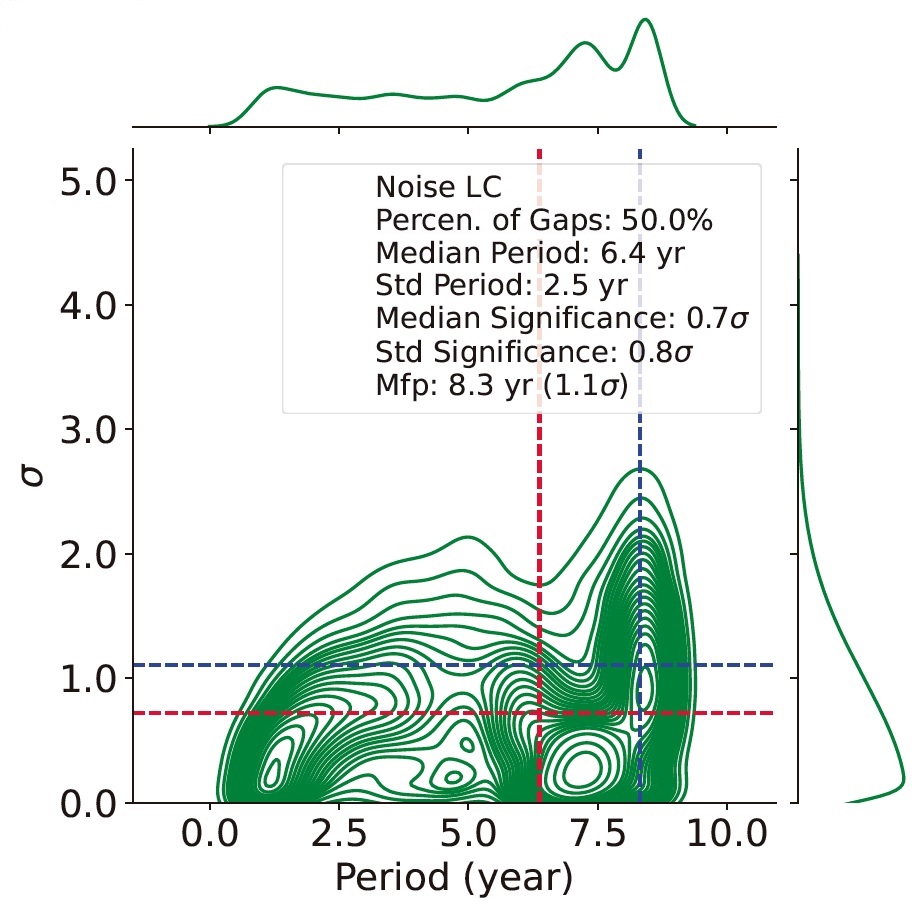}
        \includegraphics[scale=0.24]{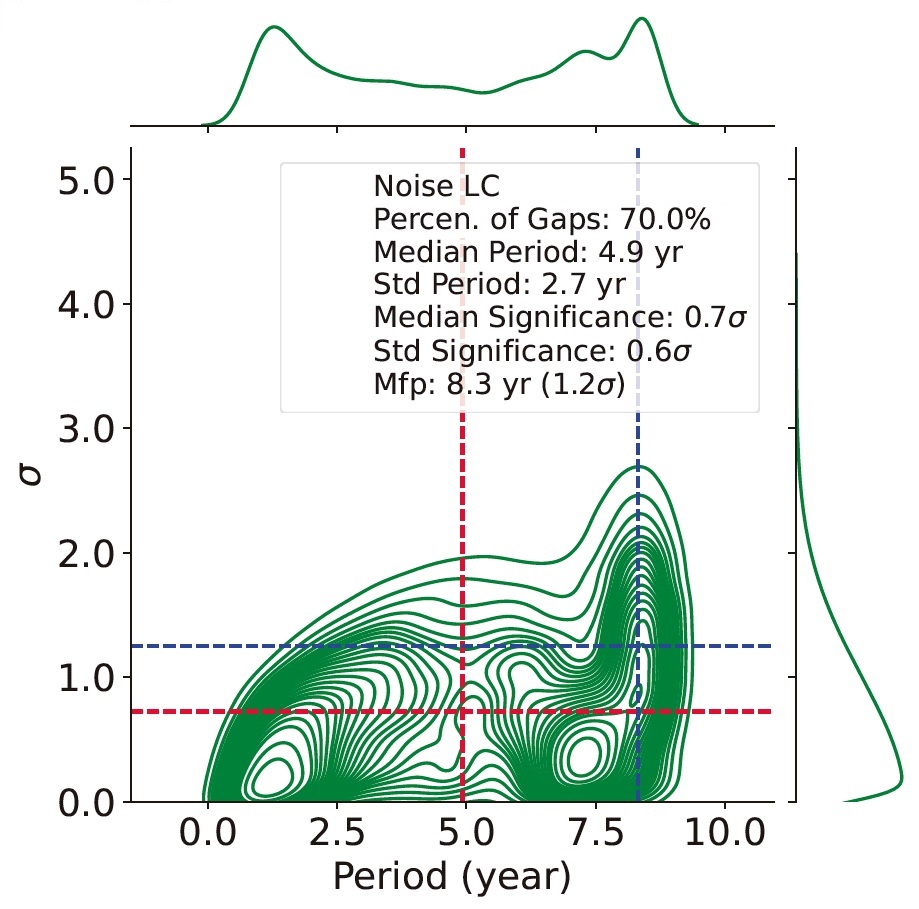}
        \caption{Distributions for the period and significance for pure noise LCs, for SSA with a window length of 40\%. \textit{Left}: no gaps. \textit{Center}: 50\% of randomly distributed gaps. \textit{Right}: 70\% of randomly distributed gaps. The results denote the evolution of the period according to the increase in the percentage of gaps. When there are no gaps, the period distribution is approximately homogeneous in the period range, with an excess in the higher period expected for the red noise LCs. The dotted red vertical and horizontal lines indicate the median values for both the period and the significance of the test. The blue dotted vertical line highlights the most frequently occurring period in the tests (and the associated significance), emphasizing its prominence in the distribution. The ``Percen. of Gaps'' refers to the percentage of gaps injected in the LC. The ``Median Period'' represents the median of all periods resulting from the test, and the ``Std Period'' is the standard deviation of such periods' distribution. The ``Median Significance'' represents the median of the significance distribution associated with the test, and the ``Std Significance'' is the standard deviation of this significance distribution. ``Mfp'' represents the most frequent period resulting from the test.} \label{fig:testing_pure_noise_ssa}
\end{figure*}

The results for the test for SSA are presented in Table \ref{tab:noise_random_ssa}. The tests conducted on the noise LCs show similar results to those obtained with LSP and PDM (Figure \ref{fig:testing_pure_noise_ssa}). 

A relevant difference appears when examining how the results evolve with increasing gap percentages. In the case of LSP and PDM, the distribution of detected periods shifts toward 1.1 years with increasing gap presence, in all the tests considered. However, this effect does not take place when SSA is applied to a random-gap distribution (see Table \ref{tab:noise_random_ssa}). In this case, the prominent peak of $\approx$8.5 years is always present in this specific test case. The shift to the prominent peak $\approx$1.1 years only occurs in the test cases of annual variability with periodic and random gap distributions. In these cases, this shift starts to happen at the same percentage of gaps as LSP and PDM, $\geq$70\%. This suggests that conducting a pre-periodicity analysis with SSA enhances robustness against the effects of observational gaps.

The detection of a $\sim$1.1-year period may be an artifact arising from the interaction between the SSA method and the sampling pattern, even when the gaps are randomly distributed. As the percentage of gaps increases, SSA becomes more susceptible to amplifying spurious periodic components that resonate with the temporal structure of the remaining data points. Although no explicit annual signature is introduced, the consistent emergence of this timescale, particularly at high gap levels (e.g., 70\%), suggests that the combination of data sparsity and fragmentation is inducing the artificial signal. In terms of significance, the distribution follows a trend similar to those observed in LSP and PDM, with a low rate of $>$3$\sigma$ detections, $\approx$0.20\% (see Figure~\ref{fig:testing_pure_noise_ssa}), and no occurrences at 5$\sigma$.

\begin{figure}
	\centering
	\includegraphics[scale=0.22]{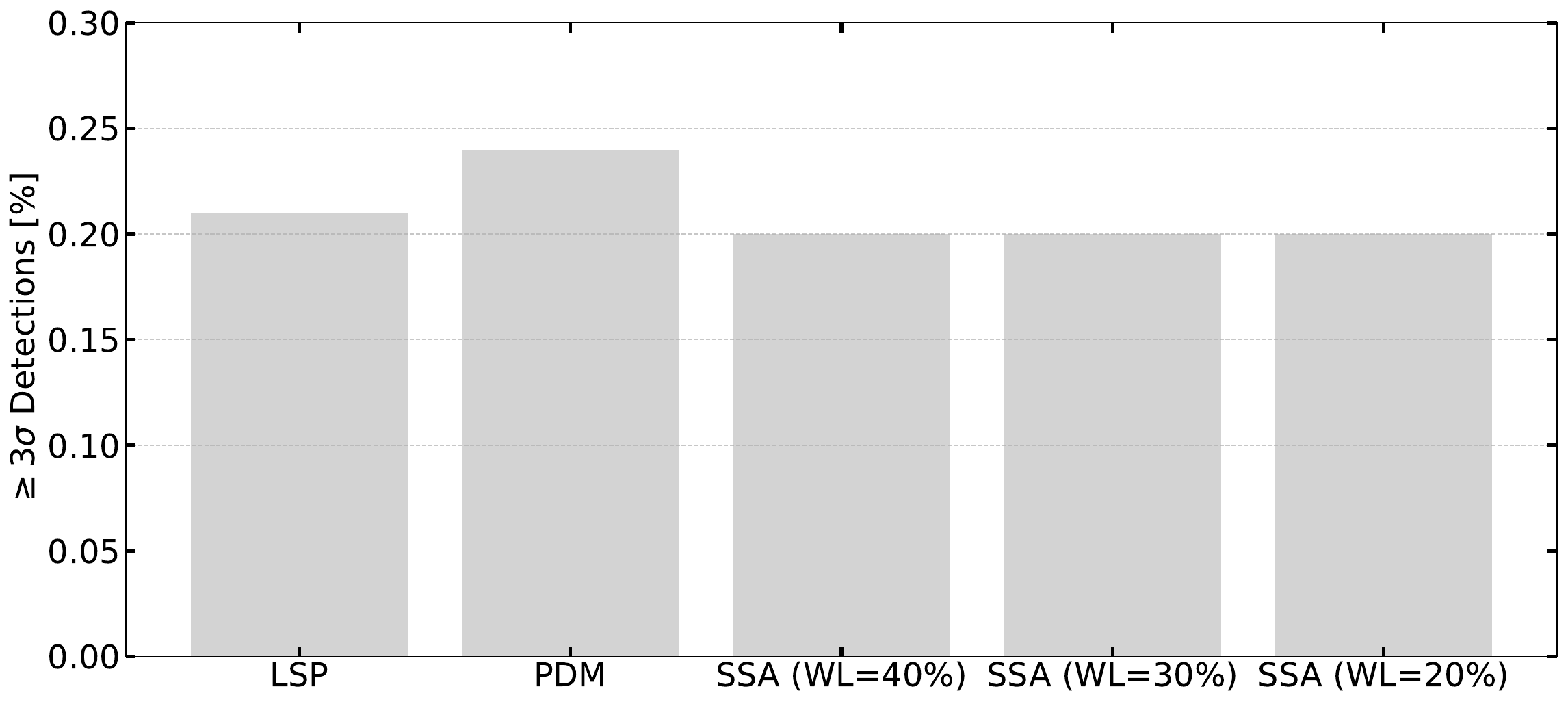}
	\caption{Representation of the different rates of significant detections ($\geq 3\sigma$) for the methods LSP, PDM, and SSA. ``WL'' denotes the window length used for the SSA decomposition (see $\S$\ref{sec:window_lengh}).} \label{fig:significant_distribution_methods}
\end{figure}

Therefore, a first conclusion of our tests is that the techniques employed are robust against producing spurious significant detections in pure noise LCs. Furthermore, this robustness persists even when observational gaps are present, indicating that they maintain reliability under realistic data conditions involving irregular sampling (Figure \ref{fig:significant_distribution_methods}).

Finally, the results from all three methods indicate that the observed shift toward a $\sim$1-year period in gap-affected LCs should be interpreted with caution. The heightened sensitivity of these techniques to a high percentage of gaps can bias the period search toward artificially short timescales, which are not characteristic of the typical long-period features expected from red-noise-dominated stochastic processes. 

\subsection{Periodic LCs: Results}\label{sec:periodic_lcs}

Here, we repeat the analysis shown in the previous section but consider cases where a true periodic signal is contaminated with noise and distorted by injecting gaps. For the injection of the gaps, we use the same gap scenarios presented in $\S$\ref{sec:types_gaps}.

\subsubsection{LSP} 
Regarding the periodic LCs, the results for LSP are presented in Table \ref{tab:period_2_years} and Table \ref{tab:period_3_years}, illustrating how the presence of gaps affects the characterization of a periodic signal. In all cases, the periodic behavior corresponding to the 2-year and 3-year cycles is successfully recovered in the analysis. However, the impact on the estimated significance associated with the detected period varies considerably (some examples of the period-significance distribution are shown in Figure \ref{fig:appendix_levels}).

The effect of data gaps is more pronounced for the 2-year period signal compared to the 3-year period signal, especially when gaps are randomly introduced into the LC. For example, in the case of the 2-year period, the estimated significance shows a substantial reduction of $\approx$50\% when the percentage of gaps is 90\%, whereas, for the 3-year period, the reduction is only about 18\%, for the same percentage of gaps.

The more substantial influence of gaps on the 2-year period signal can be attributed to several factors. First, the binning is fixed, meaning the sampling frequency is constant. As a result, the 3-year period is sampled more sparsely than the 2-year period, making the latter more sensitive to data gaps and loss of coherence. Consequently, gaps in the data disrupt this pattern more severely, making it more challenging to identify the periodic behavior. Additionally, since a 2-year period naturally results in more complete cycles over the same time span compared to a 3-year period, there are more opportunities for gaps to misalign with critical points in the signal. This misalignment further diminishes the robustness of the detected periodicity. 

Furthermore, as the percentage of gaps increases, the effective duty cycle (the percentage of usable data) decreases. Consequently, for the 2-year signal, a high percentage of gaps may leave only fragmented pieces of cycles remaining, significantly weakening the periodic structure and, thus, the associated significance. Conversely, the 3-year period signal, with fewer required oscillations within the same observational timeframe, tends to preserve a clearer periodic structure despite data gaps.

A similar pattern is observed for the other gap tests, the annual variability with periodic and aperiodic gap distributions. The decrease in significance is smaller for the maximum of the gap percentage 90\%, with reductions of approximately 18\% and 24\% for the 2-year signal and 21\% and 32\% for the 3-year signal for the annual variability with periodic and aperiodic gap distribution, respectively. Although the difference is slight, the ``periodic'' gap distribution shows better results. This suggests that a ``periodic'' distribution may be preferable for observational campaigns with regular observation schedules. The structured timing provided by periodic gaps appears to preserve recurring signals better, while not introducing spurious signals where they are not genuinely present.

In conclusion, in the presence of a periodic signal, data gaps do not distort the ability to infer the signal's period, with significant detections up to high gap percentages ($\leq$70\%). 

\subsubsection{PDM} 

Regarding the PDM analysis, the results presented in Table \ref{tab:period_2_years} and Table \ref{tab:period_3_years} are worse than those obtained using the LSP in all the gap distributions considered. In most cases, the signal's true period is detected with significant confidence until the percentage of gaps reaches 80\%.

\begin{figure}
	\centering
	\includegraphics[scale=0.26]{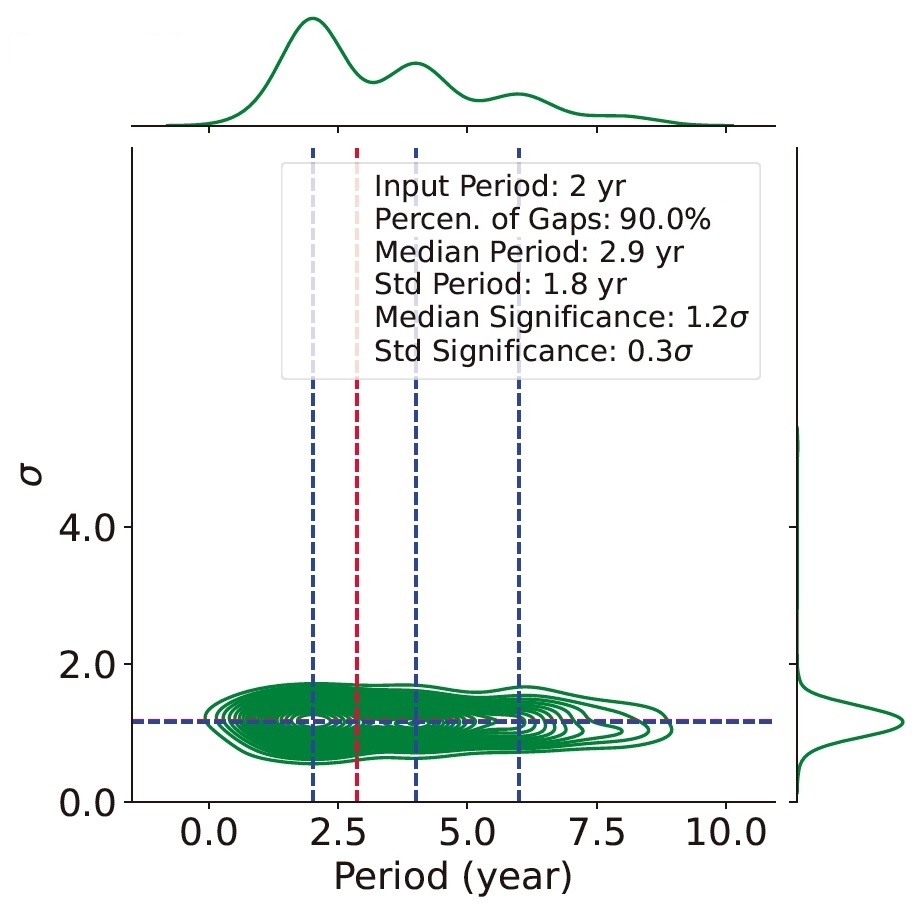}
	\caption{Example of PDM analysis for a 2-year periodic signal with random gaps, where 90\% of the data points are missing. The resulting period distribution reveals three distinct excesses near approximately 2 years, 4 years, and 6 years, indicated by the blue dotted lines. These values correspond to harmonics of the true period present in the signal, as discussed in \citet[][]{penil_2020}.} \label{fig:pdm_harmonics}
\end{figure}

Additionally, the uncertainty in the inferred period increases rapidly as the percentage of gaps rises, indicating a greater likelihood of obtaining an incorrect period. This trend is not observed in the LSP results, where the uncertainties in the mean period remain lower than those of PDM. For example, with a percentage of gaps of 50\%, the period uncertainty is about 35\% of the period for PDM, compared to only 1\% for LSP. Figure~\ref{fig:pdm_harmonics} shows an example in which harmonics of the 2-year period (e.g., 4 and 6 years) appear in the period distribution. This phenomenon, previously reported in \citet[][]{penil_2020}, becomes more frequent as the number of gaps increases, thereby reducing the reliability of the results. In contrast, this phenomenon did not appear when using LSP.

\subsubsection{SSA}
\begin{figure*}
	\centering
	\includegraphics[scale=0.26]{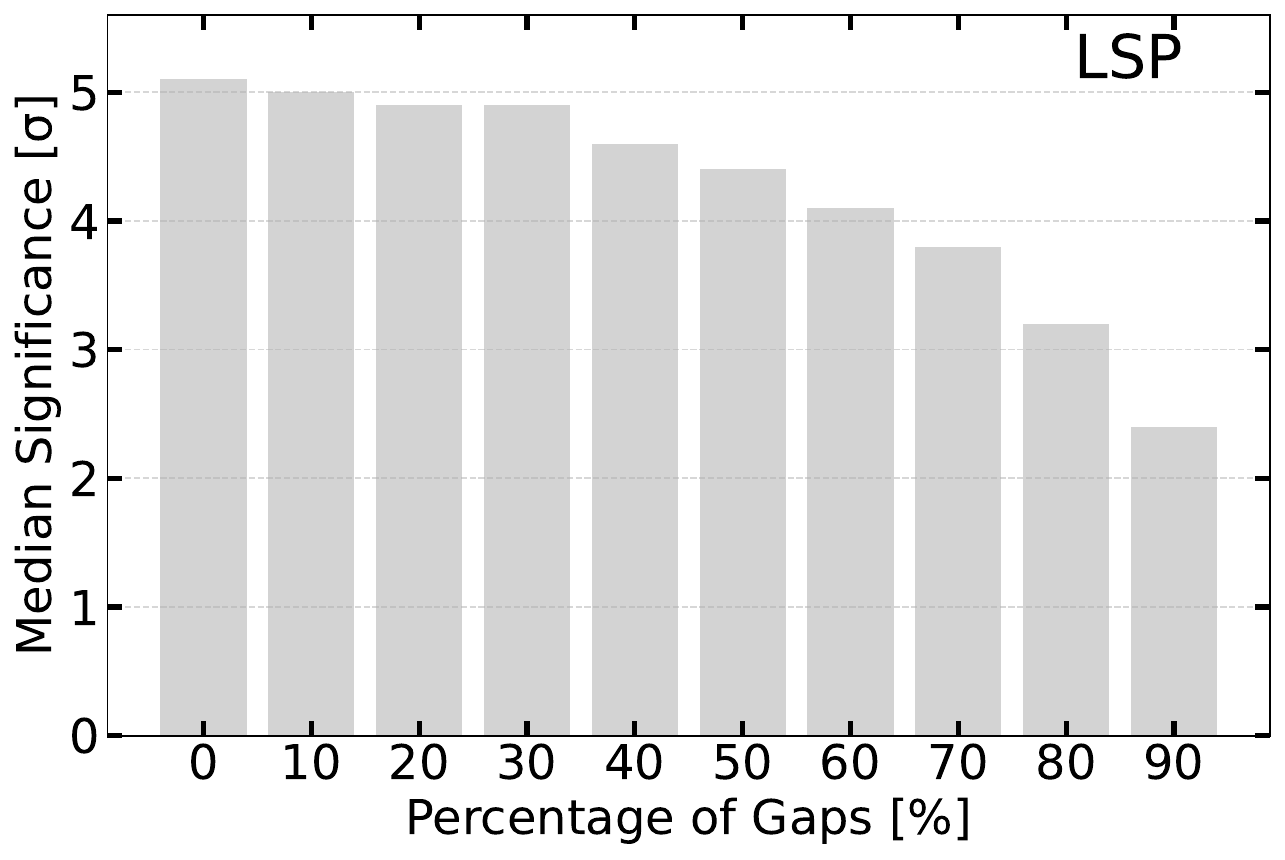}
	\includegraphics[scale=0.26]{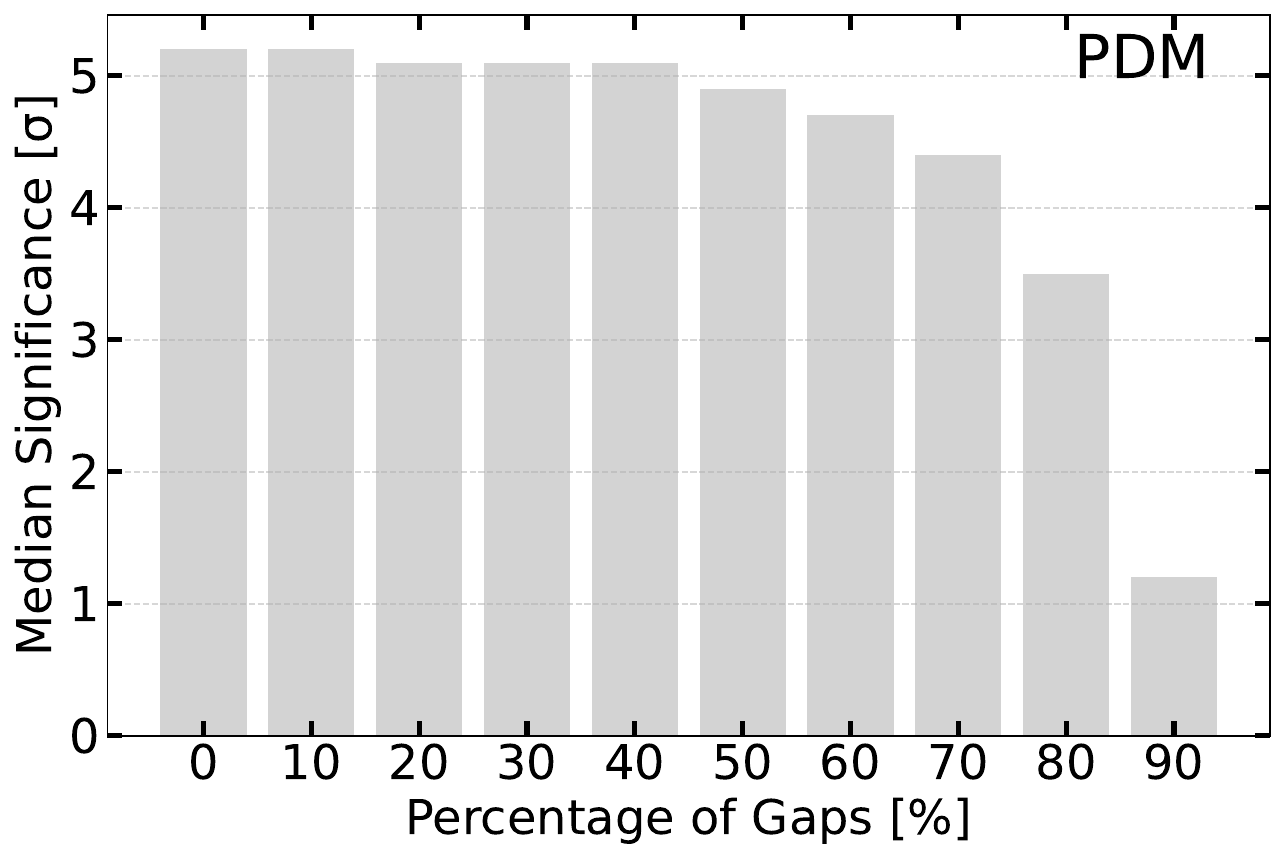}
        \includegraphics[scale=0.26]{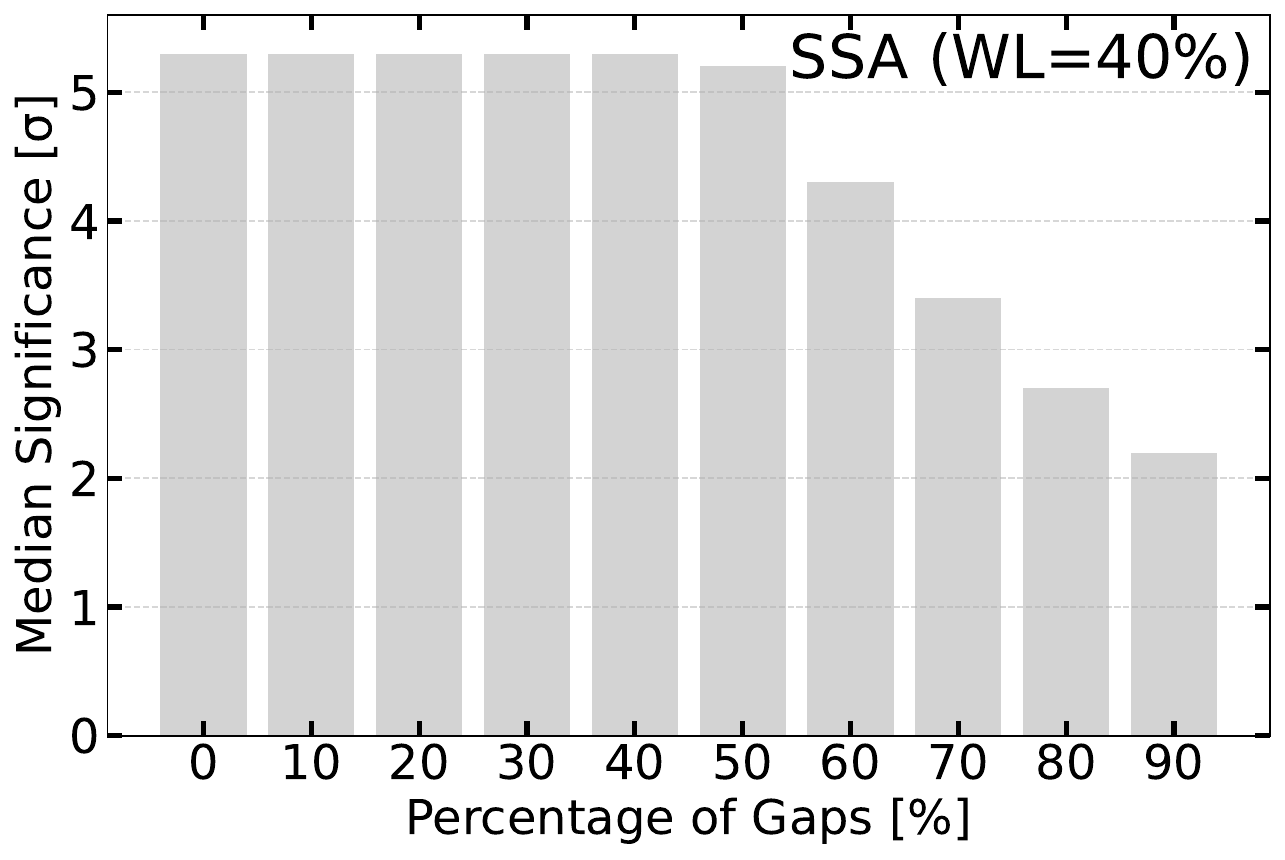}
        \includegraphics[scale=0.26]{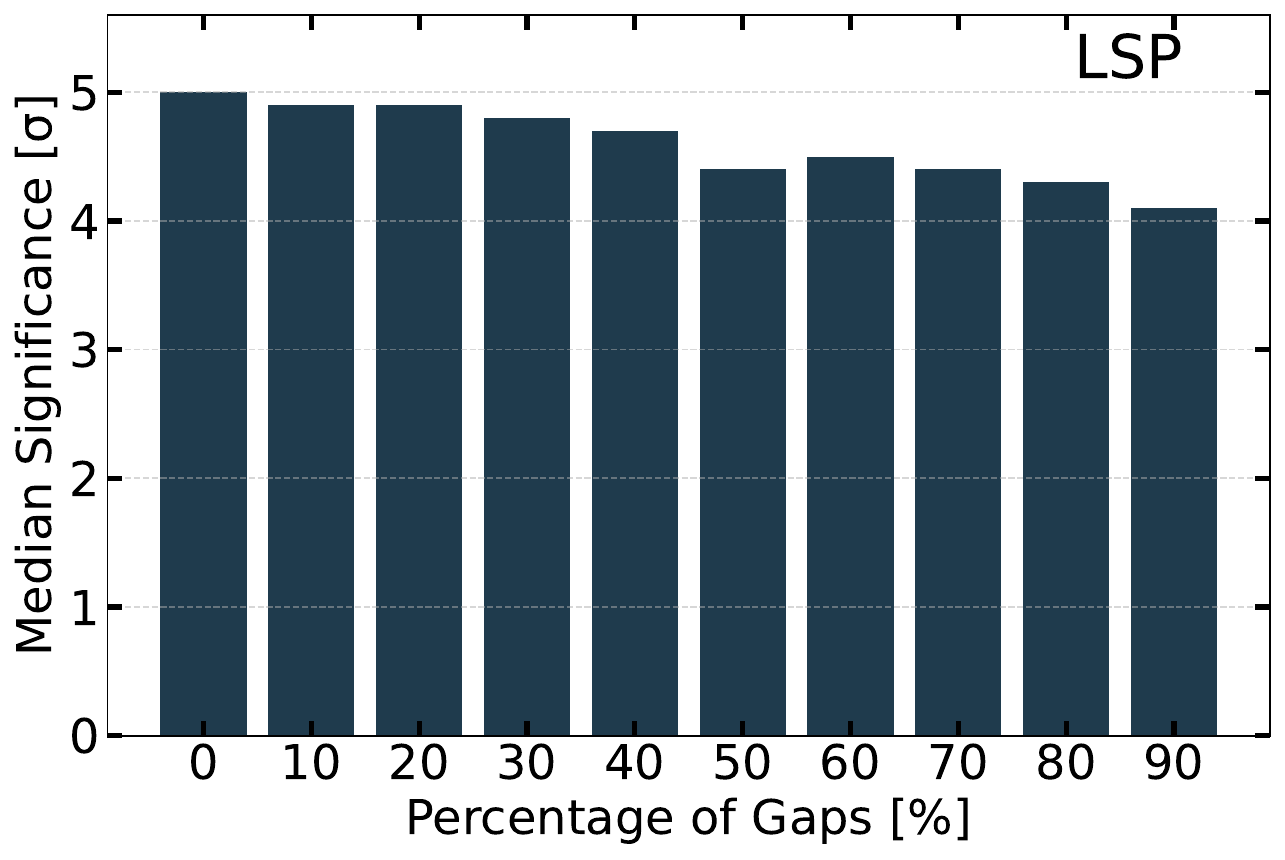}
	\includegraphics[scale=0.26]{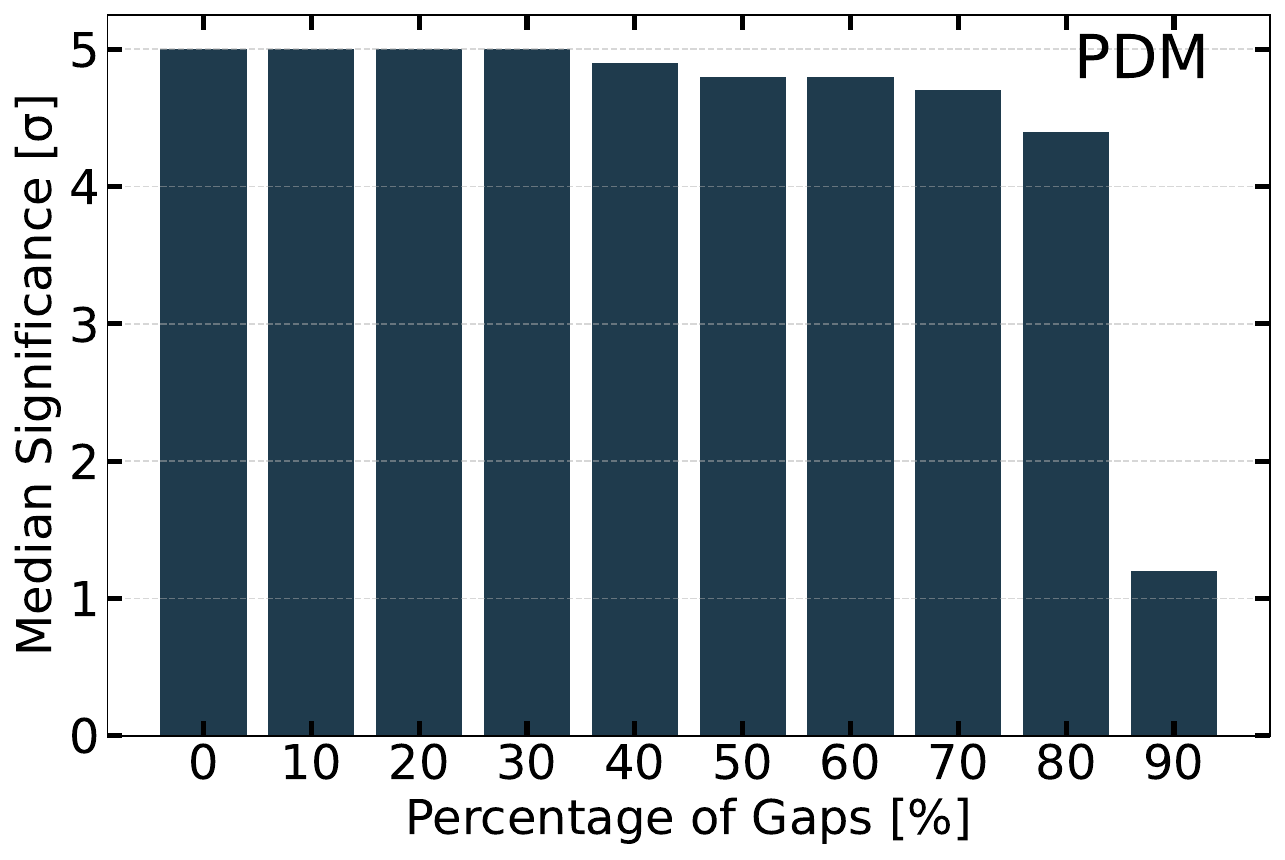}
        \includegraphics[scale=0.26]{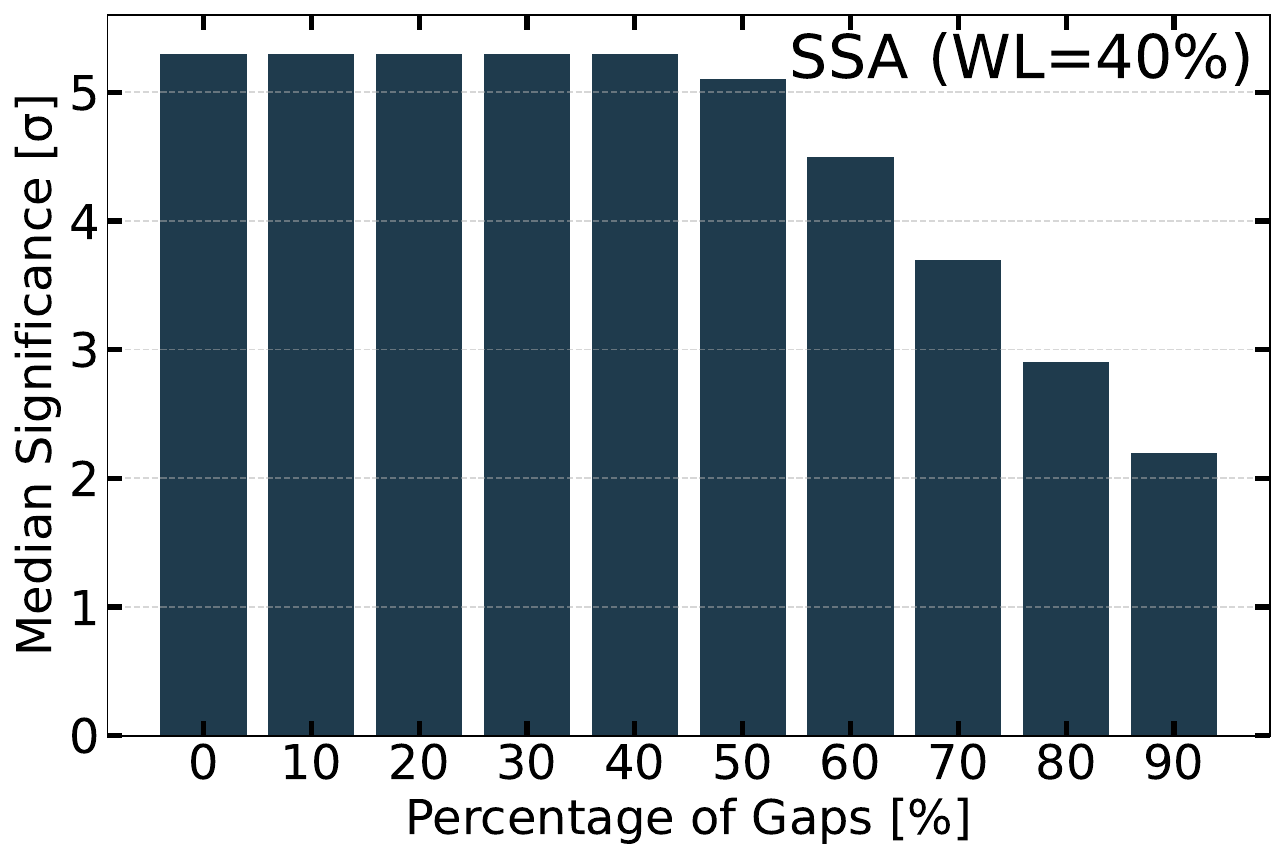}
	\caption{Distribution of the median significance according to the percentage of gaps in a random distribution. \textit{Top}: Periodic signal of 2 years. \textit{Bottom}. Periodic signal of 3 years. \textit{Left}: LSP. \textit{Center}: PDM. \textit{Right}: SSA with window length (WL) of 40\%. The plots show how the application of LSP and SSA is more sensitive to the presence of gaps than PDM. This fact could be due to the way both methods manage the analysis of the time series.} \label{fig:significant_detections_2_3_periods}
\end{figure*}

The results are shown in Table \ref{tab:period_2_years_ssa} and Table \ref{tab:period_3_years_ssa}. SSA consistently identifies the same period across all analyzed scenarios. For random gaps, the significance results are comparable to those of directly using LSP, and slightly worse ($\approx$10\%) than directly using PDM when the percentage of gaps is $\geq$60\% (see Figure \ref{fig:significant_detections_2_3_periods}). In the cases of annual gaps (both periodic and aperiodic yearly distributed), SSA shows better performance in terms of significance, maintaining a significance level of approximately 5$\sigma$ up to 80\% of gaps, compared to 50\% for LSP and for PDM. No significant differences are observed between the 2-year and 3-year signals (Figure \ref{fig:significant_detections_2_3_periods}).

The first key conclusion is that SSA can extract the oscillatory components in moderate gap LCs. Regarding the performance of LSP, PDM, and SSA, both LSP and SSA exhibit a decrease in significance by $\approx$30\% in the presence of randomly distributed gaps, particularly when the percentage of gaps exceeds 70\%, whereas PDM is less affected ($\approx$10\%) under the same conditions (Figure~\ref{fig:significant_detections_2_3_periods}). This increased sensitivity could be related to the inherent design of LSP, which aims to extract sinusoidal components from the data. The presence of gaps disrupts the continuity and phase coherence needed for accurate sinusoidal reconstruction, often leading to a significant reduction in the detection significance of genuine periodic signals. Regarding SSA, its reduced performance in the presence of large random gaps could arise from the disruption of the trajectory matrix construction. This matrix encodes the temporal structure of the time series and is essential for separating trends, oscillations, and noise components \citep[][]{alba_ssa}. 

When the percentage of gaps exceeds a critical threshold (e.g., 70\%), the lagged embedding vectors become incomplete, compromising the matrix’s structure. As a result, the decomposition into meaningful signal components breaks down, leading to a significant loss in the ability to detect quasi-periodic patterns.

In contrast, PDM could be less affected by such discontinuities, as it does not assume any specific functional form but instead identifies periodicity by minimizing the variance of phase-folded data. However, a limitation of PDM is its relatively high period uncertainty. Even at moderate gap levels (e.g., $\geq$40\%), PDM often yields less precise period estimates, increasing the likelihood of identifying an inaccurate or ambiguous periodicity. This could be due to the emergence of harmonics of the true period previously mentioned (Figure~\ref{fig:pdm_harmonics}).

When the gap distribution follows an annual pattern (whether periodic or aperiodic), SSA exhibits better performance regarding LSP and PDM, with differences in significance remaining within 10\% to 50\%. This suggests that SSA is more robust to seasonally modulated sampling patterns, preserving a larger fraction of the underlying signal’s detectability compared to the other methods. 

A conclusion is that for gap percentages of $\leq$50\%, all methods yield consistent and compatible results (Figure \ref{fig:significant_detections_2_3_periods}). Therefore, setting an upper limit of 50\% on the percentage of gaps in a $\gamma$-ray LC appears to be a reasonable threshold for conducting reliable periodicity analyses \citep[][]{penil_2020, alba_ssa, penil_2022}. Below this limit, the integrity of the periodic signal is generally preserved across methods, minimizing the risk of spurious detections or significant biases in the estimated period and significance.

\subsection{Evaluating the Window Length in SSA}\label{sec:window_lengh}
In SSA, the window length is a fundamental parameter that directly influences the method's ability to decompose a time series into characteristic components, such as trends, periodicities, and noise. The choice of window length must carefully balance two objectives: it should be large enough to capture the temporal structures of interest (e.g., underlying periodic signals or long-term trends), yet not so large that it makes the grouping of the components more difficult, introducing artificial components or overfitting the noise. Selecting an inappropriate window length may either obscure real features or amplify random fluctuations, thereby undermining the reliability of the analysis. However, there are no strict guidelines for selecting a specific window length in SSA. 

There is no universally accepted rule for selecting the window length in SSA \citep[][]{golyandina_ssa}. However, the window length can not exceed 50\% of the total length of the time series. This parameter should be selected based on the analysis objectives and the specific characteristics of the signal under study. For the purpose of detecting periodicities in LCs, it is often recommended to begin with a relatively large window length and then reduce it if necessary. Care must be taken, as a large window can lead to a more complex decomposition, making interpretation difficult. Conversely, selecting a window that is too small may overly smooth the signal, potentially obscuring important features. Therefore, it is recommended to perform the analysis with different window lengths, starting from larger values and reducing them as needed to balance resolution and interpretability. A well-chosen window length allows SSA to efficiently separate the signal from noise and the other components of interest. 

In the previous test, we used a window length of 40\%, following the approach adopted in \citet{alba_ssa}. In this section, we extend the analysis by exploring shorter window lengths, specifically 30\% and 20\% of the total LC length, to evaluate their potential influence on the outcomes of SSA. The corresponding results are presented in Table~\ref{tab:noise_random_ssa}, Table~\ref{tab:period_2_years_ssa}, and Table~\ref{tab:period_3_years_ssa}.

\begin{figure*}
	\centering
	\includegraphics[scale=0.24]{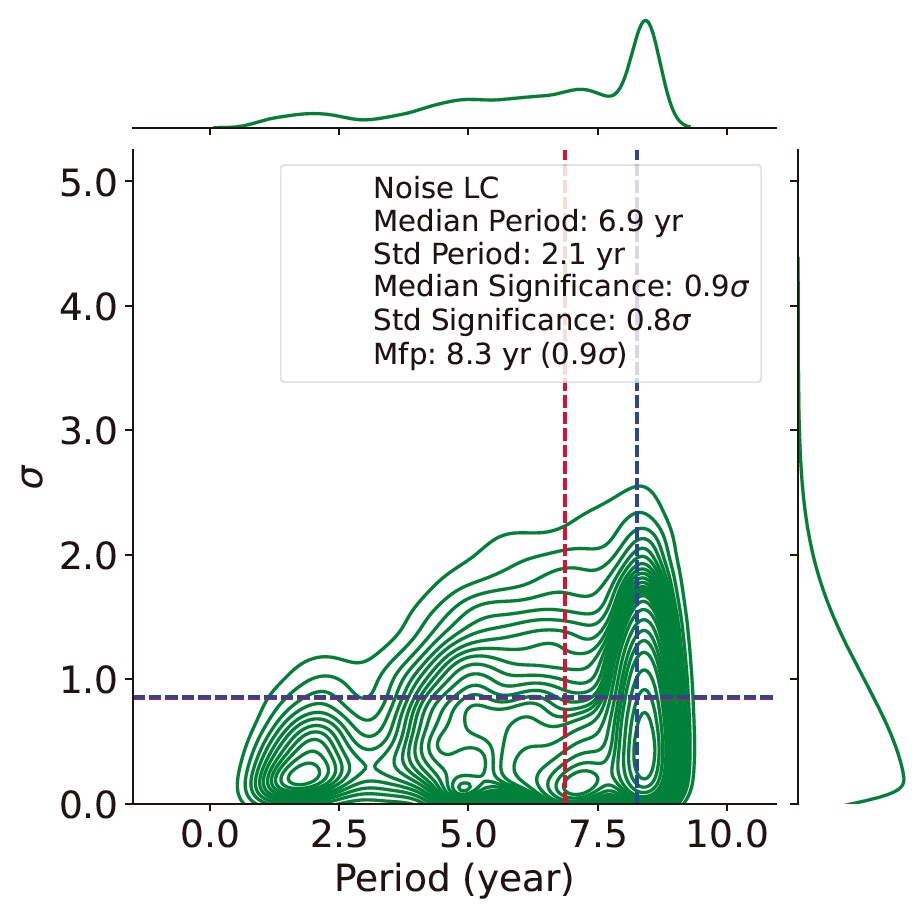}
        \includegraphics[scale=0.24]{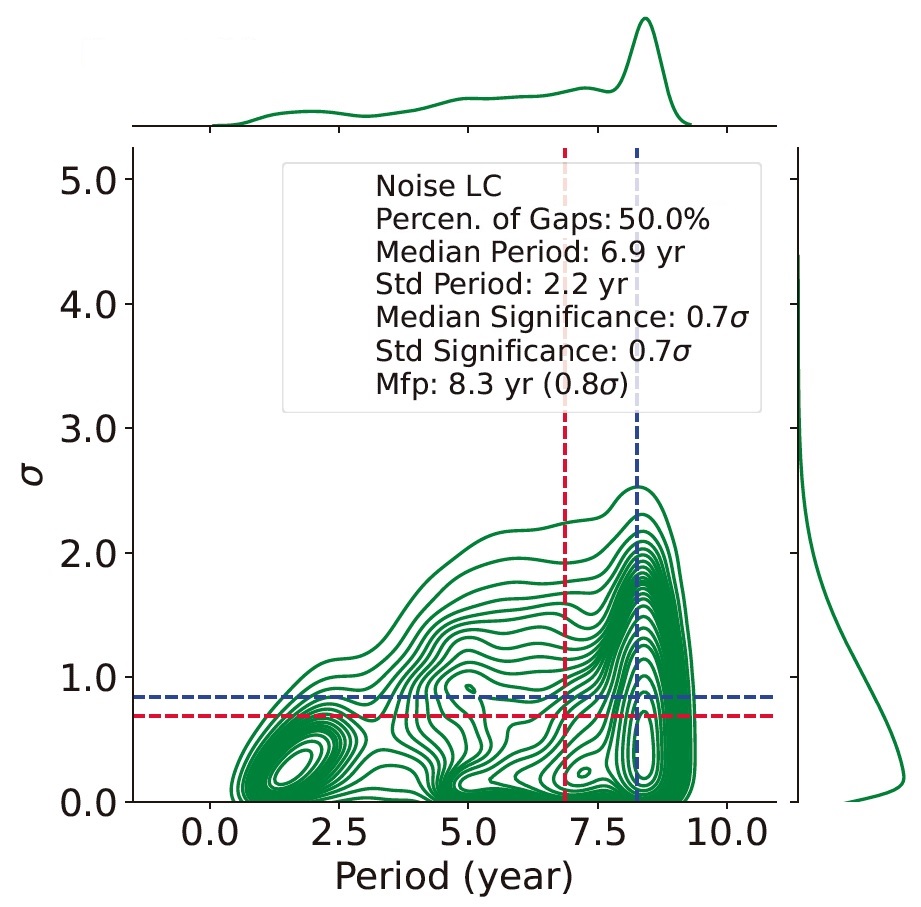}
        \includegraphics[scale=0.24]{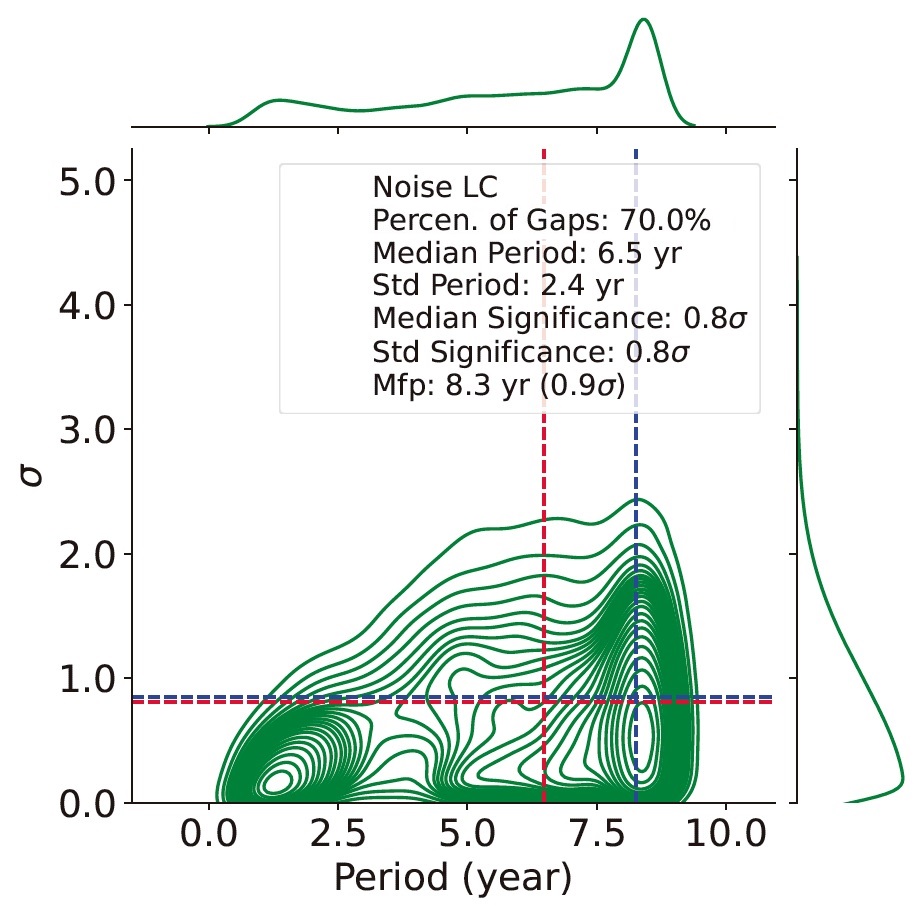}
        \caption{Distributions for the period and significance for pure noise LCs, for SSA using a window length of 20\%. \textit{Left}: no gaps. \textit{Center}: 50\% of randomly distributed gaps. \textit{Right}: 70\% of randomly distributed gaps. These results highlight key differences compared to those presented in Figure \ref{fig:testing_pure_noise_ssa}, which used a 40\% window length. Specifically, the number of significant detections ($\geq$3$\sigma$) is lower when using the 20\% window, indicating a reduced tendency to identify periodicities in purely stochastic data falsely. The spurious 1-year peak observed in the 70\% gap scenario in Figure \ref{fig:testing_pure_noise_ssa} does not appear here, further supporting the conclusion that shorter window lengths are more effective at suppressing gap-induced artifacts. Overall, the 20\% window length yields more reliable results for noise-dominated LCs and helps prevent false detections that could arise from data sampling effects.     
        The dotted red vertical and horizontal lines indicate the median values for both the period and the significance of the test. The blue dotted vertical line highlights the most frequently occurring period in the tests (and the associated significance), emphasizing its prominence in the distribution. The ``Percen. of Gaps'' refers to the percentage of gaps injected in the LC. The ``Median Period'' represents the median of all periods resulting from the test, and the ``Std Period'' is the standard deviation of such periods' distribution. The ``Median Significance'' represents the median of the significance distribution associated with the test, and the ``Std Significance'' is the standard deviation of this significance distribution. ``Mfp'' represents the most frequent period resulting from the test.} \label{fig:testing_pure_noise_ssa_window02}
\end{figure*}

For the pure noise LCs, we observe differences in the period detection behavior depending on the selected window length. With a 40\% window length, the transition from the most frequently detected 8.3-year period to a spurious $\sim$1.1-year period occurs when the percentage of gaps reaches approximately 70\%. However, when the window length is reduced to 30\% and 20\% (Figure \ref{fig:testing_pure_noise_ssa_window02}), this transition shifts to higher gap levels, specifically between 80\% and 90\%. This indicates that shorter window lengths make SSA less sensitive to the structural distortions introduced by high levels of missing data. This behavior can be understood in terms of how SSA captures oscillatory modes. A larger window length allows for the extraction of longer-period components but may also increase susceptibility to spurious periodicities caused by structured sampling patterns, especially when data gaps are prominent. In contrast, shorter window lengths constrain the decomposition to shorter segments of the time series, which may help reduce the influence of sampling artifacts, thus preserving the integrity of the analysis under heavily gapped conditions, capturing the dominant patterns of the signal. Additionally, focusing on the percentage of significant detections, we observe similar results regarding the rate of significant periods, $\approx$0.20\% with no 5$\sigma$ detections.  

For LCs containing a true periodic signal, the period detection performance of SSA remains consistent across the different window lengths tested. The detected period is accurately recovered in all cases, indicating that moderate changes in window size do not significantly alter the method’s ability to isolate the underlying oscillatory component, as long as the component decomposition and grouping are performed correctly. However, we observe that the significance of the detected signal gradually decreases as the percentage of gaps increases. Specifically, for LCs with 90\% of their data missing, the significance drops by approximately 30\% compared to the gap-free case. 

These results indicate that SSA is generally robust to variations in window length when applied to LCs containing periodic signals, even in the presence of substantial observational gaps. However, when applied to pure noise LCs, shorter window lengths, such as 20\%, demonstrate a lower trend to shift to a 1-year peak. 

\subsection{Theoretical Basis of SSA Gap Resistance}
This robustness stems from the use of low-rank approximations of the trajectory matrix: even when sparsity is introduced by missing values, the decomposition preserves the main temporal structure. A central element in understanding the robustness of SSA to missing data is the condition number of the trajectory matrix. The condition number, denoted $\kappa_{2}$, is defined as the ratio of the largest to the smallest singular value of the matrix. It measures how sensitive the outcome of the SVD is to perturbations in the input. A low $\kappa_{2}$ indicates that the decomposition is stable and reliable, while a high $\kappa_{2}$ signals numerical instability, affecting the extracted components. In this sense, $\kappa_{2}$ provides a diagnostic of the gap resistance of SSA.

To study the gap effect, we simulated condition numbers by varying the percentage of gaps and window lengths. Our simulations show that, relative to the no-gap baseline, $\kappa_{2}$ increases only modestly (by $\sim$10–20\%) when up to 20\% of gaps, meaning that SSA can still recover dominant trends and oscillations with little loss of stability. With 30-50\% gaps, the relative increase in $\kappa_{2}$ reaches $\sim$50–100\%, reflecting growing sensitivity to the percentage of gaps but still within a manageable range to recover the components of the LC. When more than half of the data are absent, however, $\kappa_{2}$ rises by several hundred percent, and above 70\% it increases by several orders of magnitude. This marks a critical threshold beyond which the low-rank approximation loses numerical stability, and SSA can no longer be expected to recover reliable signal components from the trajectory matrix.

The role of window length is twofold. For moderate gap levels, longer windows suppress the relative growth of $\kappa_{2}$, as they better capture long-range correlations and mitigate sparsity effects by averaging across more overlapping lagged vectors. However, under high percentages of gaps ($\geq$70\%), this advantage disappears because the added rows of the trajectory matrix contain too few valid entries to provide independent information, which amplifies numerical instability. In such cases, $\kappa_{2}$ grows faster for a large window length, as the matrix becomes dominated by empty structure rather than signal. Consequently, shorter windows are comparatively more robust when the series is heavily gapped, while larger windows remain preferable when gaps are moderate, since they strike a balance between capturing long-term dynamics and maintaining numerical stability. This dual behavior highlights the need to adapt the window length to the degree of data sampling rather than applying a single choice universally.

\subsection{Testing the origin of the $\sim$1-year period}
Here, we test whether the $\sim$1-year period observed in our analysis arises from the gap structure of the LC. To quantify the role of the gaps, we simulated a red-noise LC with a uniform 28-day cadence and analyzed its PSD. We then constructed the associated spectral window by computing the periodogram of a unit-amplitude series sampled at the same epochs. The spectral window provides a direct measure of how the time-sampling pattern redistributes power in the frequency domain: uniform sampling produces a flat response, while irregular sampling can introduce sidelobes and aliases that can overlap with genuine spectral features.

\begin{figure}
	\centering
	\includegraphics[scale=0.27]{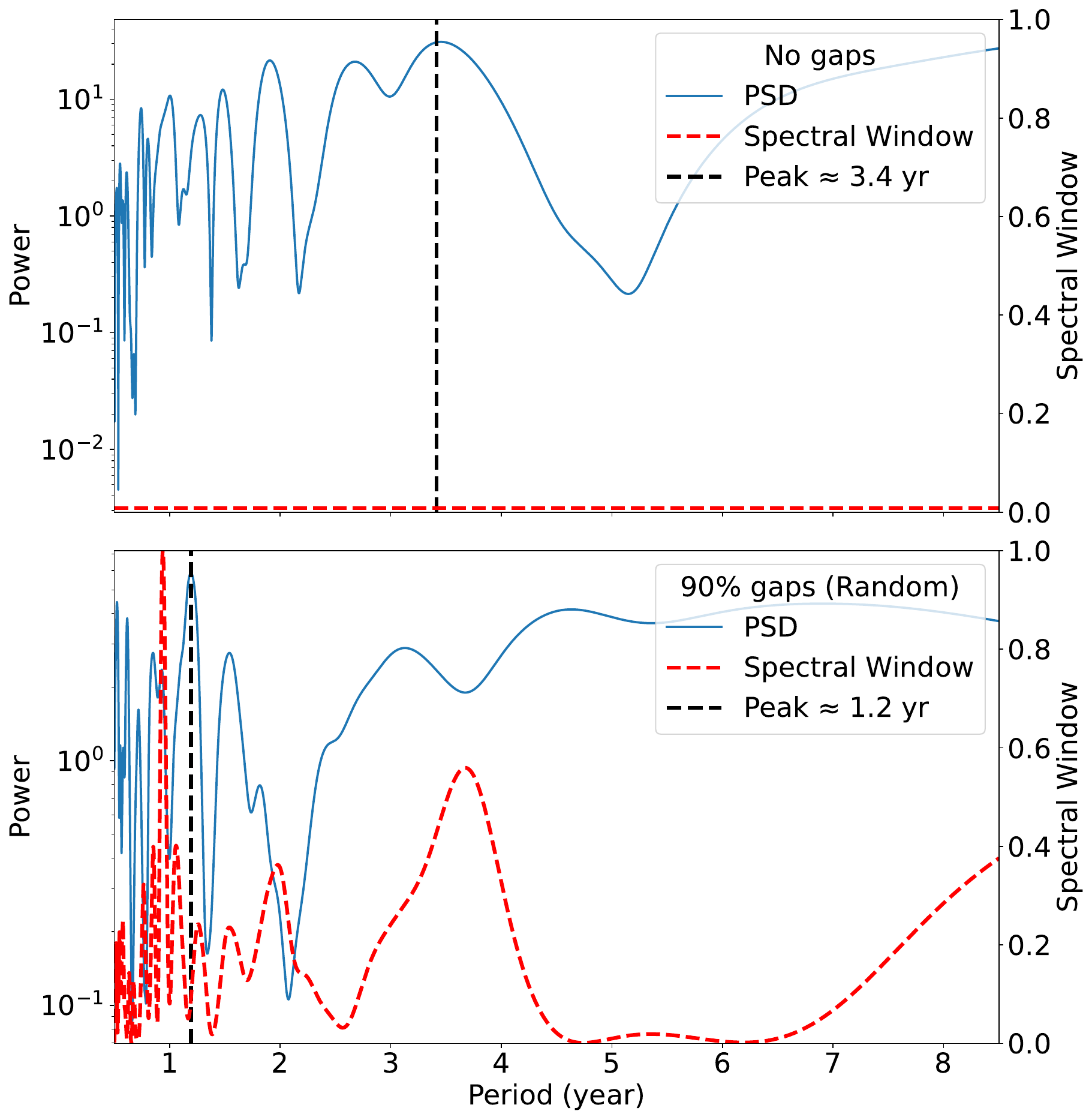}
	\caption{PSD (blue) of a red-noise LC sampled every 28 days, compared with the corresponding spectral windows (red, normalized to unity on the right axis). \textit{Top}: Uniformly sampled case with no missing data. The PSD shows the expected red-noise trend with a broad maximum at $\sim$8.5 yr, while the spectral window is nearly flat, indicating that regular sampling introduces no significant aliases. \textit{Bottom}: Case with 90\%  random gap distribution. The PSD is now dominated by a spurious peak at $\sim$1 yr, which coincides with strong structure in the spectral window. This shows that under an extreme percentage of gaps, the observed PSD is driven primarily by the sampling pattern rather than by intrinsic variability.} \label{fig:spectral_window}
\end{figure}

Figure \ref{fig:spectral_window} compares two different scenarios. In the uniformly sampled case (top panel), the PSD follows the expected red-noise slope with excess power at long timescales, and the dominant maximum occurs at $\sim$8.5 yr. This broad peak reflects stochastic fluctuations and the finite duration of the simulation rather than a meaningful periodicity. The spectral window is nearly flat, confirming that uniform coverage does not inject artificial signals into the PSD.

By contrast, when 90\% of the points are removed randomly, the PSD is dominated by a sharp peak at $\sim$1 yr. The spectral window exhibits strong power at the same period, and the coincidence between the two demonstrates that the apparent peak is an alias introduced by the sampling pattern rather than an intrinsic feature of the signal. We repeated this test for the other two gap-distribution scenarios discussed in $\S$\ref{sec:types_gaps}, obtaining consistent results in which the $\sim$1 yr peak can be explained by the gaps and their influence in the spectral window.

This test highlights that peaks at or near 1 year must be interpreted with caution in real data sets. Under high levels of gap distribution, the spectral window can dominate the PSD and generate spurious features that mimic quasi-periodic oscillations.

\section{Real Use Cases} \label{sec:usecase}
We begin by conducting a periodicity-significance study using the LSP, PDM, and SSA methodologies on objects from the sample described in $\S$\ref{sec:gaps_in_lcs}, which have a percentage of gaps $\geq$50\% randomly distributed. This results in a sample of 442 objects. To evaluate the significance of the previous sample using LSP and PDM, we simulate 100,000 artificial LCs following the procedure outlined by TK.

While we acknowledge that this procedure does not fully reproduce the statistical characteristics of real blazar LCs and may overestimate the significance of detected signals \citep[][]{jorge_2022}, it offers key practical advantages. The technique by \citet{emma_lc}, though more accurate, involves higher computational demands, making it less feasible for large-scale testing. Given the extensive size of our sample and the exploratory nature of this initial analysis, we opt for the more computationally efficient method. Our primary objective at this stage is to evaluate the impact of data gaps on periodicity detection rather than to achieve the highest possible accuracy in significance estimation. Prioritizing computational efficiency allows us to perform a broad and systematic evaluation across a wide range of test scenarios. Despite its limitations, this approach enables us to maintain methodological consistency with the main analysis described in Section~\ref{sec:results}. For the refinement of the results presented in Section~\ref{sec:usecase}, where a more detailed significance estimation is essential, we adopt the more robust simulation framework introduced by \citet{emma_lc}. This two-step strategy allows us to balance scope and precision in our investigation.

\begin{figure*}
	\centering
	\includegraphics[scale=0.335]{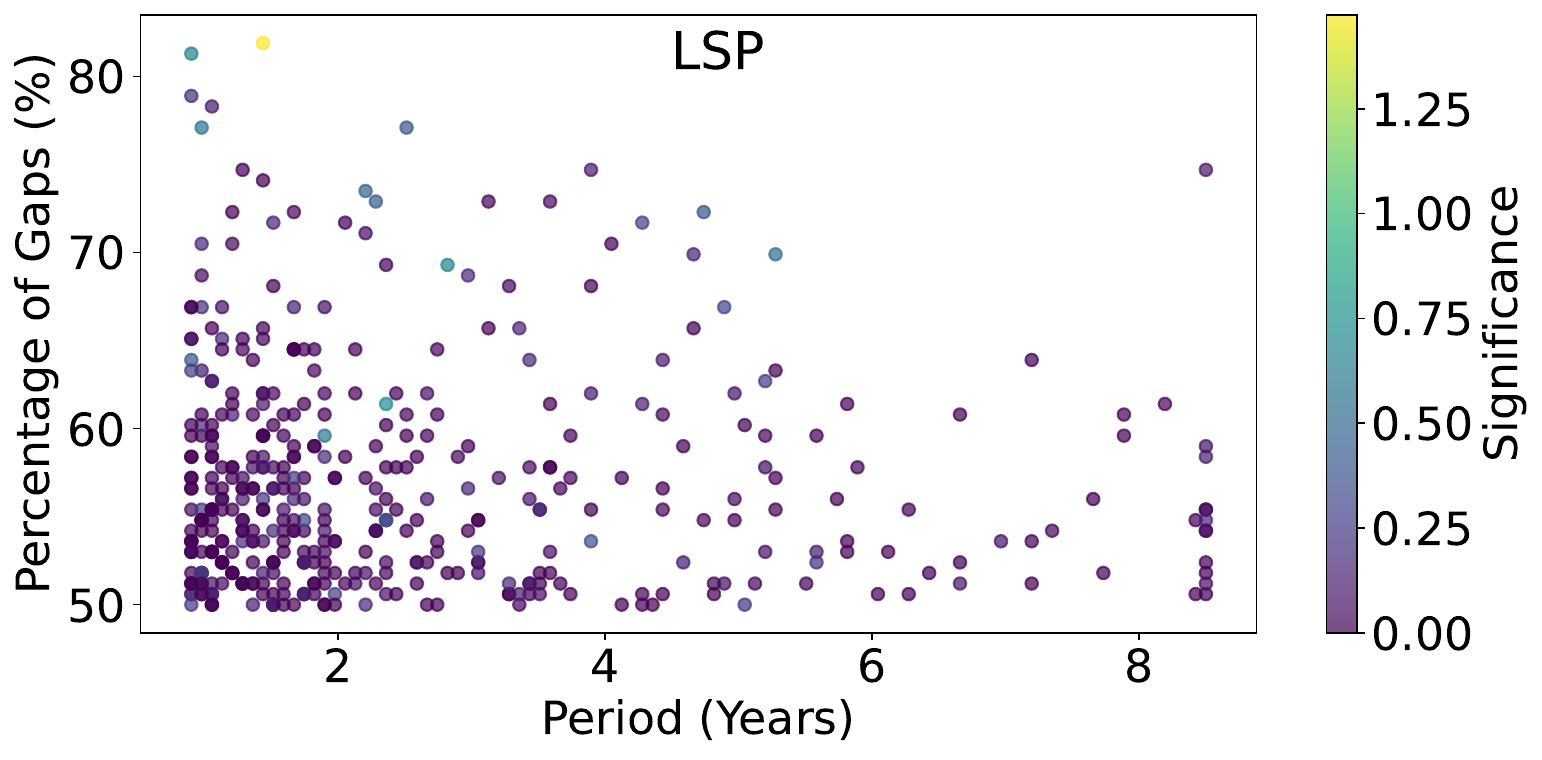}
        \includegraphics[scale=0.335]{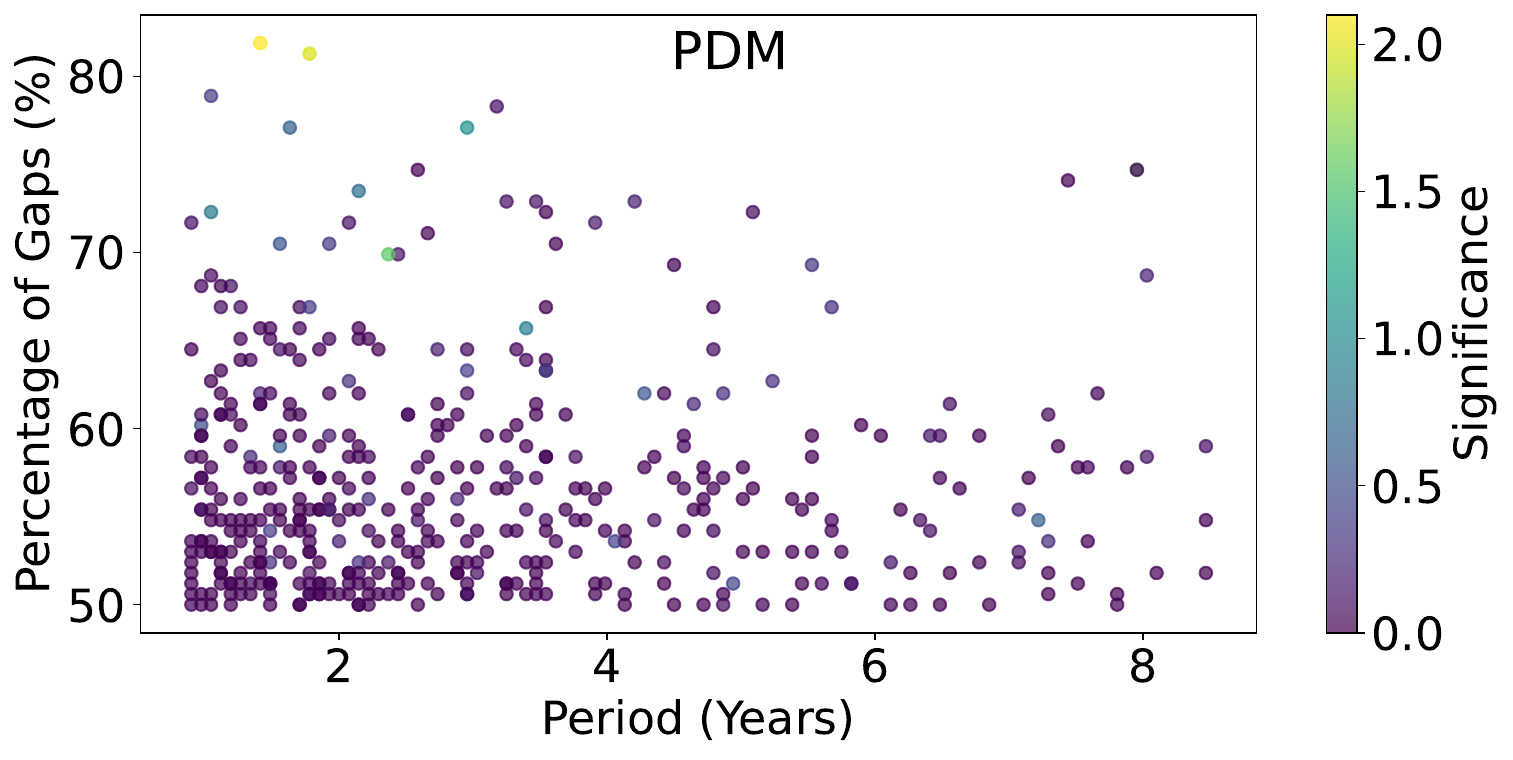}
        \includegraphics[scale=0.335]{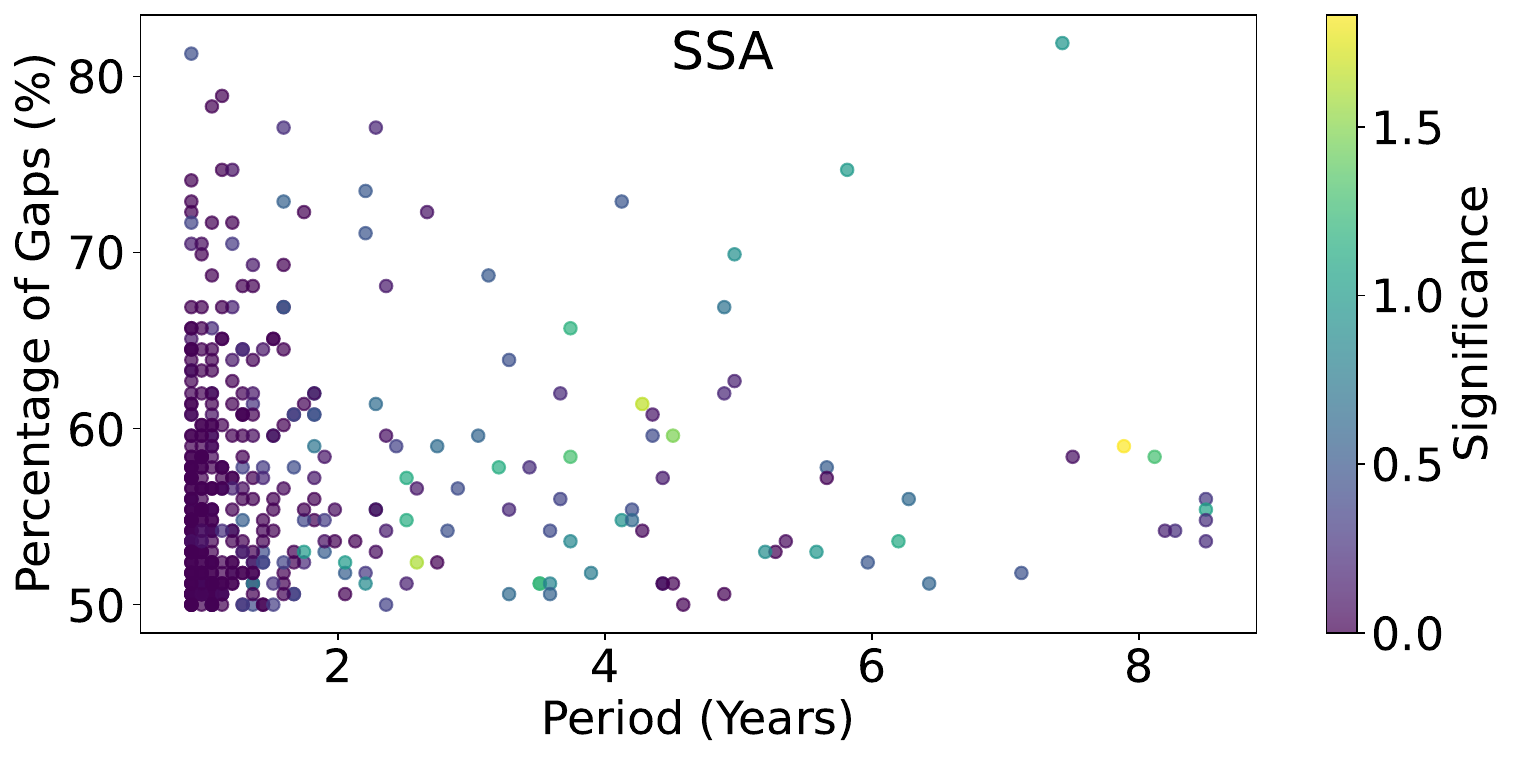}
        \caption{Distributions of the period and significance for \textit{Fermi}-LAT LCs with a gap percentage $\geq$ 50\%. The sample includes 442 objects.  \textit{Top Left}: LSP results show a tendency to detect periods in the 1-2 year range for 42\% of the sample, typically with low significance ($\approx$0.2$\sigma$). \textit{Top Right}: PDM results reveal a similar trend, with 58\% of the sample displaying periods in the 1-2 year range ($\approx$0.1$\sigma$). \textit{Bottom}: SSA results show a clear trend in periods in the range [1-2] years for 82\% of the sample ($0.1\sigma$).} \label{fig:fermi_lat_objects_distribution}
\end{figure*}

The results, shown in Figure \ref{fig:fermi_lat_objects_distribution}, reveal an excess of objects with periods in the range of 1-2 years. This excess is observed in both LSP and PDM methods, with 42\% of objects identified by LSP and 58\% by PDM. The median significance associated with the inferred periods is 0.1$\sigma$ for LSP and 0.2$\sigma$ for PDM. These results are consistent with the one obtained in $\S$\ref{sec:results}. Additionally, the percentage of significant detections ($\geq$3$\sigma$) for LSP and PDM is 0.0\%, consistent with the results of $\S$\ref{sec:results}. 

For the SSA, we adopt a window length of 20\% based on the results discussed in $\S$\ref{sec:window_lengh}. This choice is motivated by the balance it offers between preserving sensitivity to true periodic signals and minimizing the impact induced by data gaps. The analysis reveals a strong tendency to detect periods in the range of 1 to 2 years. SSA identifies periods within this range in 83\% of the sample, with a median significance of 0.1$\sigma$. This concentration around short periods suggests a potential bias in SSA, particularly in datasets with a high percentage of missing data, as previously noted in $\S$\ref{sec:alternative_methods}. Furthermore, SSA yields no significant detections.

To evaluate the sensitivity of SSA to the choice of window size, we conduct a similar test using a window length of 40\%. The results are consistent with those obtained using 20\%, with 82\% of detections falling within the 1–2 year range, and the median significance remains unchanged at 0.1$\sigma$. No significant detections are obtained.

\subsection{Evaluation of Real Cases}\label{lab:use_cases}
\begin{figure*}
	\centering
        \includegraphics[scale=0.22]{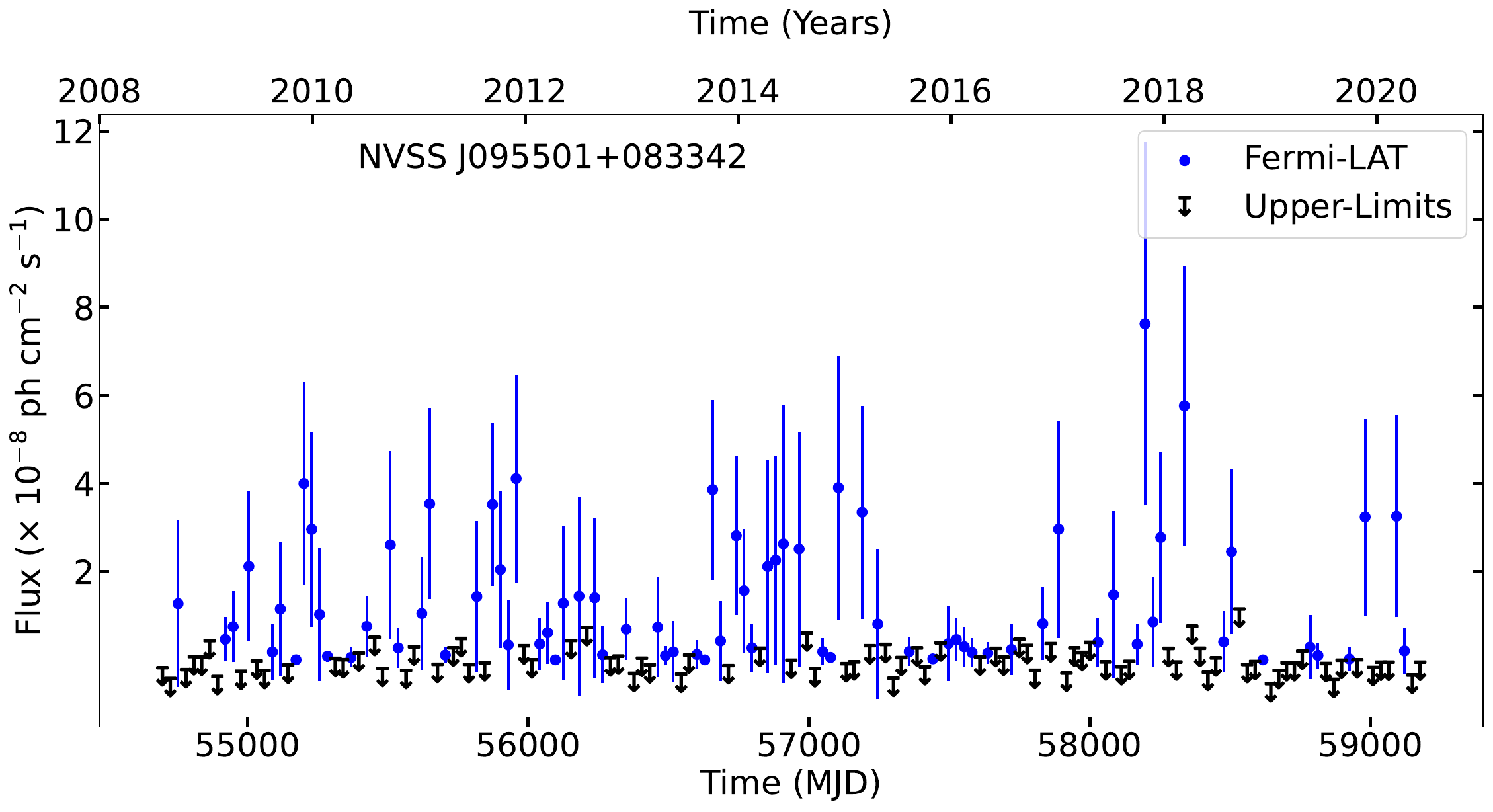}
        \includegraphics[scale=0.22]{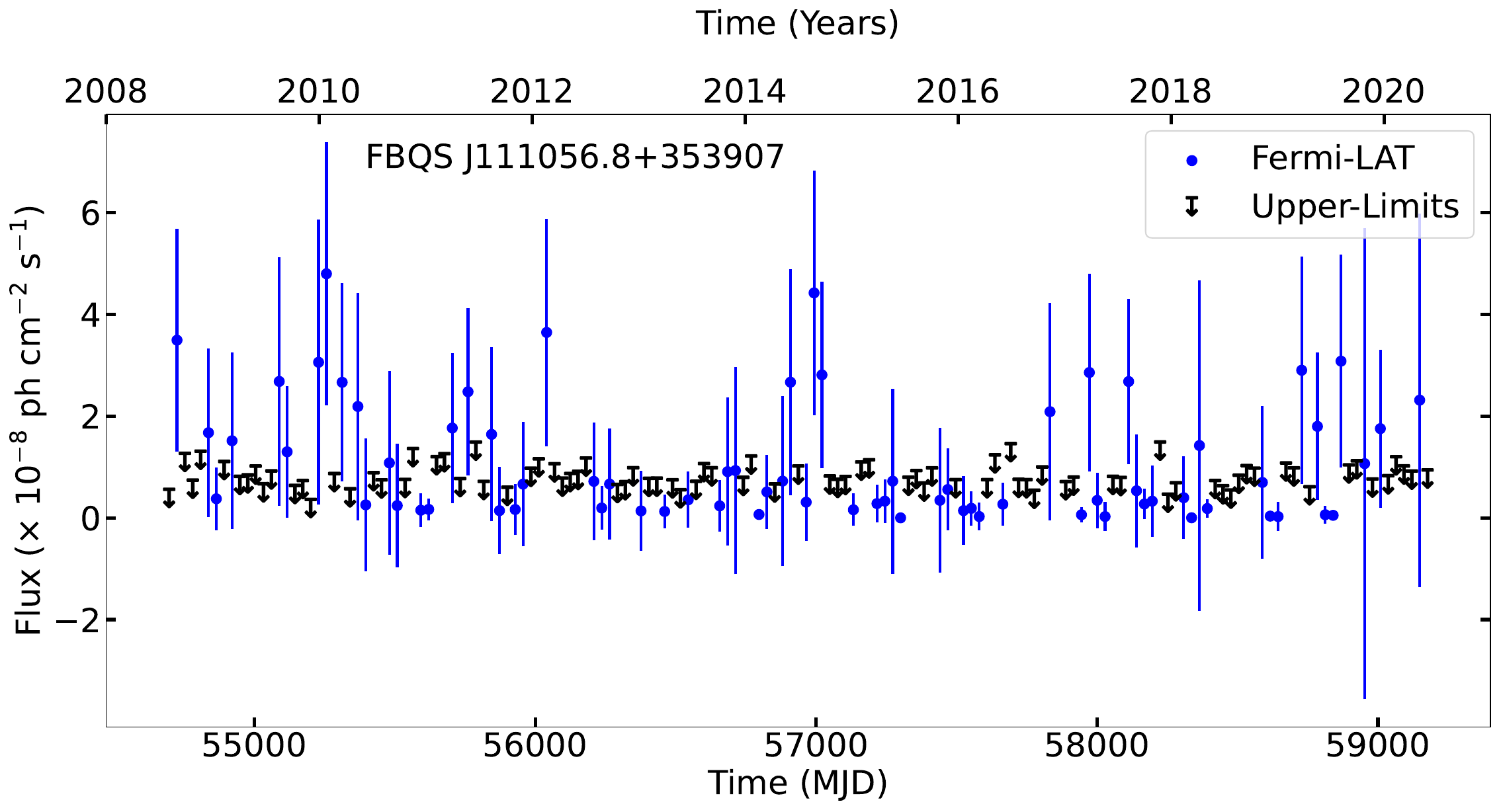}        
        \includegraphics[scale=0.22]{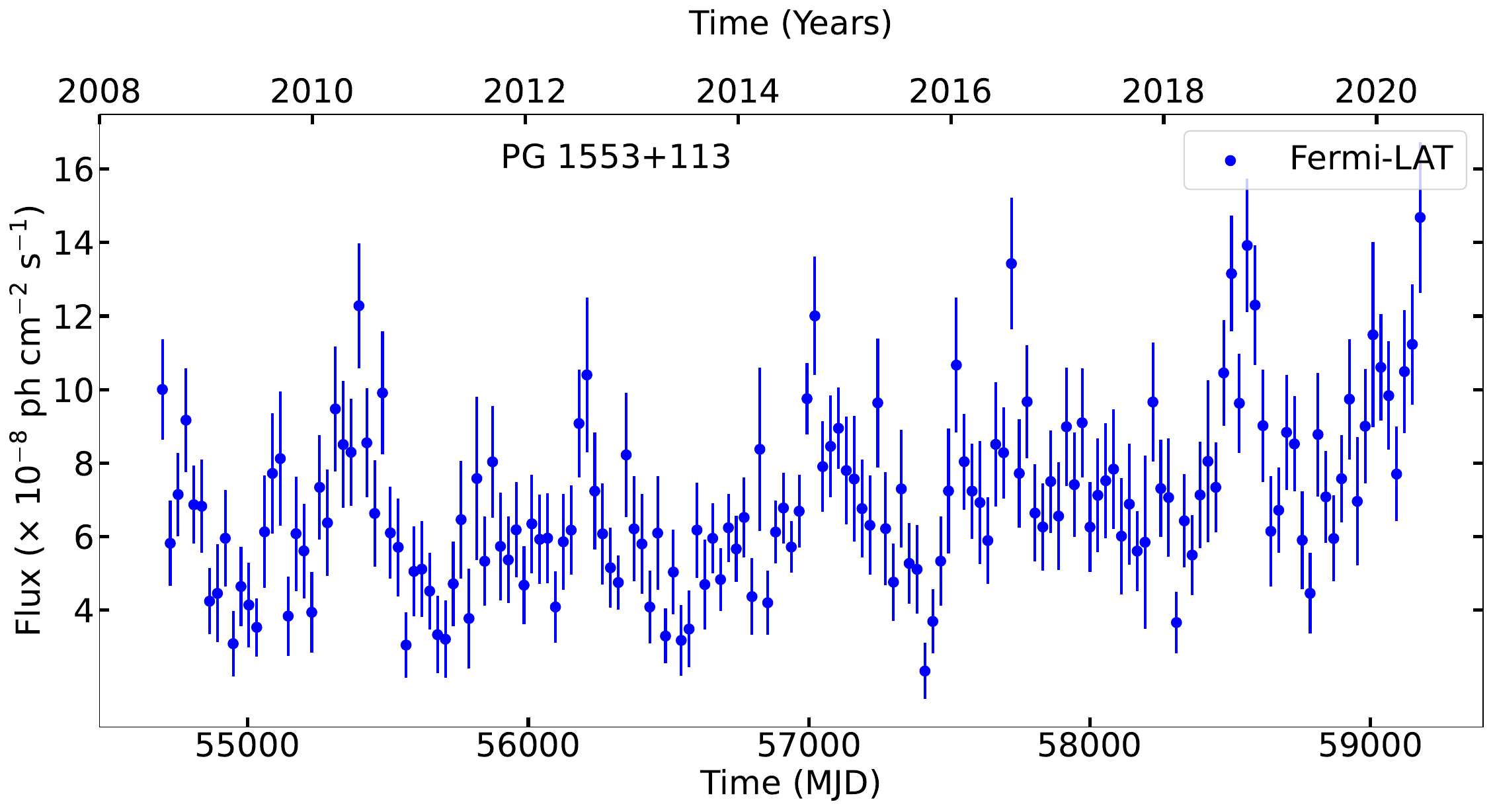}
	\caption{LCs of real data of use cases. Top. \textit{Left}: NVSS J095501+083342. \textit{Right}: FBQS J111056.8+353907. Bottom: PG 1553+113.} \label{fig:example_use_cases}
\end{figure*}

Among the previously analyzed objects, 5 exhibit at least the same period with a significance $\geq$1.5$\sigma$. From this subsample, we randomly select two objects (Figure \ref{fig:example_use_cases}): 

\begin{enumerate}
\item NVSS J095501+083342 (4FGL J0955.2+0835, BLL, $z$=unknown)\footnote{Information obtained from the 4FGL-DR2 catalog \url{https://fermi.gsfc.nasa.gov/ssc/data/access/lat/10yr_catalog/}. BLL is ``BL Lacertae'' type. BCU is ``Blazar Candidate of Uncertain type''}; period = 3.3 year; percentage of gaps = 51.2\%.
\item FBQS J111056.8+353907 (4FGL J1111.0+3542, BLL, $z$=unknown); period = 2.6 year; percentage of gaps = 52.4\%.
\end{enumerate}

\begin{figure*}
	\centering
        \includegraphics[scale=0.21]{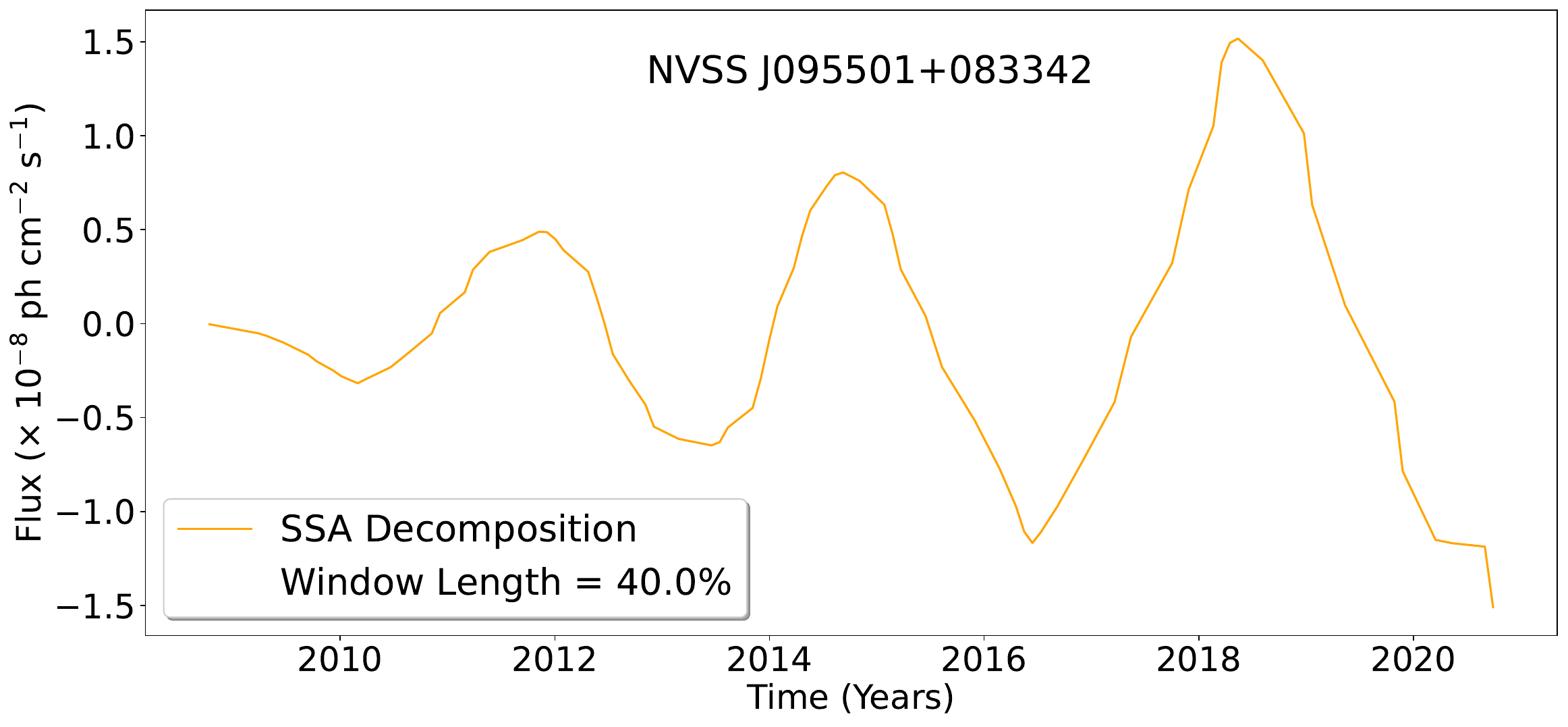}
        \includegraphics[scale=0.21]{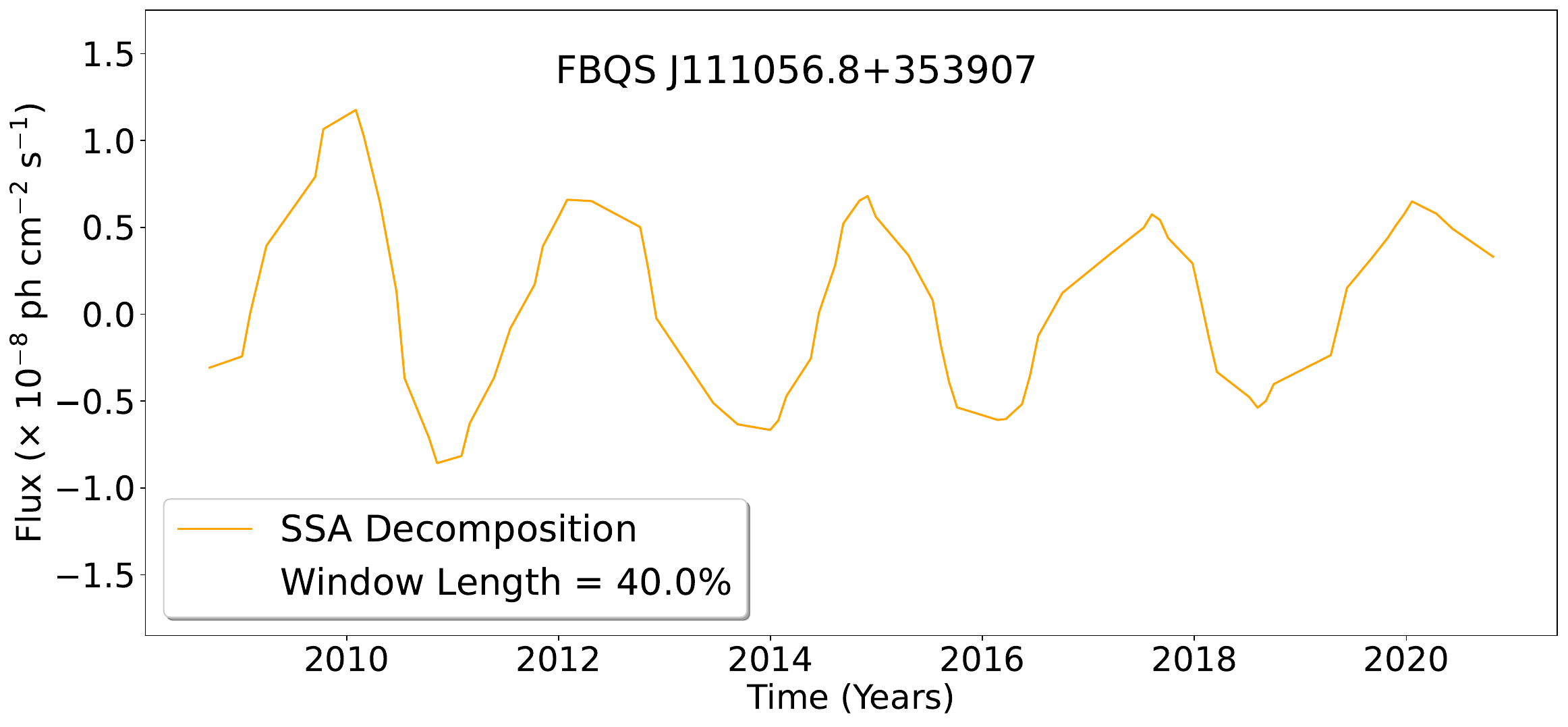}
        \includegraphics[scale=0.21]{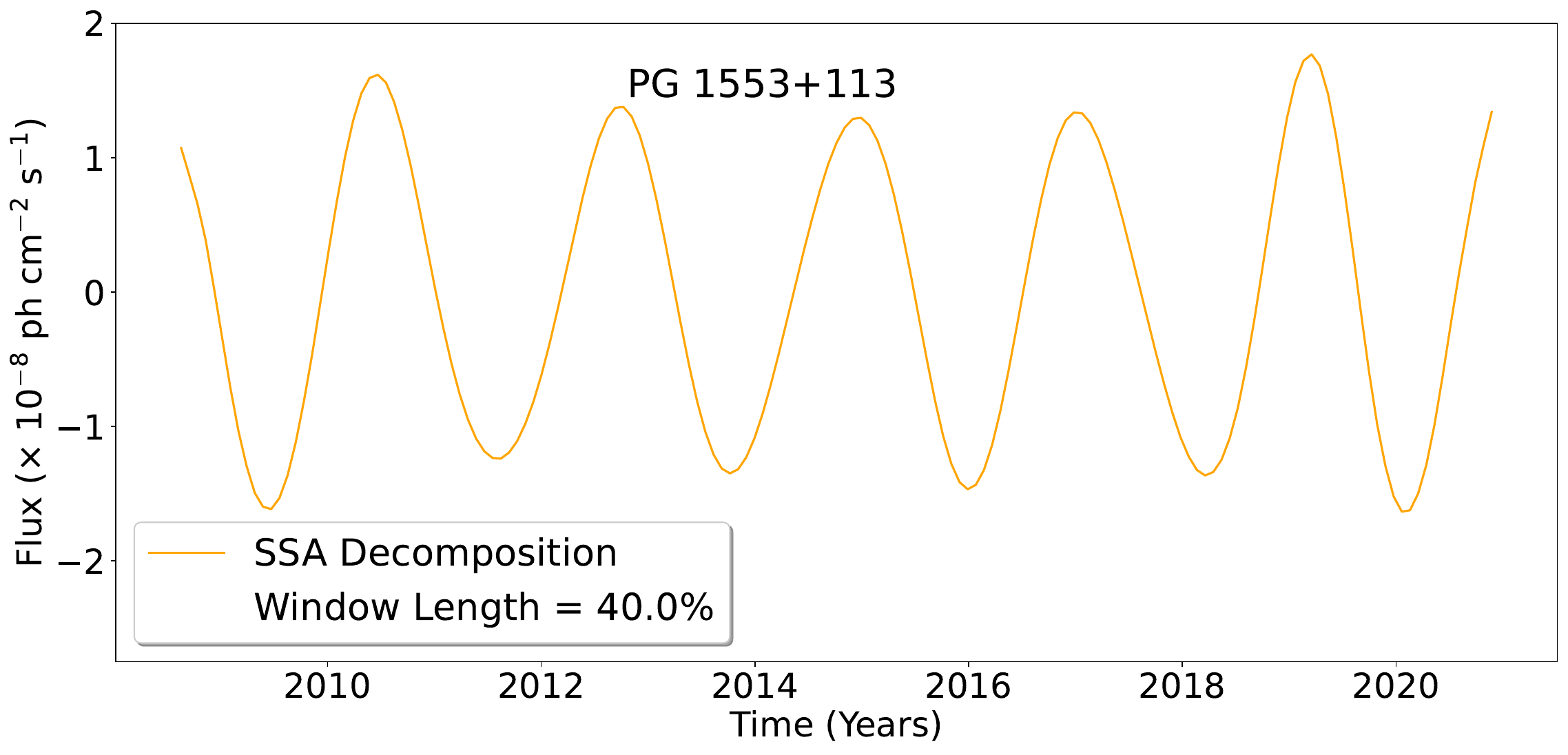} 
	\caption{SSA decomposition showing the underlying oscillatory structure using a window length of 40\%. \textit{Top}: NVSS J095501+083342 and FBQS J111056.8+353907 (see Figure \ref{fig:example_use_cases}). \textit{Bottom}: PG 1553+113 (see Figure \ref{fig:example_use_cases}). The flux axis shows negative values because it represents only the oscillatory component, excluding the overall emission behavior of the source. As a result, the oscillatory component is centered around zero, reflecting deviations from the mean rather than the total flux.} 
    \vspace{-0.3cm}
 \label{fig:ssa_blazars_04}
\end{figure*}

In these cases, we apply \citet{emma_lc} to optimize the significance estimation, using as a model for the PSD a power-law with Poisson noise. Then, we produce an artificial LC with the same properties as the original one, in terms of sampling, PSD, and probability density function (PDF) as observed in real blazar LCs. Specifically, we use the same sampling and observation uncertainties for each simulated signal as the original LC, keeping the same temporal sampling as the original LC. To ensure the reliability of this method, we estimate the significance by simulating 100,000 artificial LCs. Additionally, we perform the analysis for SSA using two different window lengths, 20\% (see  Figure \ref{fig:ssa_blazars}) and 40\% (see Figure \ref{fig:ssa_blazars_04}).  

The results are presented in Table~\ref{tab:blazar_results}. One conclusion is that SSA, when applied with a window length of 20\%, yields the highest significance values compared to LSP and PDM. In particular, SSA typically produces significance values that are approximately 50\% higher than those obtained with LSP and PDM, highlighting its enhanced sensitivity in extracting potential periodic components from gappy LCs. This improved performance is likely due to SSA's ability to isolate oscillatory patterns while effectively suppressing stochastic variations, ultimately leading to higher significance in the detection of periodic signals \citep[][]{penil_flares_2025}. However, this same capability may also introduce spurious results in the specific case of LCs with a high percentage of gaps ($\geq$50\%).  

In addition, the results show that the choice of window length in SSA impacts the estimated significance. A comparison between window lengths of 20\% and 40\% indicates that the shorter window often produces slightly higher significance values, with differences typically $\approx$10\%. However, this outcome contrasts with the findings presented in Section~\ref{sec:window_lengh}, where a window length of 40\% tended to yield higher significance. Therefore, while a 20\% window can be advantageous under certain conditions, its use should be carefully evaluated on a case-by-case basis. Adopting a fixed window length across all analyses may not be appropriate, especially when dealing with heterogeneous datasets.

Figure \ref{fig:gaps distribution} shows that most \textit{Fermi}-LAT LCs have a moderate percentage of gaps, with a peak near 30\%. We therefore analyze two representative sources within the interquartile range (25$^{\mathrm{th}}$–75$^{\mathrm{th}}$ percentiles), corresponding to $\sim$20–50\% of gaps:

\begin{enumerate}
\item NVSS J010345+132346 (4FGL J0103.8+1321, BLL, $z$=unknown); period = 1.4 year; percentage of gaps = 44.6\% (Figure \ref{fig:appendix_lc_20_50}).
\item PMN J1636$-$4101 (4FGL J1636.9$-$4103, BCU, $z$=unknown); period = 4.5 year; percentage of gaps = 24.7\% (Figure \ref{fig:appendix_lc_20_50}).
\end{enumerate}

Both cases support the same overall conclusion reached for the other two blazars: the significance values are consistent with those expected from noise LCs sharing the same gap, and the differences observed arise solely from the choice of SSA window length.

\subsection{Gap Study in a Specific LC}\label{sec:gap_study_lc}
A strategy to evaluate whether a significant period ($\geq$3$\sigma$) could result from the gaps present in the LC was employed in \citet{Adhikari2023, Adhikari2024} to assess if the presence of gaps in the analyzed LCs could produce the inferred period.

This approach involves simulating 100,000 synthetic LCs using the method described in \citet{emma_lc}, maintaining the same PSD, PDF, and gap structure (percentage and temporal coordinate) as the original data. The resulting distribution of period-significance can reveal potential biases toward specific period ranges that align with the detected period. This test helps determine whether the period significance obtained from the real LC could be attributed to a stochastic process influenced by the presence of gaps. 

Additionally, we estimate the percentage of occurrences of the same period-significance pair found in the artificial LCs than in the one obtained from the original data. Specifically, we define an occurrence as a case where the recovered period lies within the uncertainty range of the period results from the original LC and the significance exceeds that obtained from the original LC.

\begin{figure*}
	\centering
	    \includegraphics[scale=0.24]{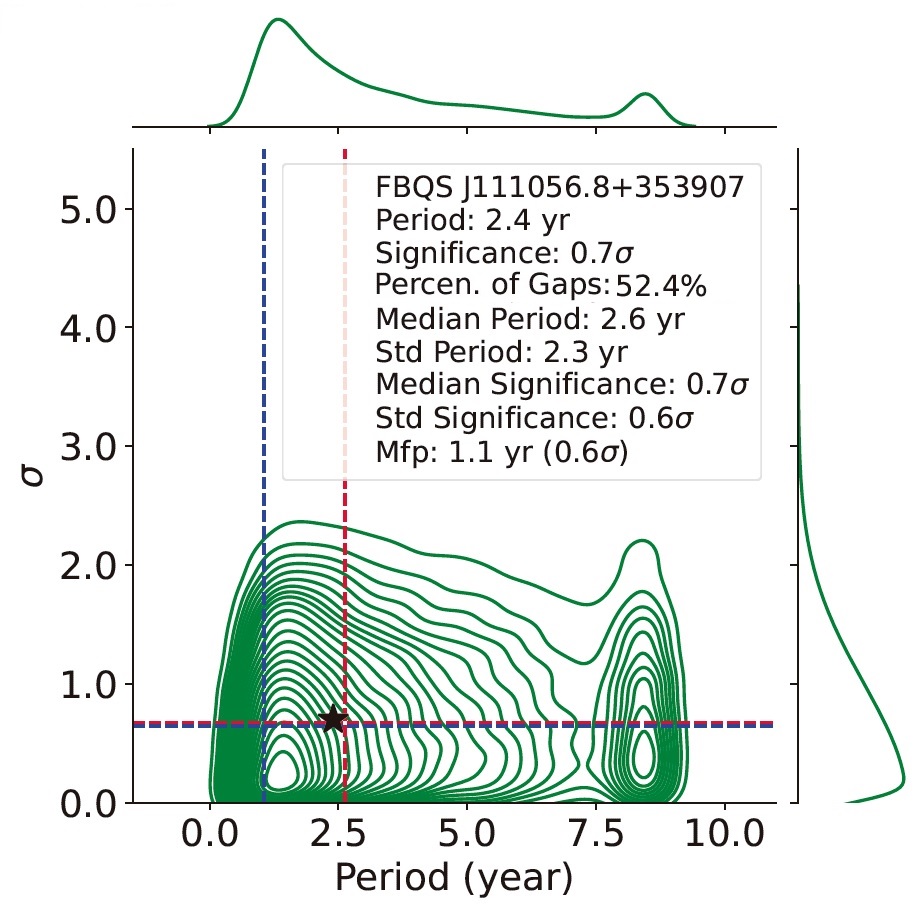}
        \includegraphics[scale=0.24]{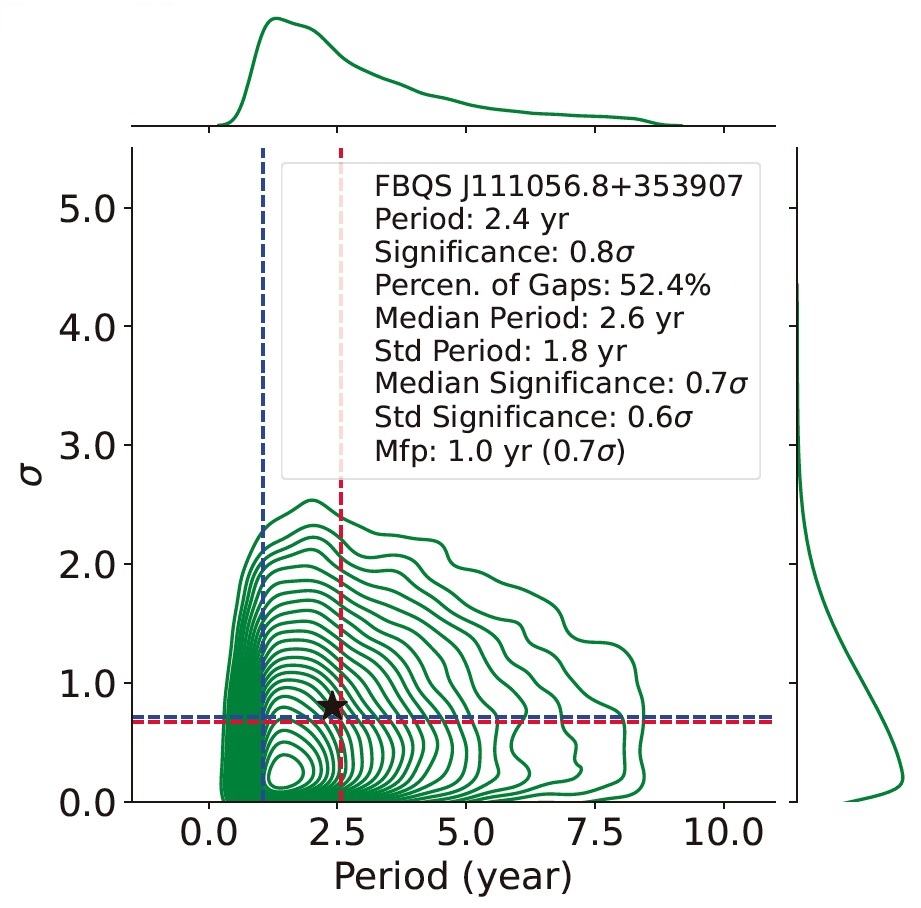}
        \includegraphics[scale=0.24]{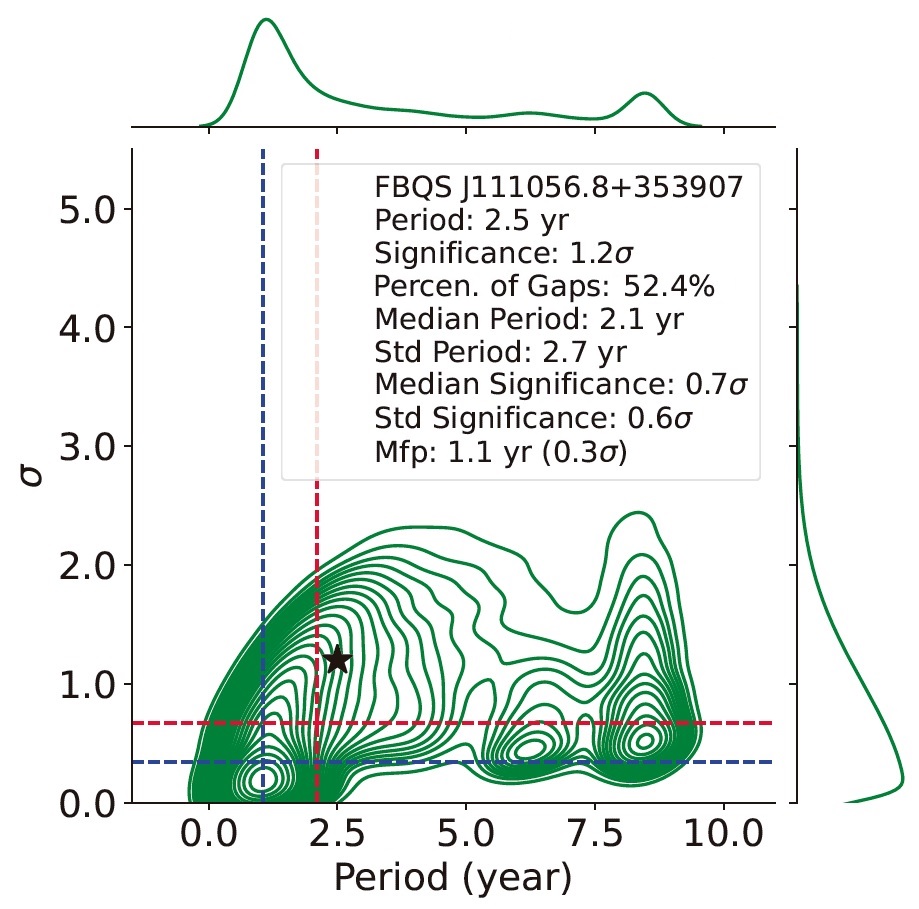}
        \caption{Distributions for the period and significance for the simulated LCs with the same properties as FBQ J111056.8+353907 using 100.000 artificial LCs with the same properties as the original LC (i.e., sampling, gap distribution, PSD, PDF). These simulations are used to assess the likelihood of obtaining spurious periodicities from noise alone under the same observational conditions as the original LC. \textit{Left}: LSP. \textit{Center}: PDM. \textit{Right}: SSA. The results denote that the period and significance obtained with the methodology of $\S$\ref{lab:use_cases} are compatible with random LCs with the same gap structure. The dotted red vertical and horizontal lines indicate the median values for both the period and the significance of the test. The blue dotted vertical line highlights the most frequently occurring period in the tests (and the associated significance), emphasizing its prominence in the distribution. The black star denotes the values period-significance estimated for the object using the specific method, and shown in Table \ref{tab:blazar_results}. The ``Percen. of Gaps'' refers to the percentage of gaps injected in the LC. The ``Median Period'' represents the median of all periods resulting from the test, and the ``Std Period'' is the standard deviation of such periods' distribution. The ``Median Significance'' represents the median of the significance distribution associated with the test, and the ``Std Significance'' is the standard deviation of this significance distribution. ``Mfp'' represents the most frequent period resulting from the test.} \label{fig:FBQ_J111056.8_random_lcs} 
\end{figure*}

The results from LSP, PDM, and SSA for FBQS J111056.8+353907 are shown in Figure \ref{fig:FBQ_J111056.8_random_lcs}. These results indicate that the period and significance values reported in Table \ref{tab:blazar_results} are consistent with those obtained from noise LC sharing the same gap structure as the original data. Specifically, the percentage of occurrences is 4.4\% for LSP, 4.7\% for PDM, and 1.9\% for SSA. These relatively high occurrence rates, particularly for LSP and PDM, suggest that the observed signal arises from the intrinsic temporal sampling and noise structure, rather than representing a statistically significant periodic component. A similar conclusion was reached for the other blazar analyzed, NVSS J095501+083342, with 3.6\%, 3.9\%, and 1.7\% for LSP, PDM, and SSA, respectively. No significant periods were obtained.  

For sources with intermediate levels of percentage of gaps (20\%–50\% gaps), the results follow the same trend. The outcomes remain consistent with those expected from noise LCs sharing the original data’s gap structure, again suggesting that the sampling function dominates the apparent periodicity. In NVSS J010345+132346, the occurrence rates are notably higher, reaching 12.2\% for LSP, 9.5\% for PDM, and 12.7\% for SSA. Similarly, PMN J1636$-$4101 yields 1.46\% for LSP, 2.48\% for PDM, and 0.87\% for SSA. These values confirm that even in cases with moderate data coverage, spurious signals can emerge with non-negligible probability.

This result suggests that hints of spurious periods can arise purely due to the presence of gaps, even when using robust detection methods, potentially leading to incorrect claims of genuine periodicity. Therefore, additional statistical tests, such as the one proposed in this section, should be incorporated into the methodology to minimize the risk of false periodicity detections.

\subsection{Test Case: PG 1553+113}
In this section, we evaluate the impact of introducing gaps in the LC of a candidate to be a periodic-emitter blazar. Specifically, we use the LC of PG 1553+113, the most promising candidate for being a periodic emitter among blazars \citep[e.g.,][]{ackermann_pg1553, alba_ssa}. A period of $\approx$2.1 years has been reported in its multiwavelength emission, with significance levels in the range of [3-5]$\sigma$ across different studies \citep[e.g.,][]{ackermann_pg1553, penil_mwl_pg1553, alba_ssa}.

Here, we used the LC from Figure \ref{fig:example_use_cases} as a real use case, introducing randomly distributed gaps in the LC following the same percentages used in our experiments (note that the LC in Figure
\ref{fig:example_use_cases} does not include any ULs). We then evaluate the evolution of the detected period and its associated significance. The same methodology described in $\S$\ref{lab:use_cases} is applied to obtain the significance values. We explore LSP, PDM, and SSA. 

The results are presented in Table \ref{tab:gap_pg1553}. For the LSP, a consistent period of 2.1 years is identified across all gap percentages, although the uncertainty in the period increases with more missing data. The significance remains roughly constant up to a gap percentage of 40\%, where the first noticeable decrease occurs with a reduction of about 20\%. The most significant drop is observed at 90\% gaps, with the significance reduced by over 100\%. Similar results are obtained for PDM. Regarding SSA, the period is approximately obtained in all cases. The significance is high ($\geq$3$\sigma$) until a percentage of 40\%. All these results are compatible with the test presented in $\S$\ref{sec:periodic_lcs}.  

Overall, the tests conducted in this study indicate that the distortion introduced by data gaps in the period–significance relationship becomes more pronounced as the percentage of gaps increases. This distortion becomes significant starting at 50\% gaps, where both LSP and PDM show a decrease in significance of at least 50\%. These findings support the 50\% gap threshold used to filter blazars in \citet{penil_2020} and \citet{penil_2022}.

Regarding SSA (with a window length of 20\%), the results demonstrate its robustness against the presence of gaps, showing less distortion as the gap percentage increases than LSP and PDM until a percentage of 50\%, with a reduction of 40\% in the significance. The period of $\approx$2.2 years is obtained until the percentage of gaps is 80\%, resulting in 1.9 years. Consequently, when combined with the findings from \citet{penil_flares_2025}, SSA appears to outperform traditional methods by mitigating the influence of both flares and data gaps in the search for periodicity in blazars.   

For a window length of 40\%, the reduction in significance with increasing gap percentage is consistent with previous results. However, the initial significance value, measured without gaps, is slightly higher, reaching 4.4$\sigma$ compared to 4.0$\sigma$ for the 20\% window length. This difference can be better understood by comparing Figure~\ref{fig:testing_pure_noise_ssa} and Figure~\ref{fig:testing_pure_noise_ssa_window02}, where the oscillatory components extracted by SSA show variations depending on the window length. Specifically, a 40\% window appears to capture larger-scale structures more effectively, resulting in a higher significance. These results demonstrate the impact of the window length on the SSA's performance, denoting the importance of adapting the window length to the specific characteristics of each analysis and LC. Those LCs dominated by lower-frequency variations may benefit from a longer window. In comparison, those more affected by gaps or exhibiting more localized variability may require a shorter window for optimal detection performance. 

\begin{figure*}
	\centering
	\includegraphics[scale=0.24]{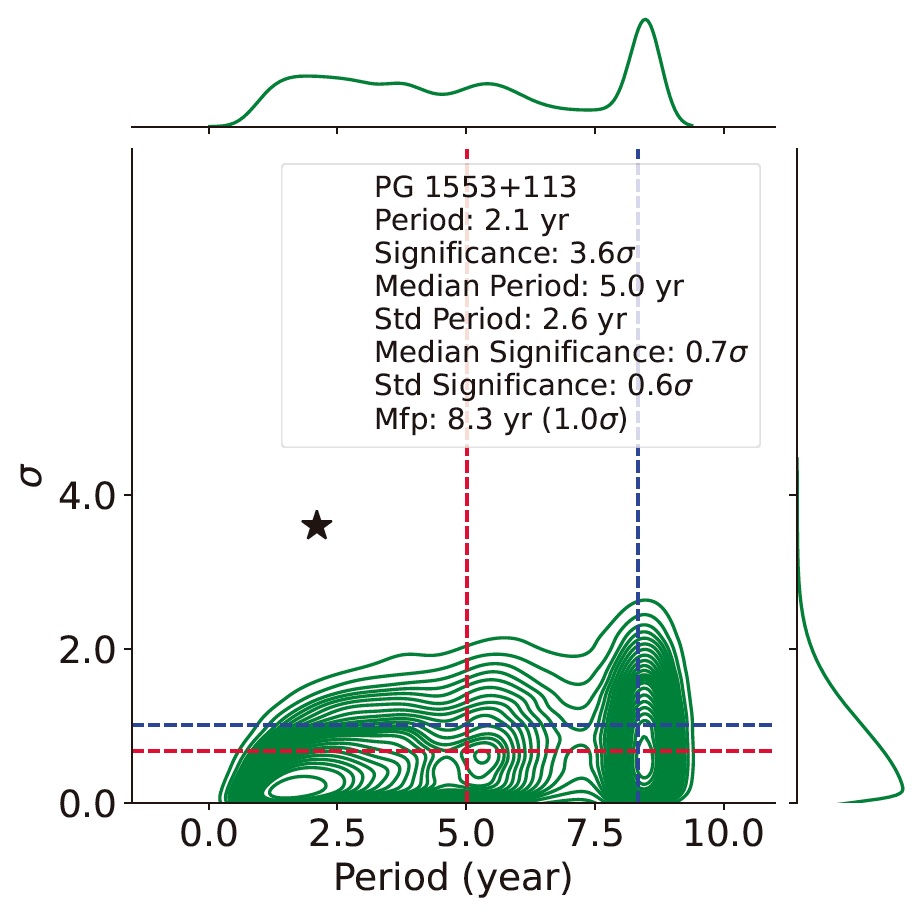}
        \includegraphics[scale=0.24]{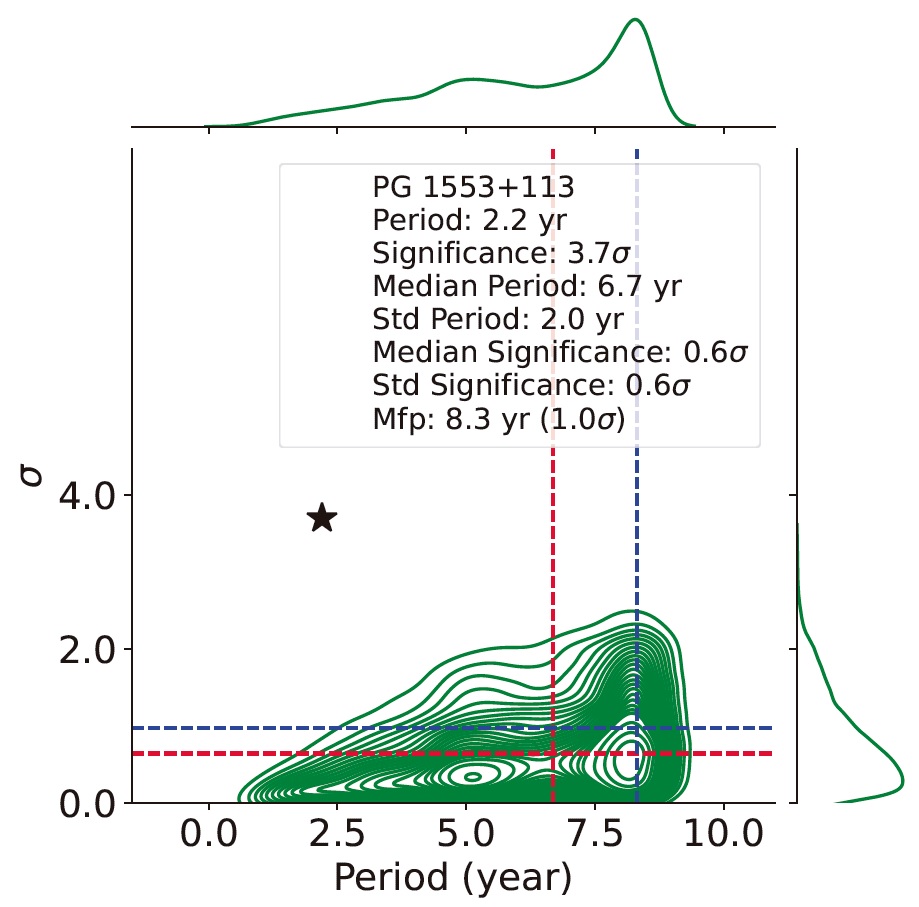}
        \includegraphics[scale=0.24]{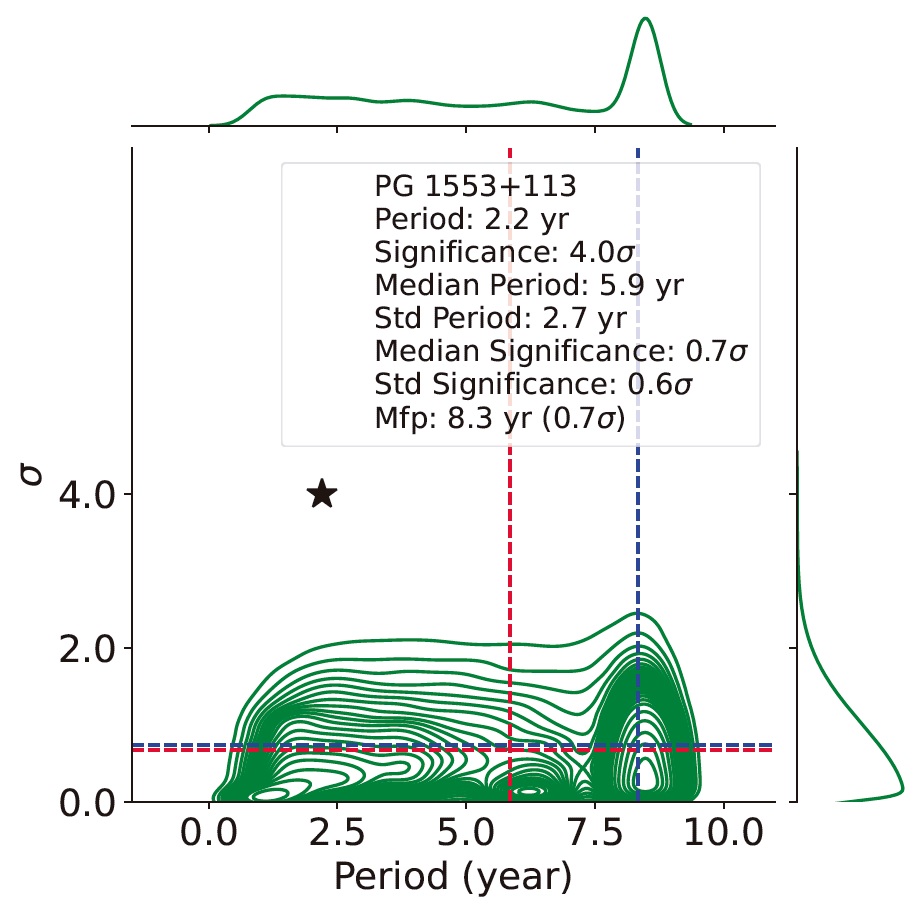}
        \caption{Distributions of the period and significance for the simulated random LCs, generated using 1,000,000 artificial LCs that replicate the observational properties of PG 1553+113, such as sampling cadence, gap distribution, PSD, and flux PDF, but without any injected periodic signal. These simulations are used to assess the likelihood of obtaining spurious periodicities from noise alone under the same observational conditions as the original LC. \textit{Left}: LSP. \textit{Center}: PDM. \textit{Right}: SSA (20\% window length). The results denote that the period and significance obtained with the methodology of $\S$\ref{lab:use_cases} are unlikely to be compatible with a stochastic origin. The dotted red vertical and horizontal lines indicate the median values for both the period and the significance of the test. The blue dotted vertical line highlights the most frequently occurring period in the tests (and the associated significance), emphasizing its prominence in the distribution. The black star denotes the values period-significance estimated for the object using the specific method, and shown in Table \ref{tab:blazar_results}. The ``Median Period'' represents the median of all periods resulting from the test, and the ``Std Period'' is the standard deviation of such periods' distribution. The ``Median Significance'' represents the median of the significance distribution associated with the test, and the ``Std Significance'' is the standard deviation of this significance distribution. ``Mfp'' represents the most frequent period resulting from the test} \label{fig:pg1553_random_lcs}. 
\end{figure*}

Finally, we conduct a test similar to that described in $\S$\ref{sec:gap_study_lc} for PG~1553+113. This source presents a distinct case within our use cases, as its LC contains no observational gaps. The results of this analysis are presented in Figure~\ref{fig:pg1553_random_lcs}. Following the approach described in $\S$\ref{sec:gap_study_lc}, we also estimate the percentage of occurrences in which the recovered period lies within the uncertainty range of the period obtained in the original LC, and the associated significance exceeds that of the original estimation. For all the methods, it is 0.0\%, suggesting that the periodic signal identified in PG~1553+113 is highly unlikely to be produced by stochastic variability. 

\section{Summary} \label{sec:summary}

In this paper, we conducted a detailed study of the potential impact of gaps in blazar LCs and their influence on periodicity-search analyses. We analyzed LSP and PDM, two of the most used methods in the search for periodicity in blazars. Our findings demonstrate that such gaps could lead to false periodicity detections. Moreover, we showed that gaps not only reduce the significance associated with genuine periodic signals but can also distort the inferred period itself. Importantly, the periods detected with $\geq$3$\sigma$ confidence are unlikely to arise from stochastic processes or from the presence of gaps.

We evaluated the performance of SSA on irregularly sampled LCs within the framework of periodicity searches. The results indicate that classical SSA can be applied reliably when the fraction of missing data is below 50\%. However, we identified a risk of spurious significant detections near one year, largely driven by seasonal observational gaps, which highlights the potential for systematic biases in LCs with a substantial percentage of gaps. Additionally, we examined the impact of the window length, a key parameter in SSA, on the significance of the results. Our findings indicate that a 20\% window length tends to obtain better performance, particularly when applied to pure noise LCs that might otherwise lead to false detections. However, this cannot be considered a universal rule, as the optimal window length may vary depending on the specific properties of each LC and the purpose of the analysis. As such, each case should be handled individually to ensure reliable and robust periodicity detection.

To further explore these effects, we analyzed \textit{Fermi}-LAT data from a sample of blazars with $\geq$50\% gaps in their LCs. Our investigation found no significant periodic detections in these objects. Additionally, we performed an experiment using the $\gamma$-ray LC of PG 1553+113, where we progressively removed increasing percentages of gaps to observe the evolution of period significance. This experiment demonstrated a distortion of the baseline results as the percentage of gaps increased, confirming the substantial influence of gaps on periodic analysis outcomes. This distortion started to be significant when the percentage of gaps was 50\%. Thus, this is the recommended threshold for selecting LCs to avoid a significant impact of the gaps. Finally, our analysis produced similar results to those obtained from real LCs.  

Our results emphasize the critical need to account for data gaps in periodicity analysis methods, as they can significantly affect the reliability of detected signals and introduce misleading periodicities. These findings underline the importance of robust testing and gap-aware methodologies when analyzing LCs with irregular sampling patterns.

\begin{table*}
\centering
\caption{Periodicity analysis results of the blazars NVSS J095501+083342 and FBQS J111056.8+353907 (Figure \ref{fig:example_use_cases}). The table shows the percentage of gaps in each LC. The analysis methods used are the LSP, PDM, and SSA methodology presented in $\S$\ref{sec:alternative_methods}. We report two sets of results separated by the two vertical lines. The first group of results is obtained in $\S$\ref{sec:usecase}, using \citet{timmer_koenig_1995}. The second set of results is obtained using \citet{emma_lc} and the procedure of $\S$\ref{lab:use_cases}. Additionally, the SSA results include the ones obtained using two different window lengths (WL), 20\% and 40\%. We show the period and its significance. A value of “--” indicates that no result is available, since these sources, with $<$50\% gaps, were not included in the test of Figure~\ref{fig:fermi_lat_objects_distribution}, and therefore no results are provided using the \citet{timmer_koenig_1995} framework. All reported periods are given in years.\label{tab:blazar_results}}
{%
\begin{tabular}{l|cccc||ccc|ccc}
\hline
\hline
Source & Gaps [\%] & LSP & PDM & SSA & LSP & PDM & SSA & SSA \\
  & & & & WL=20\% & & & WL=20\% & WL=40\% \\
\hline
NVSS J095501+083342 & 51.2 & 3.3 (0.7$\sigma$) & 3.3 (0.8$\sigma$) & 3.5 (1.5$\sigma$) & 3.3$\pm$0.6 (0.6$\sigma$) & 3.5$\pm$0.6 (0.5$\sigma$) & 3.5$\pm$0.6 (0.9$\sigma$) & 3.5$\pm$0.6 (0.4$\sigma$) \\
FBQS J111056.8+353907 & 52.4 & 2.6 (0.8$\sigma$) & 2.4 (0.6$\sigma$) & 2.6 (1.5$\sigma$) & 2.4$\pm$0.3 (0.7$\sigma$) & 2.4$\pm$0.5 (0.8$\sigma$) & 2.5$\pm$0.2 (1.2$\sigma$) & 2.5$\pm$0.2 (1.3$\sigma$) \\
\hline
\hline
NVSS J010345+132346 & 44.6 & -- & -- & -- & 1.4$\pm$0.3 (0.6$\sigma$) & 1.4$\pm$0.4 (0.6$\sigma$) & 1.4$\pm$0.2 (1.0$\sigma$) & 1.6$\pm$0.2 (1.1$\sigma$)\\
PMN J1636-4101 & 24.7 & -- & -- & -- & 4.5$\pm$0.6 (0.6$\sigma$) & 4.5$\pm$0.5 (0.5$\sigma$) & 4.6$\pm$0.4 (1.2$\sigma$) & 4.2$\pm$0.4 (0.8$\sigma$) \\
\hline
\hline
\end{tabular}%
}
\end{table*}

\section {Software}
\begin{enumerate}
        \item Astropy \citep{astropy_2013, astropy_2018, astropy_2022}
        \item emcee \citep {emcee}
	    \item PyAstronomy \citep{PyAstronomy}
	    \item SciPy \citep {SciPy}
	    \item Simulating light curves \citep{connolly_code}
        \item Singular Spectrum Analysis \url{https://www.kaggle.com/code/jdarcy/introducing-ssa-for-time-series-decomposition}
\end{enumerate}

\section{Acknowledgements}

P.P. and M.A. acknowledge funding under NASA contract 80NSSC20K1562. This work was supported by the European Research Council, ERC Starting grant MessMapp, S.B. Principal Investigator, under contract no. 949555, and by the German Science Foundation DFG, research grant “Relativistic Jets in Active Galaxies” (FOR 5195, grant No. 443220636).

\section{Data Availability}

All the data used in this work are publicly available or available on request to the responsible for the corresponding observatory/facility.

\bibliographystyle{mnras}
\bibliography{literature} 

\begin{thebibliography}{}
\makeatletter
\relax
\def\mn@urlcharsother{\let\do\@makeother \do\$\do\&\do\#\do\^\do\_\do\%\do\~}
\def\mn@doi{\begingroup\mn@urlcharsother \@ifnextchar [ {\mn@doi@} {\mn@doi@[]}}
\def\mn@doi@[#1]#2{\def\@tempa{#1}\ifx\@tempa\@empty \href {http://dx.doi.org/#2} {doi:#2}\else \href {http://dx.doi.org/#2} {#1}\fi \endgroup}
\def\mn@eprint#1#2{\mn@eprint@#1:#2::\@nil}
\def\mn@eprint@arXiv#1{\href {http://arxiv.org/abs/#1} {{\tt arXiv:#1}}}
\def\mn@eprint@dblp#1{\href {http://dblp.uni-trier.de/rec/bibtex/#1.xml} {dblp:#1}}
\def\mn@eprint@#1:#2:#3:#4\@nil{\def\@tempa {#1}\def\@tempb {#2}\def\@tempc {#3}\ifx \@tempc \@empty \let \@tempc \@tempb \let \@tempb \@tempa \fi \ifx \@tempb \@empty \def\@tempb {arXiv}\fi \@ifundefined {mn@eprint@\@tempb}{\@tempb:\@tempc}{\expandafter \expandafter \csname mn@eprint@\@tempb\endcsname \expandafter{\@tempc}}}

\bibitem[\protect\citeauthoryear{{Ackermann} et~al.,}{{Ackermann} et~al.}{2015}]{ackermann_pg1553}
{Ackermann} M.,  et~al., 2015, \mn@doi [\apjl] {10.1088/2041-8205/813/2/L41}, \href {https://ui.adsabs.harvard.edu/abs/2015ApJ...813L..41A} {813, L41}

\bibitem[\protect\citeauthoryear{{Adhikari}, {Pe{\~n}il}, {Westernacher-Schneider}, {Dom{\'\i}nguez}, {Ajello}, {Buson}, {Rico}  \& {Zrake}}{{Adhikari} et~al.}{2024}]{Adhikari2023}
{Adhikari} S.,  {Pe{\~n}il} P.,  {Westernacher-Schneider} J.~R.,  {Dom{\'\i}nguez} A.,  {Ajello} M.,  {Buson} S.,  {Rico} A.,   {Zrake} J.,  2024, \mn@doi [\apj] {10.3847/1538-4357/ad310a}, \href {https://ui.adsabs.harvard.edu/abs/2024ApJ...965..124A} {965, 124}

\bibitem[\protect\citeauthoryear{{Adhikari}, {Pe{\~n}il}, {Dom{\'\i}nguez}, {Ajello}, {Buson}  \& {Rico}}{{Adhikari} et~al.}{2025}]{Adhikari2024}
{Adhikari} S.,  {Pe{\~n}il} P.,  {Dom{\'\i}nguez} A.,  {Ajello} M.,  {Buson} S.,   {Rico} A.,  2025, \mn@doi [\mnras] {10.1093/mnras/staf783}, \href {https://ui.adsabs.harvard.edu/abs/2025MNRAS.540.1449A} {540, 1449}

\bibitem[\protect\citeauthoryear{{Ajello} et~al.,}{{Ajello} et~al.}{2020}]{4fgl_catalog}
{Ajello} M.,  et~al., 2020, \mn@doi [\apj] {10.3847/1538-4357/ab791e}, \href {https://ui.adsabs.harvard.edu/abs/2020ApJ...892..105A} {892, 105}

\bibitem[\protect\citeauthoryear{{Astropy Collaboration} et~al.,}{{Astropy Collaboration} et~al.}{2013}]{astropy_2013}
{Astropy Collaboration} et~al., 2013, \mn@doi [\aap] {10.1051/0004-6361/201322068}, \href {https://ui.adsabs.harvard.edu/abs/2013A&A...558A..33A} {558, A33}

\bibitem[\protect\citeauthoryear{{Astropy Collaboration} et~al.,}{{Astropy Collaboration} et~al.}{2018}]{astropy_2018}
{Astropy Collaboration} et~al., 2018, \mn@doi [\aj] {10.3847/1538-3881/aabc4f}, \href {https://ui.adsabs.harvard.edu/abs/2018AJ....156..123A} {156, 123}

\bibitem[\protect\citeauthoryear{{Astropy Collaboration} et~al.,}{{Astropy Collaboration} et~al.}{2022}]{astropy_2022}
{Astropy Collaboration} et~al., 2022, \mn@doi [\apj] {10.3847/1538-4357/ac7c74}, \href {https://ui.adsabs.harvard.edu/abs/2022ApJ...935..167A} {935, 167}

\bibitem[\protect\citeauthoryear{{Bhatta} \& {Dhital}}{{Bhatta} \& {Dhital}}{2020}]{bhatta_s5_0716}
{Bhatta} G.,  {Dhital} N.,  2020, \mn@doi [\apj] {10.3847/1538-4357/ab7455}, \href {https://ui.adsabs.harvard.edu/abs/2020ApJ...891..120B} {891, 120}

\bibitem[\protect\citeauthoryear{{Cavaliere} \& {Padovani}}{{Cavaliere} \& {Padovani}}{1989}]{cavaliere_1989}
{Cavaliere} A.,  {Padovani} P.,  1989, \mn@doi [\apjl] {10.1086/185425}, \href {https://ui.adsabs.harvard.edu/abs/1989ApJ...340L...5C} {340, L5}

\bibitem[\protect\citeauthoryear{{Connolly}}{{Connolly}}{2015}]{connolly_code}
{Connolly} S.~D.,  2015, arXiv e-prints, \href {https://ui.adsabs.harvard.edu/abs/2015arXiv150306676C} {p. arXiv:1503.06676}

\bibitem[\protect\citeauthoryear{Covino, Sandrinelli  \& Treves}{Covino et~al.}{2018}]{covino_negation}
Covino S.,  Sandrinelli A.,   Treves A.,  2018, \mn@doi [\mnras] {10.1093/mnras/sty2720}, 482, 1270

\bibitem[\protect\citeauthoryear{{Czesla}, {Schr{\"o}ter}, {Schneider}, {Huber}, {Pfeifer}, {Andreasen}  \& {Zechmeister}}{{Czesla} et~al.}{2019}]{PyAstronomy}
{Czesla} S.,  {Schr{\"o}ter} S.,  {Schneider} C.~P.,  {Huber} K.~F.,  {Pfeifer} F.,  {Andreasen} D.~T.,   {Zechmeister} M.,  2019, {PyA: Python astronomy-related packages}, Astrophysics Source Code Library, record ascl:1906.010 (\mn@eprint {ascl} {1906.010})

\bibitem[\protect\citeauthoryear{{Dey} et~al.,}{{Dey} et~al.}{2018}]{dey2018authenticating}
{Dey} L.,  et~al., 2018, \mn@doi [\apj] {10.3847/1538-4357/aadd95}, \href {https://ui.adsabs.harvard.edu/abs/2018ApJ...866...11D} {866, 11}

\bibitem[\protect\citeauthoryear{{Emmanoulopoulos}, {McHardy}  \& {Papadakis}}{{Emmanoulopoulos} et~al.}{2013}]{emma_lc}
{Emmanoulopoulos} D.,  {McHardy} I.~M.,   {Papadakis} I.~E.,  2013, \mn@doi [\mnras] {10.1093/mnras/stt764}, \href {https://ui.adsabs.harvard.edu/abs/2013MNRAS.433..907E} {433, 907}

\bibitem[\protect\citeauthoryear{Feigelson, Babu  \& Caceres}{Feigelson et~al.}{2018}]{feigelson_arima}
Feigelson E.~D.,  Babu G.~J.,   Caceres G.~A.,  2018, Frontiers in Physics, 6, 80

\bibitem[\protect\citeauthoryear{{Foreman-Mackey}, {Hogg}, {Lang}  \& {Goodman}}{{Foreman-Mackey} et~al.}{2013}]{emcee}
{Foreman-Mackey} D.,  {Hogg} D.~W.,  {Lang} D.,   {Goodman} J.,  2013, \mn@doi [\pasp] {10.1086/670067}, \href {https://ui.adsabs.harvard.edu/abs/2013PASP..125..306F} {125, 306}

\bibitem[\protect\citeauthoryear{{Foreman-Mackey}, {Agol}, {Ambikasaran}  \& {Angus}}{{Foreman-Mackey} et~al.}{2017}]{gaussian_process}
{Foreman-Mackey} D.,  {Agol} E.,  {Ambikasaran} S.,   {Angus} R.,  2017, \mn@doi [\aj] {10.3847/1538-3881/aa9332}, \href {https://ui.adsabs.harvard.edu/abs/2017AJ....154..220F} {154, 220}

\bibitem[\protect\citeauthoryear{{Foster}}{{Foster}}{1996}]{foster_wwz}
{Foster} G.,  1996, \mn@doi [\aj] {10.1086/118137}, \href {https://ui.adsabs.harvard.edu/abs/1996AJ....112.1709F} {112, 1709}

\bibitem[\protect\citeauthoryear{{Franchini}, {Lodato}  \& {Facchini}}{{Franchini} et~al.}{2016}]{franchini_lense}
{Franchini} A.,  {Lodato} G.,   {Facchini} S.,  2016, \mn@doi [\mnras] {10.1093/mnras/stv2417}, \href {https://ui.adsabs.harvard.edu/abs/2016MNRAS.455.1946F} {455, 1946}

\bibitem[\protect\citeauthoryear{{Giannios}, {Uzdensky}  \& {Begelman}}{{Giannios} et~al.}{2009}]{Giannios2009}
{Giannios} D.,  {Uzdensky} D.~A.,   {Begelman} M.~C.,  2009, \mn@doi [\mnras] {10.1111/j.1745-3933.2009.00635.x}, \href {https://ui.adsabs.harvard.edu/abs/2009MNRAS.395L..29G} {395, L29}

\bibitem[\protect\citeauthoryear{Golyandina \& Osipov}{Golyandina \& Osipov}{2007}]{golyandina2007}
Golyandina N.,  Osipov E.,  2007, \mn@doi [Journal of Statistical Planning and Inference] {https://doi.org/10.1016/j.jspi.2006.05.014}, 137, 2642

\bibitem[\protect\citeauthoryear{Golyandina \& Zhigljavsky}{Golyandina \& Zhigljavsky}{2020}]{golyandina2020singular}
Golyandina N.~E.,  Zhigljavsky A.~A.,  2020, Singular Spectrum Analysis for Time Series.
SpringerBriefs in Statistics, Springer, \mn@doi{10.1007/978-3-662-62436-4}

\bibitem[\protect\citeauthoryear{Golyandina, Dudnik  \& Shlemov}{Golyandina et~al.}{2023}]{golyandina_ssa}
Golyandina N.,  Dudnik P.,   Shlemov A.,  2023, \mn@doi [Algorithms] {10.3390/a16070353}, 16

\bibitem[\protect\citeauthoryear{Gracia, Peitz, Keller  \& Camenzind}{Gracia et~al.}{2003}]{gracias_modulation_disk}
Gracia J.,  Peitz J.,  Keller C.,   Camenzind M.,  2003, \mn@doi [\mnras] {10.1046/j.1365-8711.2003.06832.x}, 344, 468

\bibitem[\protect\citeauthoryear{{Greco} et~al.,}{{Greco} et~al.}{2016}]{ssa_greco}
{Greco} G.,  et~al., 2016, in {Napolitano} N.~R.,  {Longo} G.,  {Marconi} M.,  {Paolillo} M.,   {Iodice} E.,  eds,  Astrophysics and Space Science Proceedings Vol. 42, The Universe of Digital Sky Surveys. p.~105 (\mn@eprint {arXiv} {1509.03342}), \mn@doi{10.1007/978-3-319-19330-4_16}

\bibitem[\protect\citeauthoryear{{Gregory} \& {Loredo}}{{Gregory} \& {Loredo}}{1992}]{bayesian_mcmc}
{Gregory} P.~C.,  {Loredo} T.~J.,  1992, \mn@doi [\apj] {10.1086/171844}, \href {https://ui.adsabs.harvard.edu/abs/1992ApJ...398..146G} {398, 146}

\bibitem[\protect\citeauthoryear{{Gross} \& {Vitells}}{{Gross} \& {Vitells}}{2010}]{gross_vitells_trial}
{Gross} E.,  {Vitells} O.,  2010, \mn@doi [European Physical Journal C] {10.1140/epjc/s10052-010-1470-8}, \href {https://ui.adsabs.harvard.edu/abs/2010EPJC...70..525G} {70, 525}

\bibitem[\protect\citeauthoryear{{Kelly}, {Becker}, {Sobolewska}, {Siemiginowska}  \& {Uttley}}{{Kelly} et~al.}{2014}]{carma_kelly}
{Kelly} B.~C.,  {Becker} A.~C.,  {Sobolewska} M.,  {Siemiginowska} A.,   {Uttley} P.,  2014, \mn@doi [\apj] {10.1088/0004-637X/788/1/33}, \href {https://ui.adsabs.harvard.edu/abs/2014ApJ...788...33K} {788, 33}

\bibitem[\protect\citeauthoryear{{Kreutzer}, {Gillen}, {Briegal}  \& {Queloz}}{{Kreutzer} et~al.}{2023}]{acf_irregular}
{Kreutzer} L.~T.,  {Gillen} E.,  {Briegal} J.~T.,   {Queloz} D.,  2023, \mn@doi [\mnras] {10.1093/mnras/stad1223}, \href {https://ui.adsabs.harvard.edu/abs/2023MNRAS.522.5049K} {522, 5049}

\bibitem[\protect\citeauthoryear{{Liodakis}, {Romani}, {Filippenko}, {Kiehlmann}, {Max-Moerbeck}, {Readhead}  \& {Zheng}}{{Liodakis} et~al.}{2018}]{Liodakis2018}
{Liodakis} I.,  {Romani} R.~W.,  {Filippenko} A.~V.,  {Kiehlmann} S.,  {Max-Moerbeck} W.,  {Readhead} A.~C.~S.,   {Zheng} W.,  2018, \mn@doi [\mnras] {10.1093/mnras/sty2264}, \href {https://ui.adsabs.harvard.edu/abs/2018MNRAS.480.5517L} {480, 5517}

\bibitem[\protect\citeauthoryear{{Lomb}}{{Lomb}}{1976}]{lomb_1976}
{Lomb} N.~R.,  1976, \mn@doi [\apss] {10.1007/BF00648343}, \href {https://ui.adsabs.harvard.edu/abs/1976Ap&SS..39..447L} {39, 447}

\bibitem[\protect\citeauthoryear{{Nalewajko}}{{Nalewajko}}{2013}]{Nalewajko2013}
{Nalewajko} K.,  2013, \mn@doi [\mnras] {10.1093/mnras/sts711}, \href {https://ui.adsabs.harvard.edu/abs/2013MNRAS.430.1324N} {430, 1324}

\bibitem[\protect\citeauthoryear{Nina~Golyandina}{Nina~Golyandina}{2020}]{SSA_algorithm}
Nina~Golyandina A.~Z.,  2020, Singular Spectrum Analysis for Time Series, 2 edn.
S, Springer Berlin, Heidelberg, \mn@doi{https://doi.org/10.1007/978-3-662-62436-4}

\bibitem[\protect\citeauthoryear{{Oh}, {Kam}, {Lim}  \& {Kim}}{{Oh} et~al.}{2025}]{machine_learning}
{Oh} Y.,  {Kam} S.,  {Lim} D.-Y.,   {Kim} S.,  2025, \mn@doi [arXiv e-prints] {10.48550/arXiv.2508.17521}, \href {https://ui.adsabs.harvard.edu/abs/2025arXiv250817521O} {p. arXiv:2508.17521}

\bibitem[\protect\citeauthoryear{{Otero-Santos}, {Pe{\~n}il}, {Acosta-Pulido}, {Becerra Gonz{\'a}lez}, {Raiteri}, {Carnerero}  \& {Villata}}{{Otero-Santos} et~al.}{2023}]{jorge_2022}
{Otero-Santos} J.,  {Pe{\~n}il} P.,  {Acosta-Pulido} J.~A.,  {Becerra Gonz{\'a}lez} J.,  {Raiteri} C.~M.,  {Carnerero} M.~I.,   {Villata} M.,  2023, \mn@doi [\mnras] {10.1093/mnras/stac3142}, \href {https://ui.adsabs.harvard.edu/abs/2023MNRAS.518.5788O} {518, 5788}

\bibitem[\protect\citeauthoryear{{Otero-Santos} et~al.,}{{Otero-Santos} et~al.}{2024}]{otero_3C_371}
{Otero-Santos} J.,  et~al., 2024, \mn@doi [\aap] {10.1051/0004-6361/202449647}, \href {https://ui.adsabs.harvard.edu/abs/2024A&A...686A.228O} {686, A228}

\bibitem[\protect\citeauthoryear{{Pe{\~n}il} et~al.,}{{Pe{\~n}il} et~al.}{2020}]{penil_2020}
{Pe{\~n}il} P.,  et~al., 2020, \mn@doi [\apj] {10.3847/1538-4357/ab910d}, \href {https://ui.adsabs.harvard.edu/abs/2020ApJ...896..134P} {896, 134}

\bibitem[\protect\citeauthoryear{{Pe{\~n}il} et~al.,}{{Pe{\~n}il} et~al.}{2024}]{penil_mwl_pg1553}
{Pe{\~n}il} P.,  et~al., 2024, \mn@doi [\mnras] {10.1093/mnras/stad3246}, \href {https://ui.adsabs.harvard.edu/abs/2024MNRAS.52710168P} {527, 10168}

\bibitem[\protect\citeauthoryear{{Pe{\~n}il}, {Torres-Alb{\`a}}, {Rico}, {Ajello}, {Buson}  \& {Adhikari}}{{Pe{\~n}il} et~al.}{2025a}]{penil_flares_2025}
{Pe{\~n}il} P.,  {Torres-Alb{\`a}} N.,  {Rico} A.,  {Ajello} M.,  {Buson} S.,   {Adhikari} S.,  2025a, \mn@doi [\mnras] {10.1093/mnras/staf482}, \href {https://ui.adsabs.harvard.edu/abs/2025MNRAS.539..993P} {539, 993}

\bibitem[\protect\citeauthoryear{{Pe{\~n}il}, {Ajello}, {Buson}, {Dom{\'\i}nguez}, {Westernacher-Schneider}, {Rico}, {Adhikari}  \& {Zrake}}{{Pe{\~n}il} et~al.}{2025b}]{penil_2022}
{Pe{\~n}il} P.,  {Ajello} M.,  {Buson} S.,  {Dom{\'\i}nguez} A.,  {Westernacher-Schneider} J.~R.,  {Rico} A.,  {Adhikari} S.,   {Zrake} J.,  2025b, \mn@doi [\mnras] {10.1093/mnras/staf1108}, \href {https://ui.adsabs.harvard.edu/abs/2025MNRAS.541.2955P} {541, 2955}

\bibitem[\protect\citeauthoryear{{Pe{\~n}il}, {Dom{\'\i}nguez}, {Buson}, {Ajello}, {Adhikari}  \& {Rico}}{{Pe{\~n}il} et~al.}{2025c}]{penil_trends}
{Pe{\~n}il} P.,  {Dom{\'\i}nguez} A.,  {Buson} S.,  {Ajello} M.,  {Adhikari} S.,   {Rico} A.,  2025c, \mn@doi [\apj] {10.3847/1538-4357/ada4b3}, \href {https://ui.adsabs.harvard.edu/abs/2025ApJ...980...38P} {980, 38}

\bibitem[\protect\citeauthoryear{{Resconi}, {Franco}, {Gross}, {Costamante}  \& {Flaccomio}}{{Resconi} et~al.}{2009}]{Resconi2009}
{Resconi} E.,  {Franco} D.,  {Gross} A.,  {Costamante} L.,   {Flaccomio} E.,  2009, \mn@doi [\aap] {10.1051/0004-6361/200911770}, \href {https://ui.adsabs.harvard.edu/abs/2009A&A...502..499R} {502, 499}

\bibitem[\protect\citeauthoryear{{Rico}, {Dom{\'\i}nguez}, {Pe{\~n}il}, {Ajello}, {Buson}, {Adhikari}  \& {Movahedifar}}{{Rico} et~al.}{2025}]{alba_ssa}
{Rico} A.,  {Dom{\'\i}nguez} A.,  {Pe{\~n}il} P.,  {Ajello} M.,  {Buson} S.,  {Adhikari} S.,   {Movahedifar} M.,  2025, \mn@doi [\aap] {10.1051/0004-6361/202452495}, \href {https://ui.adsabs.harvard.edu/abs/2025A&A...697A..35R} {697, A35}

\bibitem[\protect\citeauthoryear{Rieger}{Rieger}{2019}]{rieger_2019}
Rieger F.~M.,  2019, \mn@doi [Galaxies] {10.3390/galaxies7010028}, 7

\bibitem[\protect\citeauthoryear{{Sarkar}, {Gupta}, {Chitnis}  \& {Wiita}}{{Sarkar} et~al.}{2021}]{sarkar_curved_jet}
{Sarkar} A.,  {Gupta} A.~C.,  {Chitnis} V.~R.,   {Wiita} P.~J.,  2021, \mn@doi [\mnras] {10.1093/mnras/staa3211}, \href {https://ui.adsabs.harvard.edu/abs/2021MNRAS.501...50S} {501, 50}

\bibitem[\protect\citeauthoryear{{Scargle}}{{Scargle}}{1982}]{scargle_1982}
{Scargle} J.~D.,  1982, \mn@doi [\apj] {10.1086/160554}, \href {https://ui.adsabs.harvard.edu/abs/1982ApJ...263..835S} {263, 835}

\bibitem[\protect\citeauthoryear{{Stellingwerf}}{{Stellingwerf}}{1978}]{pdm_stellingwerf}
{Stellingwerf} R.~F.,  1978, \mn@doi [\apj] {10.1086/156444}, \href {https://ui.adsabs.harvard.edu/abs/1978ApJ...224..953S} {224, 953}

\bibitem[\protect\citeauthoryear{{Timmer} \& {Koenig}}{{Timmer} \& {Koenig}}{1995}]{timmer_koenig_1995}
{Timmer} J.,  {Koenig} M.,  1995, \aap, \href {https://ui.adsabs.harvard.edu/abs/1995A&A...300..707T} {300, 707}

\bibitem[\protect\citeauthoryear{{Urry}}{{Urry}}{1996}]{urry_variability}
{Urry} C.~M.,  1996, in {Miller} H.~R.,  {Webb} J.~R.,   {Noble} J.~C.,  eds,  Astronomical Society of the Pacific Conference Series Vol. 110, Blazar Continuum Variability. p.~391 (\mn@eprint {arXiv} {astro-ph/9609023})

\bibitem[\protect\citeauthoryear{{Urry} \& {Padovani}}{{Urry} \& {Padovani}}{1995}]{urry1995}
{Urry} C.~M.,  {Padovani} P.,  1995, \mn@doi [\pasp] {10.1086/133630}, \href {https://ui.adsabs.harvard.edu/abs/1995PASP..107..803U} {107, 803}

\bibitem[\protect\citeauthoryear{Vaughan, Edelson, Warwick  \& Uttley}{Vaughan et~al.}{2003}]{vaughan_2003}
Vaughan S.,  Edelson R.,  Warwick R.,   Uttley P.,  2003, Monthly Notices of the Royal Astronomical Society, 345, 1271

\bibitem[\protect\citeauthoryear{Vaughan, Uttley, Markowitz, Huppenkothen, Middleton, Alston, Scargle  \& Farr}{Vaughan et~al.}{2016}]{vaughan_criticism}
Vaughan S.,  Uttley P.,  Markowitz A.~G.,  Huppenkothen D.,  Middleton M.~J.,  Alston W.~N.,  Scargle J.~D.,   Farr W.~M.,  2016, \mn@doi [\mnras] {10.1093/mnras/stw1412}, 461, 3145

\bibitem[\protect\citeauthoryear{{Villata} \& {Raiteri}}{{Villata} \& {Raiteri}}{1999}]{villata_helical_jet}
{Villata} M.,  {Raiteri} C.~M.,  1999, \aap, \href {https://ui.adsabs.harvard.edu/abs/1999A&A...347...30V} {347, 30}

\bibitem[\protect\citeauthoryear{Virtanen et~al.,}{Virtanen et~al.}{2020}]{SciPy}
Virtanen P.,  et~al., 2020, Nature methods, 17, 261

\bibitem[\protect\citeauthoryear{{Wiita}}{{Wiita}}{2006}]{wiita_lecture}
{Wiita} P.~J.,  2006, arXiv e-prints, \href {https://ui.adsabs.harvard.edu/abs/2006astro.ph..3728W} {pp astro--ph/0603728}

\makeatother
\end{thebibliography}

\clearpage

\appendix

\renewcommand{\thetable}{A\arabic{table}}
\renewcommand{\thefigure}{A\arabic{figure}}
\section*{Appendix}\label{sec:appendix}

\subsection{Tables}
This section includes the tables with the results of the different tests performed in this study. 
\begin{table*}
\centering
\caption{Periodicity analysis results from the tests conducted using pure noise LCs are presented. The gap range explored in these tests spans from 0\% to 90\%. Three distinct gap scenarios are considered: randomly injected gaps throughout the LC (``Random''), periodic gaps introduced annually in the LC (``Periodic''), and randomly distributed gaps introduced annually in the LC (``Aperiodic''). 
The analysis methods used are the LSP and PDM techniques. We report two key results: the period (and its significance) corresponding to the median of all test results (denoted as ``Median'') and the period (and its significance) corresponding to the most frequently detected period in the tests (denoted as ``Mfp''). All the periods are in years.\label{tab:noise_random}}
{%
\begin{tabular}{l|ccccccc}
\hline
\hline
Type LC & Type of Gaps & Percentage of Gaps [\%] & LSP (Median) & LSP (Mfp) & PDM (Media) & PDM (Mfp)\\
\hline
		\multirow{3}{*}{Noise} & Random & 
                            \makecell{0 \\ 10 \\ 20 \\ 30 \\ 40 \\ 50 \\ 60 \\ 70 \\ 80 \\ 90} &
                            \makecell{4.7$\pm$2.5 (0.9$\pm$0.9$\sigma$) \\ 4.6$\pm$2.6 (0.9$\pm$0.8$\sigma$) \\ 4.4$\pm$2.6 (0.8$\pm$0.8$\sigma$) \\ 4.2$\pm$2.6 (0.8$\pm$0.8$\sigma$) \\ 4.0$\pm$2.6 (0.8$\pm$0.9$\sigma$) \\ 3.6$\pm$2.5 (0.9$\pm$0.9$\sigma$)\\ 3.3$\pm$2.6 (0.7$\pm$0.8$\sigma$) \\ 2.8$\pm$2.4 (0.7$\pm$0.7$\sigma$) \\ 2.5$\pm$2.3 (0.7$\pm$0.8$\sigma$) \\ 2.1$\pm$2.1 (0.8$\pm$0.8$\sigma$)} &
                            \makecell{8.3 (1.0$\sigma$) \\ 8.3 (1.0$\sigma$) \\ 8.3 (1.0$\sigma$) \\ 8.3 (1.0$\sigma$) \\ 8.3 (1.0$\sigma$) \\ 8.3 (1.0$\sigma$) \\ 8.3 (1.0$\sigma$) \\ 1.2 (0.9$\sigma$) \\ 1.1 (0.8$\sigma$) \\ 1.1 (0.6$\sigma$)} &
                            \makecell{6.5$\pm$2.0 (0.9$\pm$0.8$\sigma$) \\ 6.5$\pm$2.0 (0.8$\pm$0.7$\sigma$) \\ 6.4$\pm$2.1 (0.9$\pm$0.8$\sigma$) \\ 6.0$\pm$2.2 (0.9$\pm$0.7$\sigma$) \\ 5.8$\pm$2.4 (0.9$\pm$0.7$\sigma$) \\ 5.3$\pm$2.4 (0.9$\pm$0.6$\sigma$) \\ 4.5$\pm$2.5 (0.7$\pm$0.9$\sigma$) \\ 4.1$\pm$2.3 (0.8$\pm$0.9$\sigma$) \\ 3.1$\pm$2.4 (0.9$\pm$0.8$\sigma$) \\ 2.9$\pm$2.2 (0.8$\pm$0.7$\sigma$)} &
                            \makecell{8.3 (1.0$\sigma$) \\ 8.3 (1.0$\sigma$) \\ 8.3 (1.0$\sigma$) \\ 8.3 (1.0$\sigma$) \\ 8.3 (1.0$\sigma$) \\ 8.3 (0.9$\sigma$) \\ 8.3 (0.9$\sigma$) \\ 1.2 (0.6$\sigma$) \\ 1.1 (0.7$\sigma$) \\ 1.1 (0.7$\sigma$)} \\\cline{2-7}
                           & Periodic & 
                            \makecell{0 \\ 10 \\ 20 \\ 30 \\ 40 \\ 50 \\ 60 \\ 70 \\ 80 \\ 90} &
                            \makecell{4.6$\pm$2.9 (0.8$\pm$0.8$\sigma$) \\ 4.6$\pm$2.6 (0.8$\pm$0.8$\sigma$) \\ 4.6$\pm$2.6 (0.8$\pm$0.6$\sigma$) \\ 4.4$\pm$2.6 (0.8$\pm$0.7$\sigma$) \\ 4.2$\pm$2.6 (0.8$\pm$0.8$\sigma$) \\ 3.7$\pm$2.6 (0.7$\pm$0.8$\sigma$) \\ 3.6$\pm$2.6 (0.7$\pm$0.6$\sigma$) \\ 3.2$\pm$2.6 (0.8$\pm$0.7$\sigma$) \\ 2.7$\pm$2.5 (0.8$\pm$0.9$\sigma$) \\ 2.5$\pm$2.3 (0.7$\pm$0.7$\sigma$)} &
                            \makecell{8.3 (0.9$\sigma$) \\ 8.3 (0.9$\sigma$) \\ 8.3 (0.9$\sigma$) \\ 8.3 (0.9$\sigma$) \\ 8.3 (1.0$\sigma$) \\ 8.3 (1.0$\sigma$) \\ 8.3 (1.0$\sigma$) \\ 1.2 (0.7$\sigma$) \\ 1.1 (0.8$\sigma$) \\ 1.1 (0.9$\sigma$)} &
                            \makecell{6.5$\pm$2.0 (0.8$\pm$0.8$\sigma$) \\ 6.5$\pm$2.0 (0.8$\pm$0.9$\sigma$) \\ 6.1$\pm$2.1 (0.9$\pm$0.9$\sigma$) \\ 5.8$\pm$2.2 (0.7$\pm$0.9$\sigma$) \\ 6.5$\pm$2.3 (0.8$\pm$0.7$\sigma$) \\ 6.0$\pm$2.3 (0.8$\pm$0.8$\sigma$) \\ 5.5$\pm$2.3 (0.9$\pm$0.7$\sigma$) \\ 4.3$\pm$2.5 (0.7$\pm$0.9$\sigma$) \\ 4.2$\pm$2.4 (0.7$\pm$0.9$\sigma$) \\ 4.1$\pm$2.2 (0.7$\pm$0.8$\sigma$)} &
                            \makecell{8.3 (1.0$\sigma$) \\ 8.3 (1.0$\sigma$) \\ 8.3 (1.0$\sigma$) \\ 8.3 (1.0$\sigma$) \\ 8.3 (0.9$\sigma$) \\ 8.3 (0.9$\sigma$) \\ 8.3 (0.9$\sigma$) \\ 8.3 (1.0$\sigma$) \\ 2.2 (0.7$\sigma$) \\ 1.4 (0.8$\sigma$)} \\\cline{2-7}
                            & Aperiodic & 
                            \makecell{0 \\ 10 \\ 20 \\ 30 \\ 40 \\ 50 \\ 60 \\ 70 \\ 80 \\ 90} &
                            \makecell{4.6$\pm$2.9 (0.9$\pm$0.8$\sigma$) \\ 4.7$\pm$2.6 (0.9$\pm$0.8$\sigma$) \\ 4.4$\pm$2.6 (0.7$\pm$0.9$\sigma$) \\ 4.3$\pm$2.6 (0.7$\pm$0.7$\sigma$) \\ 4.3$\pm$2.6 (0.8$\pm$0.8$\sigma$) \\ 3.7$\pm$2.6 (0.8$\pm$0.9$\sigma$) \\ 3.5$\pm$2.6 (0.7$\pm$0.8$\sigma$) \\ 3.2$\pm$2.6 (0.9$\pm$0.8$\sigma$) \\ 2.8$\pm$2.6 (0.8$\pm$0.7$\sigma$) \\ 2.3$\pm$2.3 (0.8$\pm$0.8$\sigma$)} &
                            \makecell{8.3 (0.9$\sigma$) \\ 8.3 (0.9$\sigma$) \\ 8.3 (0.9$\sigma$) \\ 8.3 (1.0$\sigma$) \\ 8.3 (0.9$\sigma$) \\ 8.3 (1.0$\sigma$) \\ 8.3 (1.0$\sigma$) \\ 8.3 (1.1$\sigma$) \\ 1.1 (0.8$\sigma$) \\ 1.1 (0.8$\sigma$)} &
                            \makecell{6.5$\pm$2.0 (0.8$\pm$0.7$\sigma$) \\ 6.6$\pm$2.1 (0.8$\pm$0.7$\sigma$) \\ 6.6$\pm$2.1 (0.9$\pm$0.8$\sigma$) \\ 6.2$\pm$2.2 (0.7$\pm$0.7$\sigma$) \\ 5.8$\pm$2.3 (0.8$\pm$0.7$\sigma$) \\ 5.5$\pm$2.4 (0.7$\pm$0.8$\sigma$) \\ 5.0$\pm$2.4 (0.7$\pm$0.8$\sigma$) \\ 5.1$\pm$2.4 (0.7$\pm$0.7$\sigma$) \\ 4.0$\pm$2.4 (0.7$\pm$0.7$\sigma$) \\ 3.4$\pm$2.3 (0.8$\pm$0.7$\sigma$)} &
                            \makecell{8.3 (0.9$\sigma$) \\ 8.3 (0.9$\sigma$) \\ 8.3 (0.9$\sigma$) \\ 8.3 (0.9$\sigma$) \\ 8.3 (0.8$\sigma$) \\ 8.3 (0.8$\sigma$) \\ 8.3 (0.9$\sigma$) \\ 8.3 (0.8$\sigma$) \\ 2.2 (0.7$\sigma$) \\ 1.1 (0.8$\sigma$)} \\
\hline
\hline
\end{tabular}%
}
\end{table*}
\begin{table*}
\centering
\caption{Periodicity analysis results from tests using pure noise LCs with SSA are presented. They are shown the results using three different window lengths (WL): 40\%, 30\%, and 20\%. The gap range explored in these tests spans from 0\% to 90\%. Three distinct gap scenarios are considered: randomly injected gaps throughout the LC (``Random''), periodic gaps introduced annually in the LC (``Periodic''), and randomly distributed gaps introduced annually in the LC (``Aperiodic''). We report two key results: the period (and its significance) corresponding to the median of all test results (denoted as ``Median'') and the period (and its significance) corresponding to the most frequently detected period in the tests (denoted as ``Mfp''). All the periods are in years.\label{tab:noise_random_ssa}}
{%
\begin{tabular}{l|cc|cc|cc|cc|ccc}
\hline
\hline
Type LC & Type of Gaps & Percentage  & SSA (Median) & SSA (Mfp) & SSA (Median) & SSA (Mfp) & SSA (Median) & SSA (Mfp) \\
 & & of Gaps [\%] & WL=40\% & WL=40\% & WL=30\% & WL=30\% & WL=20\% & WL=20\% \\
\hline
		\multirow{3}{*}{Noise} & Random & 
                            \makecell{0 \\ 10 \\ 20 \\ 30 \\ 40 \\ 50 \\ 60 \\ 70 \\ 80 \\ 90} &
                            \makecell{6.6$\pm$2.4 (0.9$\pm$0.6$\sigma$) \\ 6.7$\pm$2.3 (0.8$\pm$0.7$\sigma$)\\ 6.6$\pm$2.4 (0.7$\pm$0.7$\sigma$) \\ 6.6$\pm$2.4 (0.9$\pm$0.8$\sigma$) \\ 6.6$\pm$2.5 (0.8$\pm$0.7$\sigma$) \\ 6.4$\pm$2.5 (0.7$\pm$0.8$\sigma$) \\ 5.6$\pm$2.6 (0.7$\pm$0.7$\sigma$) \\ 4.9$\pm$2.7 (0.7$\pm$0.6$\sigma$) \\ 3.6$\pm$2.7 (0.7$\pm$0.7$\sigma$) \\ 2.5$\pm$2.4 (0.8$\pm$0.7$\sigma$)} & 
                            \makecell{8.3 (0.9$\sigma$) \\ 8.3 (0.9$\sigma$) \\ 8.3 (0.9$\sigma$) \\ 8.3 (0.9$\sigma$) \\ 8.3 (0.9$\sigma$) \\ 8.3 (1.1$\sigma$) \\ 8.3 (1.0$\sigma$) \\ 8.3 (1.2$\sigma$) \\ 8.3 (1.1$\sigma$) \\ 8.3 (1.1$\sigma$)} &
                            \makecell{7.1$\pm$2.4 (0.8$\pm$0.6$\sigma$) \\ 7.1$\pm$2.4 (0.9$\pm$0.8$\sigma$) \\ 7.1$\pm$2.4 (0.9$\pm$0.9$\sigma$) \\ 7.1$\pm$2.4 (0.8$\pm$0.7$\sigma$) \\ 7.1$\pm$2.4 (0.8$\pm$0.6$\sigma$) \\ 7.0$\pm$2.5 (0.8$\pm$0.8$\sigma$) \\ 6.7$\pm$2.7 (0.7$\pm$0.8$\sigma$) \\ 6.0$\pm$2.8 (0.7$\pm$0.7$\sigma$) \\ 3.6$\pm$2.9 (0.8$\pm$0.7$\sigma$) \\ 2.2$\pm$2.5 (0.8$\pm$0.7$\sigma$)} &
                            \makecell{8.3 (0.9$\sigma$) \\ 8.3 (0.9$\sigma$) \\ 8.3 (0.9$\sigma$) \\ 8.3 (0.8$\sigma$) \\ 8.3 (0.9$\sigma$) \\ 8.3 (0.8$\sigma$) \\ 8.3 (0.9$\sigma$) \\ 8.3 (1.0$\sigma$) \\ 8.3 (0.9$\sigma$) \\ 8.3 (0.9$\sigma$)} &
                            \makecell{6.9$\pm$2.1 (0.9$\pm$0.8$\sigma$) \\ 6.9$\pm$2.2 (0.8$\pm$0.8$\sigma$) \\ 6.9$\pm$2.2 (0.8$\pm$0.7$\sigma$) \\ 6.9$\pm$2.1 (0.8$\pm$0.9$\sigma$) \\ 7.0$\pm$2.2 (0.9$\pm$0.8$\sigma$) \\ 6.9$\pm$2.2 (0.7$\pm$0.7$\sigma$) \\ 6.8$\pm$2.3 (0.7$\pm$0.7$\sigma$) \\ 6.5$\pm$2.5 (0.8$\pm$0.8$\sigma$) \\ 5.0$\pm$2.7 (0.7$\pm$0.7$\sigma$) \\ 2.4$\pm$2.3 (0.7$\pm$0.9$\sigma$)} &
                            \makecell{8.3 (0.9$\sigma$) \\ 8.3 (0.9$\sigma$) \\ 8.3 (0.9$\sigma$) \\ 8.3 (0.9$\sigma$) \\ 8.3 (0.8$\sigma$) \\ 8.3 (0.8$\sigma$) \\ 8.3 (0.8$\sigma$) \\ 8.3 (0.9$\sigma$) \\ 8.3 (0.8$\sigma$) \\ 8.3 (1.0$\sigma$)} \\\cline{2-9}
                           & Periodic & 
                            \makecell{0 \\ 10 \\ 20 \\ 30 \\ 40 \\ 50 \\ 60 \\ 70 \\ 80 \\ 90} &
                            \makecell{6.6$\pm$2.4 (0.9$\pm$0.8$\sigma$) \\ 6.6$\pm$2.4 (0.9$\pm$0.8$\sigma$) \\ 6.7$\pm$2.4 (0.9$\pm$0.8$\sigma$) \\ 6.7$\pm$2.4 (0.8$\pm$0.8$\sigma$) \\ 6.5$\pm$2.5 (0.8$\pm$0.8$\sigma$) \\ 6.1$\pm$2.8 (0.7$\pm$0.8$\sigma$) \\ 5.0$\pm$2.9 (0.7$\pm$0.7$\sigma$) \\ 3.4$\pm$2.9 (0.8$\pm$0.8$\sigma$) \\ 2.2$\pm$2.8 (0.7$\pm$0.7$\sigma$) \\ 1.6$\pm$2.4 (0.9$\pm$0.8$\sigma$)} &
                            \makecell{8.3 (0.9$\sigma$) \\ 8.3 (0.9$\sigma$) \\ 8.3 (0.9$\sigma$) \\ 8.3 (0.9$\sigma$) \\ 8.3 (0.9$\sigma$) \\ 8.3 (1.0$\sigma$) \\ 8.3 (0.9$\sigma$) \\ 1.1 (0.6$\sigma$) \\ 1.1 (0.7$\sigma$) \\ 1.1 (0.5$\sigma$)}&
                            \makecell{7.1$\pm$2.4 (0.9$\pm$0.9$\sigma$) \\ 7.1$\pm$2.4 (0.9$\pm$0.8$\sigma$) \\ 7.1$\pm$2.5 (0.8$\pm$0.8$\sigma$) \\ 7.1$\pm$2.5 (0.9$\pm$0.8$\sigma$) \\ 7.1$\pm$2.6 (0.8$\pm$0.7$\sigma$) \\ 6.7$\pm$2.8 (0.8$\pm$0.7$\sigma$) \\ 6.5$\pm$2.9 (0.8$\pm$0.8$\sigma$) \\ 4.7$\pm$3.0 (0.9$\pm$0.8$\sigma$) \\ 2.9$\pm$3.0 (0.7$\pm$0.8$\sigma$) \\ 1.7$\pm$2.7 (0.7$\pm$0.8$\sigma$)} &
                            \makecell{8.3 (0.9$\sigma$) \\ 8.3 (0.9$\sigma$) \\ 8.3 (0.9$\sigma$) \\ 8.3 (0.9$\sigma$) \\ 8.3 (0.9$\sigma$) \\ 8.3 (0.9$\sigma$) \\ 8.3 (0.9$\sigma$) \\ 8.3 (0.9$\sigma$) \\ 1.1 (0.8$\sigma$) \\ 1.1 (0.8$\sigma$)} &
                            \makecell{7.0$\pm$2.1 (0.8$\pm$0.9$\sigma$) \\ 7.0$\pm$2.1 (0.8$\pm$0.9$\sigma$) \\ 7.0$\pm$2.2 (0.9$\pm$0.9$\sigma$) \\ 7.0$\pm$2.1 (0.9$\pm$0.8$\sigma$) \\ 6.9$\pm$2.3 (0.8$\pm$0.7$\sigma$) \\ 6.9$\pm$2.5 (0.8$\pm$0.8$\sigma$) \\ 6.6$\pm$2.6 (0.7$\pm$0.8$\sigma$) \\ 6.1$\pm$2.8 (0.8$\pm$0.8$\sigma$) \\ 5.3$\pm$2.9 (0.7$\pm$0.7$\sigma$) \\ 1.9$\pm$2.8 (0.7$\pm$0.7$\sigma$)} &
                            \makecell{8.3 (0.9$\sigma$) \\ 8.3 (0.9$\sigma$) \\ 8.3 (0.9$\sigma$) \\ 8.3 (0.9$\sigma$) \\ 8.3 (0.9$\sigma$) \\ 8.3 (0.9$\sigma$) \\ 8.3 (0.9$\sigma$) \\ 8.3 (0.9$\sigma$) \\ 8.3 (0.9$\sigma$) \\ 1.1 (0.6$\sigma$)} \\\cline{2-9}
                            & Aperiodic & 
                            \makecell{0 \\ 10 \\ 20 \\ 30 \\ 40 \\ 50 \\ 60 \\ 70 \\ 80 \\ 90} &
                            \makecell{6.6$\pm$2.4 (0.9$\pm$0.7$\sigma$) \\ 6.6$\pm$2.4 (0.9$\pm$0.8$\sigma$) \\ 6.6$\pm$2.4 (0.9$\pm$0.8$\sigma$) \\ 6.6$\pm$2.5 (0.9$\pm$0.7$\sigma$) \\ 6.5$\pm$2.6 (0.8$\pm$0.8$\sigma$) \\ 6.1$\pm$2.7 (0.7$\pm$0.7$\sigma$) \\ 6.1$\pm$2.7 (0.8$\pm$0.7$\sigma$) \\ 4.5$\pm$2.8 (0.7$\pm$0.9$\sigma$) \\ 2.9$\pm$2.8 (0.7$\pm$0.9$\sigma$) \\ 2.2$\pm$2.6 (0.8$\pm$0.6$\sigma$)} &
                            \makecell{8.3 (0.9$\sigma$) \\ 8.3 (0.9$\sigma$) \\ 8.3 (0.9$\sigma$) \\ 8.3 (0.9$\sigma$) \\ 8.3 (0.9$\sigma$) \\ 8.3 (0.9$\sigma$) \\ 8.3 (0.9$\sigma$) \\ 8.3 (0.9$\sigma$) \\ 1.1 (0.6$\sigma$) \\ 1.1 (0.6$\sigma$)} &
                            \makecell{7.1$\pm$2.4 (0.8$\pm$0.9$\sigma$) \\ 7.1$\pm$2.4 (0.8$\pm$0.8$\sigma$) \\ 7.1$\pm$2.4 (0.9$\pm$0.8$\sigma$) \\ 7.1$\pm$2.5 (0.7$\pm$0.9$\sigma$) \\ 7.1$\pm$2.5 (0.8$\pm$0.7$\sigma$) \\ 7.1$\pm$2.6 (0.7$\pm$0.8$\sigma$) \\ 6.7$\pm$2.8 (0.8$\pm$0.7$\sigma$) \\ 6.2$\pm$2.9 (0.9$\pm$0.8$\sigma$) \\ 3.9$\pm$3.0 (0.7$\pm$0.7$\sigma$) \\ 2.4$\pm$2.8 (0.7$\pm$0.7$\sigma$)} &
                            \makecell{8.3 (0.9$\sigma$) \\ 8.3 (0.9$\sigma$) \\ 8.3 (0.9$\sigma$) \\ 8.3 (0.9$\sigma$) \\ 8.3 (0.9$\sigma$) \\ 8.3 (0.9$\sigma$) \\ 8.3 (0.9$\sigma$) \\ 8.3 (0.9$\sigma$) \\ 8.3 (0.9$\sigma$) \\ 1.1 (0.7$\sigma$)} &
                            \makecell{6.9$\pm$2.2 (0.9$\pm$0.7$\sigma$) \\ 6.9$\pm$2.2 (0.9$\pm$0.8$\sigma$) \\ 6.9$\pm$2.2 (0.8$\pm$0.6$\sigma$) \\ 6.9$\pm$2.2 (0.8$\pm$0.6$\sigma$) \\ 6.9$\pm$2.3 (0.8$\pm$0.9$\sigma$) \\ 6.9$\pm$2.4 (0.8$\pm$0.8$\sigma$) \\ 6.9$\pm$2.4 (0.8$\pm$0.9$\sigma$) \\ 6.6$\pm$2.5 (0.8$\pm$0.6$\sigma$) \\ 6.0$\pm$2.7 (0.7$\pm$0.9$\sigma$) \\ 4.0$\pm$2.9 (0.8$\pm$0.7$\sigma$)} &
                            \makecell{8.3 (0.9$\sigma$) \\ 8.3 (0.9$\sigma$) \\ 8.3 (0.9$\sigma$) \\ 8.3 (0.9$\sigma$) \\ 8.3 (0.9$\sigma$) \\ 8.3 (0.9$\sigma$) \\ 8.3 (0.9$\sigma$) \\ 8.3 (0.9$\sigma$) \\ 8.3 (0.9$\sigma$) \\ 1.1 (0.6$\sigma$)} \\
\hline
\hline
\end{tabular}%
}
\end{table*}
\begin{table*}
\centering
\caption{Periodicity analysis results from the tests conducted for a periodic signal of 2 years. The gap range explored in these tests spans from 0\% to 90\%. Three distinct gap scenarios are considered: randomly injected gaps throughout the LC (``Random''), periodic gaps introduced annually in the LC (``Periodic''), and randomly distributed gaps introduced annually in the LC (``Aperiodic''). The analysis methods used are the LSP and PDM techniques. All the periods are in years.\label{tab:period_2_years}}
{%
\begin{tabular}{l|ccccccc}
\hline
\hline
Type LC & Type of Gaps & Percentage of Gaps [\%] & LSP & PDM \\
\hline
		\multirow{3}{*}{2-year } & Random & 
                            \makecell{0 \\ 10 \\ 20 \\ 30 \\ 40 \\ 50 \\ 60 \\ 70 \\ 80 \\ 90} &
                            \makecell{2.0 (5.1$\sigma$) \\ 2.0$\pm$0.01 (5.0$\pm$0.4$\sigma$) \\ 2.0$\pm$0.01 (4.9$\pm$0.4$\sigma$) \\ 2.0$\pm$0.02 (4.9$\pm$0.4$\sigma$) \\ 2.0$\pm$0.02 (4.6$\pm$0.5$\sigma$) \\ 2.0$\pm$0.02 (4.4$\pm$0.5$\sigma$) \\ 2.0$\pm$0.1 (4.1$\pm$0.5$\sigma$) \\ 2.0$\pm$0.2 (3.8$\pm$0.6$\sigma$) \\ 2.0$\pm$0.5 (3.2$\pm$0.6$\sigma$) \\ 2.0$\pm$1.6 (2.4$\pm$0.7$\sigma$)} &                      
                            \makecell{2.0 (5.2$\sigma$) \\ 2.0$\pm$0.1 (5.2$\pm$0.4$\sigma$) \\ 2.0$\pm$0.4 (5.1$\pm$0.9$\sigma$) \\ 2.0$\pm$0.4 (5.1$\pm$0.9$\sigma$) \\ 2.0$\pm$0.7 (5.1$\pm$1.0$\sigma$) \\ 2.0$\pm$1.0 (4.9$\pm$1.0$\sigma$) \\ 2.0$\pm$1.2 (4.7$\pm$1.2$\sigma$) \\ 2.0$\pm$1.4 (4.4$\pm$1.2$\sigma$) \\ 2.0$\pm$1.7 (3.6$\pm$1.2$\sigma$) \\ 2.5$\pm$2.2 (1.2$\pm$0.2$\sigma$)} \\\cline{2-5}
                           & Periodic & 
                            \makecell{0 \\ 10 \\ 20 \\ 30 \\ 40 \\ 50 \\ 60 \\ 70 \\ 80 \\ 90} &
                            \makecell{2.0 (5.1$\sigma$) \\ 2.0$\pm$0.01 (5.0$\pm$0.4$\sigma$) \\ 2.0$\pm$0.01 (5.0$\pm$0.4$\sigma$) \\ 2.0$\pm$0.01 (5.0$\pm$0.5$\sigma$) \\ 2.0$\pm$0.02 (4.9$\pm$0.6$\sigma$) \\ 2.0$\pm$0.02 (4.9$\pm$0.6$\sigma$) \\ 2.0$\pm$0.1 (4.7$\pm$0.6$\sigma$) \\ 2.0$\pm$0.3 (4.5$\pm$0.7$\sigma$) \\ 2.0$\pm$0.4 (4.3$\pm$0.8$\sigma$) \\ 2.0$\pm$0.7 (3.9$\pm$0.9$\sigma$)} &
                            \makecell{2.0 (5.2$\sigma$) \\ 2.0$\pm$0.1 (5.2$\pm$0.4$\sigma$) \\ 2.0$\pm$0.5 (5.2$\pm$0.5$\sigma$) \\ 2.0$\pm$0.6 (5.2$\pm$0.6$\sigma$) \\ 2.0$\pm$0.7 (5.2$\pm$0.8$\sigma$) \\ 2.0$\pm$0.9 (5.1$\pm$1.1$\sigma$) \\ 2.0$\pm$1.2 (4.4$\pm$1.1$\sigma$) \\ 2.0$\pm$1.2 (4.4$\pm$1.1$\sigma$) \\ 4.0$\pm$2.0 (3.2$\pm$1.4$\sigma$) \\ 3.9$\pm$2.1 (1.7$\pm$1.5$\sigma$)} \\\cline{2-5}
                            & Aperiodic & 
                            \makecell{0 \\ 10 \\ 20 \\ 30 \\ 40 \\ 50 \\ 60 \\ 70 \\ 80 \\ 90} &
                            \makecell{2.0 (5.1$\sigma$) \\ 2.0$\pm$0.01 (5.0$\pm$0.4$\sigma$) \\ 2.0$\pm$0.01 (5.0$\pm$0.4$\sigma$) \\ 2.0$\pm$0.01 (5.0$\pm$0.4$\sigma$) \\ 2.0$\pm$0.02 (4.9$\pm$0.4$\sigma$) \\ 2.0$\pm$0.02 (4.6$\pm$0.5$\sigma$) \\ 2.0$\pm$0.02 (4.5$\pm$0.5$\sigma$) \\ 2.0$\pm$0.1 (4.3$\pm$0.6$\sigma$) \\ 2.0$\pm$0.3 (4.0$\pm$0.6$\sigma$) \\ 2.0$\pm$0.7 (3.4$\pm$0.6$\sigma$)} &
                            \makecell{2.0 (5.2$\sigma$) \\ 2.0$\pm$0.1 (5.2$\pm$0.4$\sigma$) \\ 2.0$\pm$0.5 (5.2$\pm$0.5$\sigma$) \\ 2.0$\pm$0.5 (5.2$\pm$0.6$\sigma$) \\ 2.0$\pm$0.5 (5.2$\pm$0.6$\sigma$) \\ 2.0$\pm$0.9 (5.1$\pm$1.0$\sigma$) \\ 2.0$\pm$1.0 (5.1$\pm$1.1$\sigma$) \\ 2.0$\pm$1.2 (4.8$\pm$1.1$\sigma$) \\ 2.0$\pm$1.5 (4.5$\pm$1.1$\sigma$) \\ 3.8$\pm$1.7 (2.6$\pm$1.2$\sigma$)} \\
\hline
\hline
\end{tabular}%
}
\end{table*}
\begin{table*}
\centering
\caption{Periodicity analysis results from the tests conducted for a periodic signal of 3 years. The gap range explored in these tests spans from 0\% to 90\%. Three distinct gap scenarios are considered: randomly injected gaps throughout the LC (``Random''), periodic gaps introduced annually in the LC (``Periodic''), and randomly distributed gaps introduced annually in the LC (``Aperiodic''). The analysis methods used are the LSP and PDM techniques. All the periods are in years.\label{tab:period_3_years}}
{%
\begin{tabular}{l|ccccccc}
\hline
\hline
Type LC & Type of Gaps & Percentage of Gaps [\%] & LSP & PDM \\
\hline
		\multirow{3}{*}{3-year } & Random & 
                            \makecell{0 \\ 10 \\ 20 \\ 30 \\ 40 \\ 50 \\ 60 \\ 70 \\ 80 \\ 90} &
                            \makecell{3.0 (5.0$\sigma$) \\ 3.0$\pm$0.04 (4.9$\pm$0.3$\sigma$) \\ 3.0$\pm$0.04 (4.9$\pm$0.3$\sigma$) \\ 3.0$\pm$0.04 (4.8$\pm$0.3$\sigma$) \\ 3.0$\pm$0.04 (4.7$\pm$0.3$\sigma$) \\ 3.0$\pm$0.05 (4.4$\pm$0.3$\sigma$) \\ 3.0$\pm$0.05 (4.5$\pm$0.3$\sigma$) \\ 3.0$\pm$0.05 (4.4$\pm$0.3$\sigma$) \\ 3.0$\pm$0.1 (4.3$\pm$0.5$\sigma$) \\ 3.0$\pm$0.2 (4.1$\pm$0.5$\sigma$)} &                        
                            \makecell{3.0 (5.0$\sigma$) \\ 3.0$\pm$0.1 (5.0$\pm$0.2$\sigma$) \\ 3.0$\pm$0.2 (5.0$\pm$0.3$\sigma$) \\ 3.0$\pm$0.3 (5.0$\pm$0.4$\sigma$) \\ 3.0$\pm$0.4 (4.9$\pm$0.5$\sigma$) \\ 3.0$\pm$0.6 (4.8$\pm$0.6$\sigma$) \\ 3.0$\pm$0.7 (4.8$\pm$0.7$\sigma$) \\ 3.0$\pm$1.0 (4.7$\pm$0.9$\sigma$) \\ 3.0$\pm$1.2 (4.4$\pm$0.9$\sigma$) \\ 3.1$\pm$1.8 (1.2$\pm$0.2$\sigma$)} \\\cline{2-5}
                            & Periodic & 
                            \makecell{0 \\ 10 \\ 20 \\ 30 \\ 40 \\ 50 \\ 60 \\ 70 \\ 80 \\ 90} &
                            \makecell{3.0 (5.0$\sigma$) \\ 3.0$\pm$0.04 (4.9$\pm$0.3$\sigma$) \\ 3.0$\pm$0.04 (4.9$\pm$0.3$\sigma$) \\ 3.0$\pm$0.04 (4.8$\pm$0.3$\sigma$) \\ 3.0$\pm$0.04 (4.7$\pm$0.3$\sigma$) \\ 3.0$\pm$0.05 (4.6$\pm$0.3$\sigma$) \\ 3.0$\pm$0.05 (4.5$\pm$0.3$\sigma$) \\ 3.0$\pm$0.05 (4.4$\pm$0.3$\sigma$) \\ 3.0$\pm$0.1 (4.3$\pm$0.4$\sigma$) \\ 3.0$\pm$0.2 (4.1$\pm$0.5$\sigma$)} &
                            \makecell{3.0 (5.0$\sigma$) \\ 3.0$\pm$0.1 (5.0$\pm$0.3$\sigma$) \\ 3.0$\pm$0.2 (5.0$\pm$0.3$\sigma$) \\ 3.0$\pm$0.3 (5.0$\pm$0.4$\sigma$) \\ 3.0$\pm$0.5 (5.0$\pm$0.5$\sigma$) \\ 3.0$\pm$0.7 (5.0$\pm$0.6$\sigma$) \\ 3.0$\pm$0.8 (4.8$\pm$0.7$\sigma$) \\ 3.0$\pm$1.0 (4.5$\pm$0.9$\sigma$) \\ 3.0$\pm$1.4 (4.2$\pm$1.0$\sigma$) \\ 3.2$\pm$2.3 (3.2$\pm$1.4$\sigma$)} \\\cline{2-5}
                            & Aperiodic & 
                            \makecell{0 \\ 10 \\ 20 \\ 30 \\ 40 \\ 50 \\ 60 \\ 70 \\ 80 \\ 90} &
                            \makecell{3.0 (5.0$\sigma$) \\ 3.0$\pm$0.04 (4.9$\pm$0.3$\sigma$) \\ 3.0$\pm$0.04 (4.9$\pm$0.3$\sigma$) \\ 3.0$\pm$0.04 (4.8$\pm$0.3$\sigma$) \\ 3.0$\pm$0.04 (4.7$\pm$0.3$\sigma$) \\ 3.0$\pm$0.05 (4.6$\pm$0.3$\sigma$) \\ 3.0$\pm$0.05 (4.5$\pm$0.3$\sigma$) \\ 3.0$\pm$0.05 (4.3$\pm$0.4$\sigma$) \\ 3.0$\pm$0.06 (4.2$\pm$0.4$\sigma$) \\ 3.0$\pm$0.2 (3.8$\pm$0.4$\sigma$)} &
                            \makecell{3.0 (5.0$\sigma$) \\ 3.0$\pm$0.1 (5.0$\pm$0.3$\sigma$) \\ 3.0$\pm$0.1 (5.0$\pm$0.3$\sigma$) \\ 3.0$\pm$0.3 (5.0$\pm$0.4$\sigma$) \\ 3.0$\pm$0.5 (5.0$\pm$0.6$\sigma$) \\ 3.0$\pm$0.5 (5.0$\pm$0.6$\sigma$) \\ 3.0$\pm$0.6 (4.9$\pm$0.6$\sigma$) \\ 3.0$\pm$0.9 (4.9$\pm$0.7$\sigma$) \\ 3.0$\pm$1.1 (4.9$\pm$0.7$\sigma$) \\ 3.0$\pm$1.4 (3.7$\pm$1.1$\sigma$)} \\
\hline
\hline
\end{tabular}%
}
\end{table*}
\begin{table*}
\centering
\caption{Periodicity analysis results from the tests conducted for a periodic signal of 2 years using SSA. They are shown the results using three different window lengths (WL): 40\%, 30\%, and 20\%. The gap range explored in these tests spans from 0\% to 90\%. Three distinct gap scenarios are considered: randomly injected gaps throughout the LC (``Random''), periodic gaps introduced annually in the LC (``Periodic''), and randomly distributed gaps introduced annually in the LC (``Aperiodic''). All the periods are in years.\label{tab:period_2_years_ssa}}
{%
\begin{tabular}{l|ccccccc}
\hline
\hline
Type LC & Type of Gaps & Percentage of Gaps [\%] & SSA (WL=40\%) & SSA (WL=30\%) & SSA (WL=20\%) \\
\hline
		\multirow{3}{*}{2-year } & Random & 
                            \makecell{0 \\ 10 \\ 20 \\ 30 \\ 40 \\ 50 \\ 60 \\ 70 \\ 80 \\ 90} &
                            \makecell{2.0 (5.3$\sigma$) \\ 2.0$\pm$0.02 (5.3$\pm$0.1$\sigma$) \\ 2.0$\pm$0.03 (5.3$\pm$0.1$\sigma$) \\ 2.0$\pm$0.2 (5.3$\pm$0.2$\sigma$) \\ 2.0$\pm$0.5 (5.3$\pm$0.5$\sigma$) \\ 2.0$\pm$0.8 (5.2$\pm$0.9$\sigma$) \\ 2.0$\pm$1.1 (4.3$\pm$1.1$\sigma$) \\ 2.0$\pm$1.5 (3.4$\pm$1.0$\sigma$) \\ 2.0$\pm$2.0 (2.6$\pm$0.7$\sigma$) \\ 2.0$\pm$2.1 (2.2$\pm$0.5$\sigma$)} &
                            \makecell{2.0 (5.3$\sigma$) \\ 2.0$\pm$0.02 (5.3$\pm$0.1$\sigma$) \\ 2.0$\pm$0.04 (5.3$\pm$0.1$\sigma$) \\ 2.0$\pm$0.2 (5.3$\pm$0.2$\sigma$) \\ 2.0$\pm$0.5 (5.3$\pm$0.4$\sigma$) \\ 2.0$\pm$0.8 (5.3$\pm$0.9$\sigma$) \\ 2.0$\pm$1.5 (4.5$\pm$1.2$\sigma$) \\ 2.0$\pm$1.9 (3.5$\pm$1.1$\sigma$) \\ 2.0$\pm$2.2 (2.7$\pm$0.8$\sigma$) \\ 2.0$\pm$2.0 (2.2$\pm$0.4$\sigma$)} &
                            \makecell{2.0 (5.3$\sigma$) \\ 2.0$\pm$0.02 (5.3$\pm$0.1$\sigma$) \\ 2.0$\pm$0.03 (5.3$\pm$0.1$\sigma$) \\ 2.0$\pm$0.2 (5.3$\pm$0.2$\sigma$) \\ 2.0$\pm$0.5 (5.3$\pm$0.4$\sigma$) \\ 2.0$\pm$0.8 (5.3$\pm$0.9$\sigma$) \\ 2.0$\pm$1.5 (4.8$\pm$1.2$\sigma$) \\ 2.1$\pm$1.9 (3.6$\pm$1.2$\sigma$) \\ 2.0$\pm$2.2 (2.7$\pm$0.7$\sigma$) \\ 2.0$\pm$1.8 (2.3$\pm$0.4$\sigma$)} \\\cline{2-6}
                           & Periodic & 
                            \makecell{0 \\ 10 \\ 20 \\ 30 \\ 40 \\ 50 \\ 60 \\ 70 \\ 80 \\ 90} &
                            \makecell{2.0 (5.3$\sigma$) \\ 2.0$\pm$0.02 (5.3$\pm$0.1$\sigma$) \\ 2.0$\pm$0.02 (5.3$\pm$0.1$\sigma$) \\ 2.0$\pm$0.2 (5.3$\pm$0.2$\sigma$) \\ 2.0$\pm$0.2 (5.3$\pm$0.2$\sigma$) \\ 2.0$\pm$0.1 (5.3$\pm$0.3$\sigma$) \\ 2.0$\pm$0.1 (5.3$\pm$0.3$\sigma$) \\ 2.0$\pm$0.4 (5.3$\pm$0.6$\sigma$) \\ 2.0$\pm$0.7 (5.3$\pm$0.9$\sigma$) \\ 2.0$\pm$1.1 (5.0$\pm$1.2$\sigma$)} &
                            \makecell{2.0 (5.3$\sigma$) \\ 2.0$\pm$0.03 (5.3$\pm$0.1$\sigma$) \\ 2.0$\pm$0.03 (5.3$\pm$0.1$\sigma$) \\ 2.0$\pm$0.03 (5.3$\pm$0.1$\sigma$) \\ 2.0$\pm$0.04 (5.3$\pm$0.1$\sigma$) \\ 2.0$\pm$0.4 (5.3$\pm$0.3$\sigma$) \\ 2.0$\pm$0.4 (5.3$\pm$0.4$\sigma$) \\ 2.0$\pm$0.5 (5.3$\pm$0.6$\sigma$) \\ 2.0$\pm$0.7 (5.3$\pm$0.8$\sigma$) \\ 2.0$\pm$0.8 (5.3$\pm$1.1$\sigma$)} &
                            \makecell{2.0 (5.3$\sigma$) \\ 2.0$\pm$0.02 (5.3$\pm$0.1$\sigma$) \\ 2.0$\pm$0.03 (5.3$\pm$0.1$\sigma$) \\ 2.0$\pm$0.03 (5.3$\pm$0.1$\sigma$) \\ 2.0$\pm$0.03 (5.3$\pm$0.1$\sigma$) \\ 2.0$\pm$0.05 (5.3$\pm$0.3$\sigma$) \\ 2.0$\pm$0.3 (5.3$\pm$0.4$\sigma$) \\ 2.0$\pm$0.4 (5.3$\pm$0.5$\sigma$) \\ 2.0$\pm$0.6 (5.3$\pm$0.8$\sigma$) \\ 2.0$\pm$0.8 (5.2$\pm$1.2$\sigma$)} \\\cline{2-6}
                            & Aperiodic & 
                            \makecell{0 \\ 10 \\ 20 \\ 30 \\ 40 \\ 50 \\ 60 \\ 70 \\ 80 \\ 90} &
                            \makecell{2.0 (5.3$\sigma$) \\ 2.0$\pm$0.02 (5.3$\pm$0.1$\sigma$) \\ 2.0$\pm$0.02 (5.3$\pm$0.1$\sigma$) \\ 2.0$\pm$0.02 (5.3$\pm$0.1$\sigma$) \\ 2.0$\pm$0.03 (5.3$\pm$0.1$\sigma$) \\ 2.0$\pm$0.5 (5.3$\pm$0.4$\sigma$) \\ 2.0$\pm$0.5 (5.3$\pm$0.4$\sigma$) \\ 2.0$\pm$0.6 (5.3$\pm$0.8$\sigma$) \\ 2.0$\pm$1.2 (5.2$\pm$1.1$\sigma$) \\ 2.0$\pm$1.5 (3.9$\pm$1.2$\sigma$)} & 
                            \makecell{2.0 (5.3$\sigma$) \\ 2.0$\pm$0.03 (5.3$\pm$0.1$\sigma$) \\ 2.0$\pm$0.03 (5.3$\pm$0.1$\sigma$) \\ 2.0$\pm$0.04 (5.3$\pm$0.1$\sigma$) \\ 2.0$\pm$0.04 (5.3$\pm$0.1$\sigma$) \\ 2.0$\pm$0.2 (5.3$\pm$0.4$\sigma$) \\ 2.0$\pm$0.2 (5.3$\pm$0.4$\sigma$) \\ 2.0$\pm$0.7 (5.3$\pm$0.7$\sigma$) \\ 2.0$\pm$1.3 (5.0$\pm$1.2$\sigma$) \\ 2.0$\pm$1.3 (4.0$\pm$1.4$\sigma$)} & 
                            \makecell{2.0 (5.3$\sigma$) \\ 2.0$\pm$0.02 (5.3$\pm$0.1$\sigma$) \\ 2.0$\pm$0.02 (5.3$\pm$0.1$\sigma$) \\ 2.0$\pm$0.03 (5.3$\pm$0.1$\sigma$) \\ 2.0$\pm$0.04 (5.3$\pm$0.1$\sigma$) \\ 2.0$\pm$0.04 (5.3$\pm$0.1$\sigma$) \\ 2.0$\pm$0.3 (5.3$\pm$0.4$\sigma$) \\ 2.0$\pm$0.4 (5.3$\pm$0.7$\sigma$) \\ 2.0$\pm$0.8 (5.3$\pm$1.0$\sigma$) \\ 2.1$\pm$1.4 (3.6$\pm$1.2$\sigma$)} \\
\hline
\hline
\end{tabular}%
}
\end{table*}
\begin{table*}
\centering
\caption{Periodicity analysis results from the tests conducted for a periodic signal of 3 years using SSA. They are shown the results using three different window lengths (WL): 40\%, 30\%, and 20\%. The gap range explored in these tests spans from 0\% to 90\%. Three distinct gap scenarios are considered: randomly injected gaps throughout the LC (``Random''), periodic gaps introduced annually in the LC (``Periodic''), and randomly distributed gaps introduced annually in the LC (``Aperiodic''). All the periods are in years.\label{tab:period_3_years_ssa}}
{%
\begin{tabular}{l|ccccccc}
\hline
\hline
Type LC & Type of Gaps & Percentage of Gaps [\%] & SSA (WL=40\%) & SSA (WL=30\%) & SSA (WL=20\%) \\
\hline
		\multirow{3}{*}{3-year } & Random & 
                            \makecell{0 \\ 10 \\ 20 \\ 30 \\ 40 \\ 50 \\ 60 \\ 70 \\ 80 \\ 90} &
                            \makecell{3.0 (5.3$\sigma$) \\ 3.0$\pm$0.07 (5.3$\pm$0.3$\sigma$) \\ 3.0$\pm$0.07 (5.3$\pm$0.3$\sigma$) \\ 3.0$\pm$0.2 (5.3$\pm$0.5$\sigma$) \\ 3.0$\pm$0.3 (5.3$\pm$0.6$\sigma$) \\ 3.0$\pm$0.4 (5.1$\pm$0.8$\sigma$) \\ 3.0$\pm$0.8 (4.5$\pm$0.9$\sigma$) \\ 3.0$\pm$1.2 (3.7$\pm$1.0$\sigma$) \\ 3.0$\pm$1.6 (2.9$\pm$0.7$\sigma$) \\ 2.9$\pm$2.0 (2.2$\pm$0.5$\sigma$)} &
                            \makecell{3.0 (5.3$\sigma$) \\ 3.0$\pm$0.06 (5.3$\pm$0.2$\sigma$) \\ 3.0$\pm$0.06 (5.3$\pm$0.3$\sigma$) \\ 3.0$\pm$0.2 (5.3$\pm$0.3$\sigma$) \\ 3.0$\pm$0.3 (5.3$\pm$0.4$\sigma$) \\ 3.0$\pm$0.5 (5.0$\pm$0.7$\sigma$) \\ 3.0$\pm$0.9 (4.7$\pm$0.9$\sigma$) \\ 3.0$\pm$1.5 (4.0$\pm$1.0$\sigma$) \\ 3.0$\pm$1.9 (2.9$\pm$1.0$\sigma$) \\ 2.9$\pm$2.0 (2.2$\pm$0.4$\sigma$)} &
                            \makecell{3.0 (5.3$\sigma$) \\ 3.0$\pm$0.05 (5.3$\pm$0.3$\sigma$) \\ 3.0$\pm$0.05 (5.3$\pm$0.3$\sigma$) \\ 3.0$\pm$0.3 (5.3$\pm$0.6$\sigma$) \\ 3.0$\pm$0.5 (5.2$\pm$0.7$\sigma$) \\ 3.0$\pm$0.9 (5.0$\pm$0.8$\sigma$) \\ 3.0$\pm$1.0 (4.4$\pm$0.8$\sigma$) \\ 3.0$\pm$1.4 (3.9$\pm$0.9$\sigma$) \\ 3.0$\pm$1.7 (3.0$\pm$0.8$\sigma$) \\ 3.0$\pm$1.7 (2.2$\pm$0.4$\sigma$)} \\\cline{2-6}
                           & Periodic & 
                            \makecell{0 \\ 10 \\ 20 \\ 30 \\ 40 \\ 50 \\ 60 \\ 70 \\ 80 \\ 90} &
                            \makecell{3.0 (5.3$\sigma$) \\ 3.0$\pm$0.06 (5.3$\pm$0.2$\sigma$) \\ 3.0$\pm$0.06 (5.3$\pm$0.2$\sigma$) \\ 3.0$\pm$0.06 (5.3$\pm$0.2$\sigma$) \\ 3.0$\pm$0.1 (5.3$\pm$0.2$\sigma$) \\ 3.0$\pm$0.2 (5.3$\pm$0.3$\sigma$) \\ 3.0$\pm$0.3 (5.3$\pm$0.5$\sigma$) \\ 3.0$\pm$0.4 (5.3$\pm$0.7$\sigma$) \\ 3.0$\pm$0.6 (5.3$\pm$1.0$\sigma$) \\ 3.0$\pm$1.0 (4.8$\pm$1.1$\sigma$)} &
                            \makecell{3.0 (5.3$\sigma$) \\ 3.0$\pm$0.05 (5.3$\pm$0.2$\sigma$) \\ 3.0$\pm$0.06 (5.3$\pm$0.2$\sigma$) \\ 3.0$\pm$0.06 (5.3$\pm$0.2$\sigma$) \\ 3.0$\pm$0.1 (5.3$\pm$0.3$\sigma$) \\ 3.0$\pm$0.2 (5.3$\pm$0.5$\sigma$) \\ 3.0$\pm$0.4 (5.3$\pm$0.5$\sigma$) \\ 3.0$\pm$0.4 (5.3$\pm$0.7$\sigma$) \\ 3.0$\pm$0.7 (5.1$\pm$0.7$\sigma$) \\ 3.0$\pm$1.0 (4.3$\pm$1.0$\sigma$)} &
                            \makecell{3.0 (5.3$\sigma$) \\ 3.0$\pm$0.05 (5.3$\pm$0.3$\sigma$) \\ 3.0$\pm$0.06 (5.3$\pm$0.3$\sigma$) \\ 3.0$\pm$0.2 (5.3$\pm$0.3$\sigma$) \\ 3.0$\pm$0.5 (5.3$\pm$0.6$\sigma$) \\ 3.0$\pm$0.7 (5.3$\pm$0.8$\sigma$) \\ 3.0$\pm$0.9 (5.3$\pm$0.9$\sigma$) \\ 3.0$\pm$1.1 (5.2$\pm$1.0$\sigma$) \\ 3.0$\pm$1.4 (4.7$\pm$1.2$\sigma$) \\ 3.0$\pm$1.8 (3.9$\pm$1.2$\sigma$)} \\\cline{2-6}
                            & Aperiodic & 
                            \makecell{0 \\ 10 \\ 20 \\ 30 \\ 40 \\ 50 \\ 60 \\ 70 \\ 80 \\ 90} &
                            \makecell{3.0 (5.3$\sigma$) \\ 3.0$\pm$0.06 (5.3$\pm$0.2$\sigma$) \\ 3.0$\pm$0.06 (5.3$\pm$0.2$\sigma$) \\ 3.0$\pm$0.06 (5.3$\pm$0.2$\sigma$) \\ 3.0$\pm$0.1 (5.3$\pm$0.3$\sigma$) \\ 3.0$\pm$0.2 (5.3$\pm$0.4$\sigma$) \\ 3.0$\pm$0.4 (5.3$\pm$0.6$\sigma$) \\ 3.0$\pm$0.5 (5.3$\pm$0.8$\sigma$) \\ 3.0$\pm$0.7 (5.0$\pm$1.0$\sigma$) \\ 3.0$\pm$1.2 (4.2$\pm$1.0$\sigma$)} &
                            \makecell{3.0 (5.3$\sigma$) \\ 3.0$\pm$0.06 (5.3$\pm$0.2$\sigma$) \\ 3.0$\pm$0.06 (5.3$\pm$0.2$\sigma$) \\ 3.0$\pm$0.1 (5.3$\pm$0.3$\sigma$) \\ 3.0$\pm$0.1 (5.3$\pm$0.4$\sigma$) \\ 3.0$\pm$0.3 (5.3$\pm$0.4$\sigma$) \\ 3.0$\pm$0.4 (5.2$\pm$0.7$\sigma$) \\ 3.0$\pm$0.7 (5.2$\pm$0.9$\sigma$) \\ 3.0$\pm$0.9 (4.7$\pm$1.0$\sigma$) \\ 2.9$\pm$1.3 (3.8$\pm$1.1$\sigma$)} &
                            \makecell{3.0 (5.3$\sigma$) \\ 3.0$\pm$0.05 (5.3$\pm$0.2$\sigma$) \\ 3.0$\pm$0.06 (5.3$\pm$0.2$\sigma$) \\ 3.0$\pm$0.3 (5.3$\pm$0.5$\sigma$) \\ 3.0$\pm$0.4 (5.3$\pm$0.6$\sigma$) \\ 3.0$\pm$0.7 (5.3$\pm$0.8$\sigma$) \\ 3.0$\pm$0.9 (5.2$\pm$0.9$\sigma$) \\ 3.0$\pm$1.1 (5.0$\pm$1.0$\sigma$) \\ 3.0$\pm$1.5 (4.2$\pm$1.1$\sigma$) \\ 3.0$\pm$1.9 (3.5$\pm$1.1$\sigma$)} \\
\hline
\hline
\end{tabular}%
}
\end{table*}
\begin{table*}
\centering
\caption{Periodicity analysis results from the tests conducted using the LC of PG 1553+113 of Figure \ref{fig:example_use_cases}. The gap range explored in these tests spans from 0\% to 90\%. We use the gap scenario (``Random''), where the gaps are ejected randomly throughout the LC. 
The analysis methods used are the LSP, PDM, and SSA for two different window lengths (WL). We report the period in years and its significance. \label{tab:gap_pg1553}}
{%
\begin{tabular}{l|ccccccc}
\hline
\hline
Source & Type of Gaps & Gaps & LSP & PDM & SSA & SSA\\
 & & & & & WL=20\% & WL=40\% \\
\hline
		\multirow{3}{*}{PG 1553+113} & Random & 
                            \makecell{0 \\ 10 \\ 20 \\ 30 \\ 40 \\ 50 \\ 60 \\ 70 \\ 80 \\ 90} &
                            \makecell{2.1$\pm$0.1 (3.6$\sigma$) \\ 2.1$\pm$0.01 (3.6$\pm$0.2$\sigma$) \\ 2.1$\pm$0.01 (3.6$\pm$0.3$\sigma$) \\ 2.1$\pm$0.01 (3.6$\pm$0.4$\sigma$) \\ 2.1$\pm$0.1 (3.0$\pm$0.5$\sigma$) \\ 2.1$\pm$0.2 (2.7$\pm$0.7$\sigma$) \\ 2.1$\pm$0.3 (2.5$\pm$0.7$\sigma$) \\ 2.1$\pm$0.4 (2.0$\pm$0.8$\sigma$) \\ 2.1$\pm$0.4 (1.8$\pm$0.8$\sigma$) \\ 2.1$\pm$0.6 (1.3$\pm$0.8$\sigma$)} &
                            \makecell{2.2$\pm$0.2 (3.7$\sigma$) \\ 2.2$\pm$0.03 (3.7$\pm$0.2$\sigma$) \\ 2.2$\pm$0.03 (3.7$\pm$0.3$\sigma$) \\ 2.2$\pm$0.03 (3.6$\pm$0.3$\sigma$) \\ 2.2$\pm$0.1 (3.2$\pm$0.6$\sigma$) \\ 2.2$\pm$0.3 (2.6$\pm$0.7$\sigma$) \\ 2.2$\pm$0.3 (2.0$\pm$0.7$\sigma$) \\ 2.2$\pm$0.5 (1.8$\pm$0.7$\sigma$) \\ 2.2$\pm$0.5 (1.4$\pm$0.6$\sigma$) \\ 2.2$\pm$0.5 (1.1$\pm$0.6$\sigma$)} &
                            \makecell{2.2$\pm$0.2 (4.0$\sigma$) \\ 2.2$\pm$0.02 (4.0$\pm$0.2$\sigma$) \\ 2.2$\pm$0.02 (3.9$\pm$0.3$\sigma$) \\ 2.2$\pm$0.04 (3.6$\pm$0.5$\sigma$) \\ 2.2$\pm$0.1 (3.0$\pm$0.5$\sigma$) \\ 2.2$\pm$0.3 (2.3$\pm$0.8$\sigma$) \\ 2.1$\pm$0.5 (1.9$\pm$1.0$\sigma$) \\ 2.1$\pm$0.7 (1.8$\pm$1.0$\sigma$) \\ 2.0$\pm$0.8 (0.8$\pm$0.8$\sigma$) \\ 2.0$\pm$0.7 (0.7$\pm$0.8$\sigma$)} & 
                            \makecell{2.2$\pm$0.2 (4.4$\sigma$) \\ 2.2$\pm$0.02 (4.4$\pm$0.2$\sigma$) \\ 2.2$\pm$0.02 (4.3$\pm$0.3$\sigma$) \\ 2.2$\pm$0.04 (3.9$\pm$0.5$\sigma$) \\ 2.2$\pm$0.1 (3.5$\pm$0.7$\sigma$) \\ 2.2$\pm$0.2 (2.6$\pm$0.9$\sigma$) \\ 2.1$\pm$0.3 (2.2$\pm$1.3$\sigma$) \\ 2.1$\pm$0.6 (1.8$\pm$1.2$\sigma$) \\ 2.0$\pm$0.9 (0.8$\pm$0.8$\sigma$) \\ 2.0$\pm$0.7 (0.7$\pm$0.8$\sigma$)} \\\cline{2-6}
\hline
\hline
\end{tabular}%
}
\end{table*}

\clearpage

\subsection{Figures}
\begin{figure*}
	\centering
	\includegraphics[scale=0.24]      {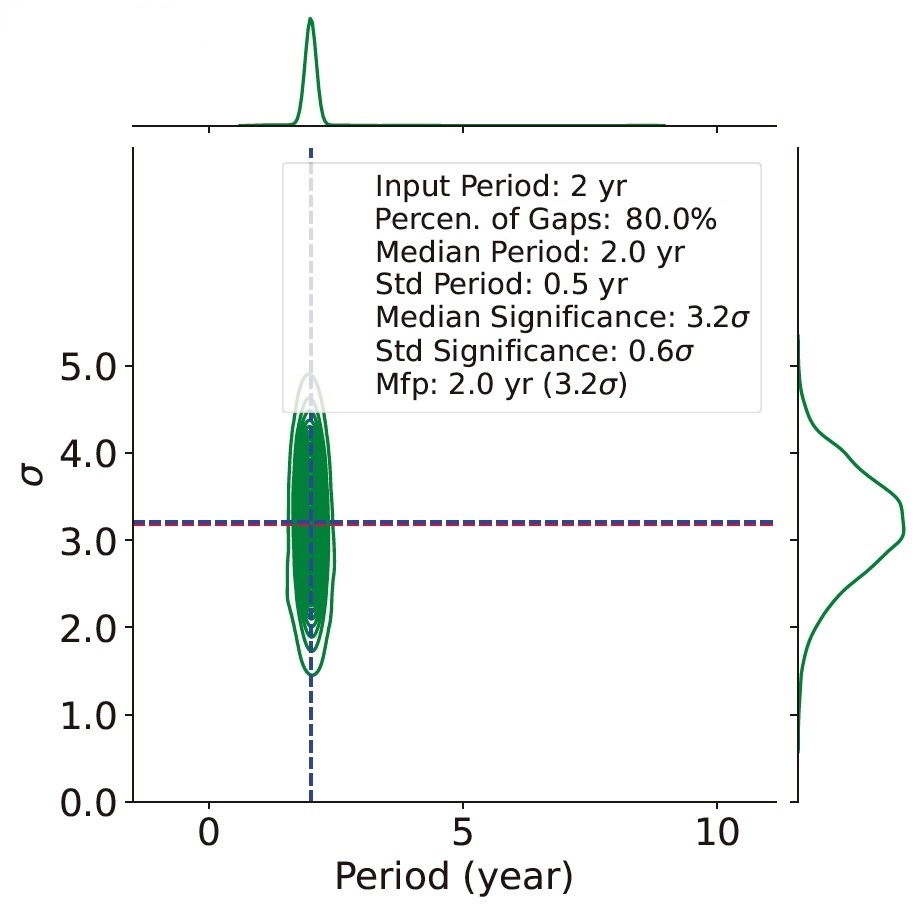}
        \includegraphics[scale=0.24]{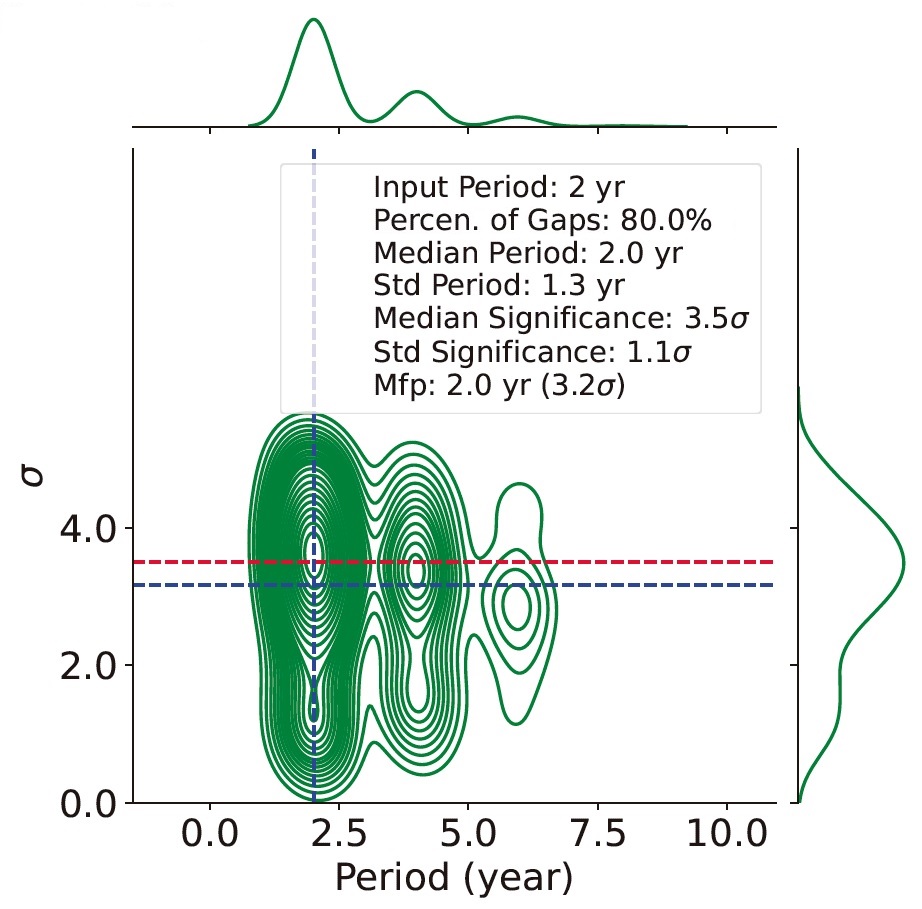}
        \includegraphics[scale=0.24]{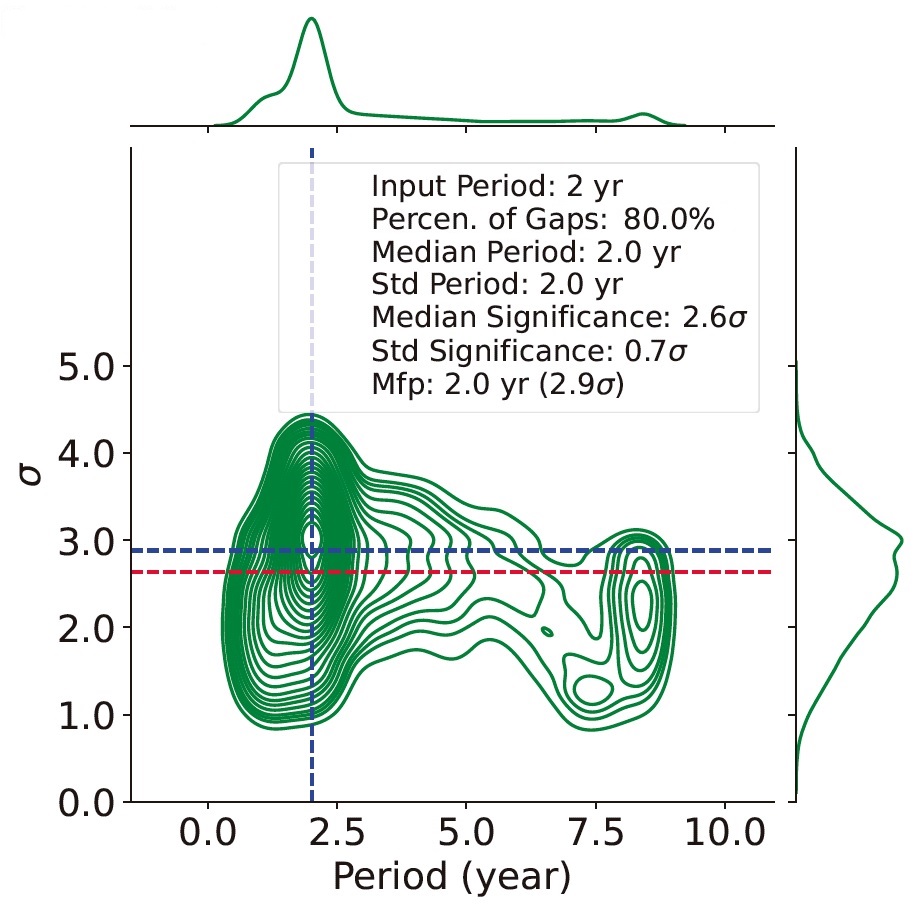}
        \includegraphics[scale=0.24]{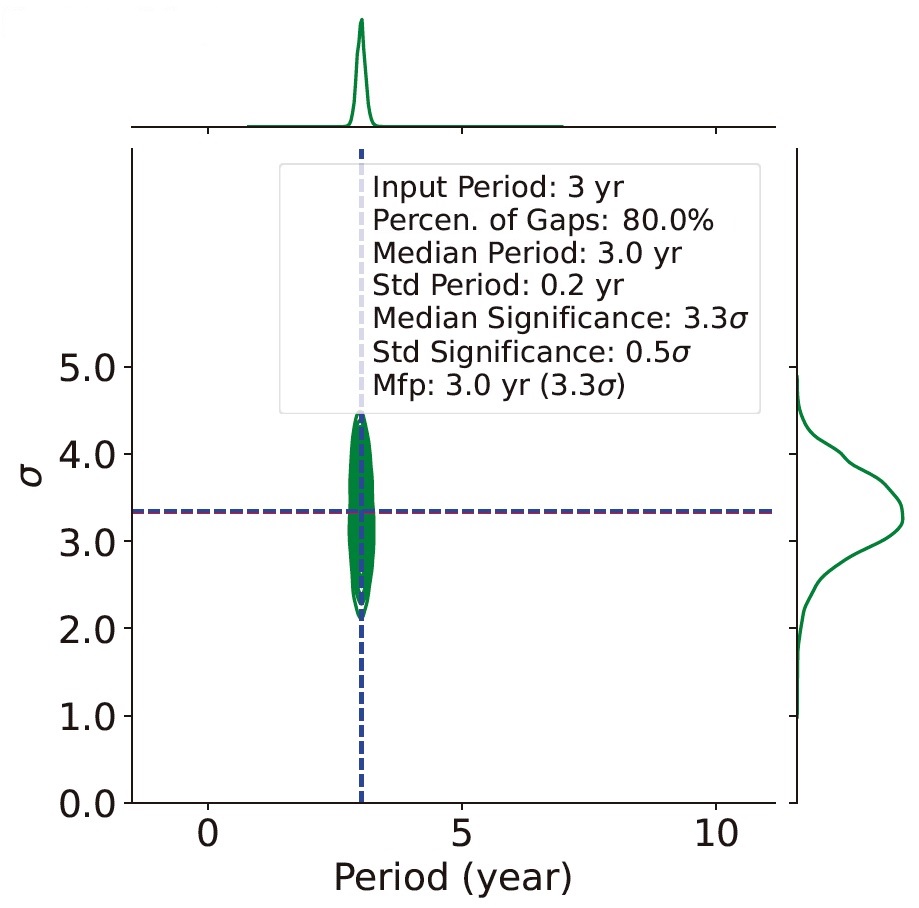}
        \includegraphics[scale=0.24]{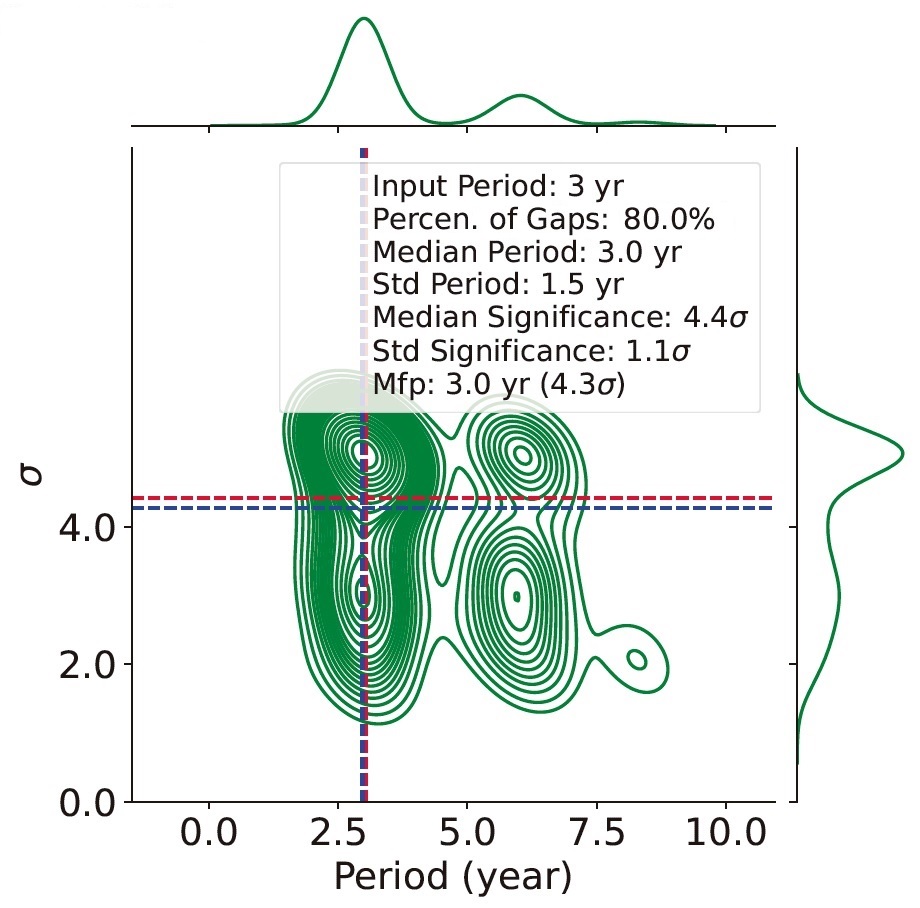}
        \includegraphics[scale=0.24]{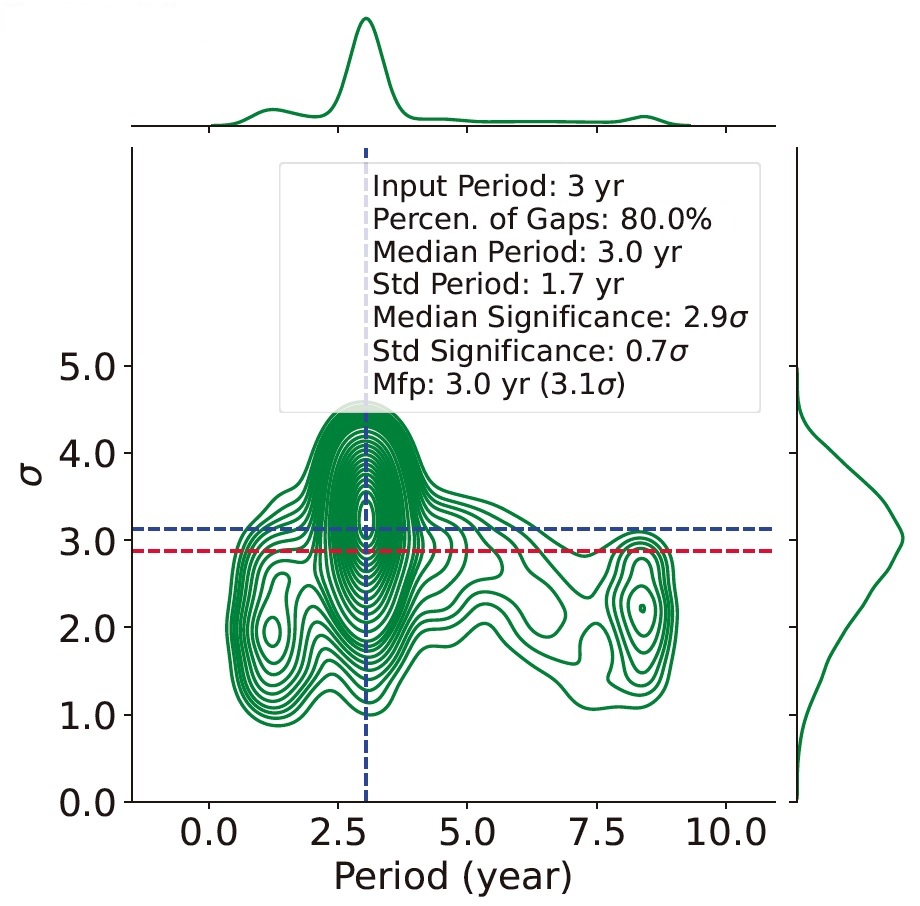}
        \caption{Examples of distributions period-significance for some test of a periodic signal. Top: 2-year periodic signal with a gap percentage of 80\%. \textit{Left}: LSP. \textit{Center}: PDM. \textit{Right}: SSA (40\% window lenght). Bottom: 3-year periodic signal with a gap percentage of 80\%. \textit{Left}: LSP. \textit{Center}: PDM. \textit{Right}: SSA (40\% window lenght).     
        The dotted red vertical and horizontal lines indicate the median values for both the period and the significance of the test. The blue dotted vertical line highlights the most frequently occurring period in the tests (and the associated significance), emphasizing its prominence in the distribution. The ``Percen. of Gaps'' refers to the percentage of gaps injected in the LC. The ``Median Period'' represents the median of all periods resulting from the test, and the ``Std Period'' is the standard deviation of such periods' distribution. The ``Median Significance'' represents the median of the significance distribution associated with the test, and the ``Std Significance'' is the standard deviation of this significance distribution. ``Mfp'' represents the most frequent period resulting from the test.} \label{fig:appendix_levels}
\end{figure*}

\begin{figure*}
	\centering
	\includegraphics[scale=0.23]{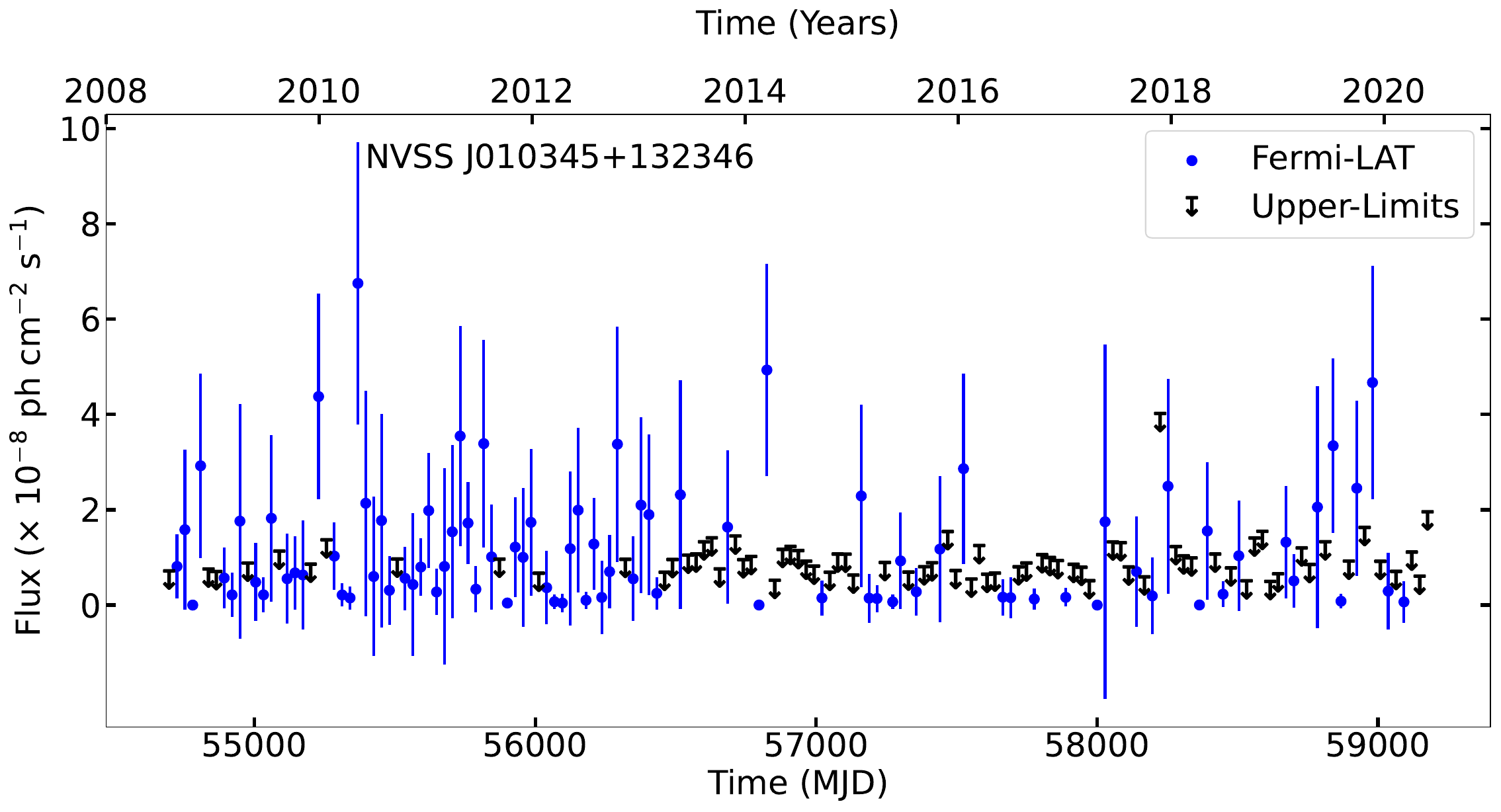} 
    \includegraphics[scale=0.23]{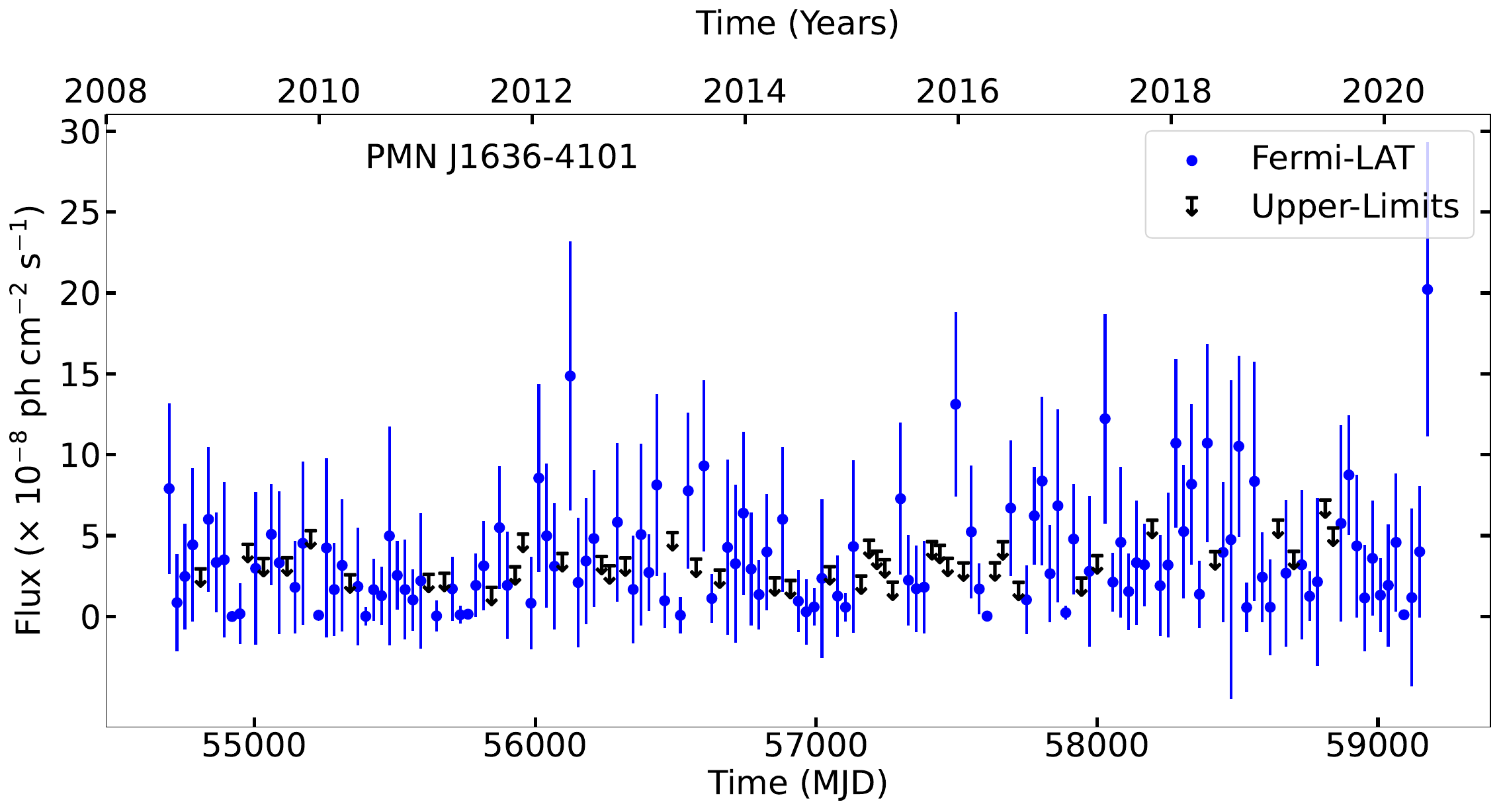}
    \caption{LCs of real data of NVSS J010345+132346 and PMN J1636$-$4101, soruces with a gap distribution in the 25$^{\mathrm{th}}\,\text{--}\,75^{\mathrm{th}}$ percentiles, corresponding to $\sim$20\%–50\% percentage of gaps, used for the test of $\S$\ref{sec:usecase}.}
\label{fig:appendix_lc_20_50}
\end{figure*}

In this section, we show some figures to support the findings presented in the paper. Figure \ref{fig:appendix_levels}  reports the LCs of blazars, and Figure \ref{fig:appendix_lc_20_50} shows the LCs of two use cases of $\S$\ref{sec:usecase}.

\bsp	
\label{lastpage}
\end{document}